\documentclass[11pt]{article}

\DeclareMathAlphabet{\scr}{U}{rsfs}{m}{n}

\pdfoutput=1
\usepackage{latexsym}
\usepackage{epsfig}
\usepackage[mathscr]{eucal}
\usepackage{amsfonts}
\usepackage{amscd}
\usepackage{cite}
\usepackage{array}   
\usepackage{amssymb}
\usepackage{colordvi}
\usepackage[centertags]{amsmath}
\usepackage{enumerate}
\usepackage{graphicx}
\usepackage{booktabs}
\usepackage{theorem}
\usepackage[footnotesize]{caption}
\usepackage{soul}
\usepackage{mcite}
\usepackage{slashed}
\usepackage{color}
\usepackage{ulem}
\usepackage{hyperref}
\usepackage{bbm}
\usepackage{multirow}
\usepackage[utf8]{inputenc}
\newcommand{\newc}{\newcommand}
\newc{\be}{\begin{equation}}
\newc{\ee}{\end{equation}}
\newc{\bea}{\begin{eqnarray}}
\newc{\eea}{\end{eqnarray}}
\newc{\ol}{\overline}
\newc{\wt}{\widetilde}
\newc{\bs}{\boldsymbol}
\newc{\m}{\mathcal}
\newc{\la}{\langle}
\newc{\ra}{\rangle}

\newcommand{\beq}{\begin{eqnarray}} 
\newcommand{\eeq}{\end{eqnarray}} 
\newcommand{\bpmatrix}{\begin{pmatrix}}
\newcommand{\epmatrix}{\end{pmatrix}}
\newcommand{\ba}{\begin{array}}
\newcommand{\ea}{\end{array}}

\renewcommand{\ol}{\text{1l}}

\renewcommand{\eqref}[1]{Eq.~(\ref{#1})}

\newcommand{\bc}{\begin{center}}
\newcommand{\ec}{\end{center}}

\newcommand{\gsim}{\raisebox{-0.13cm}{~\shortstack{$>$ \\[-0.07cm]
      $\sim$}}~}

\newcommand{\s}{\newline \vspace*{-3.5mm}}

\newcommand{\hone}{\ensuremath{H_1}}
\newcommand{\htwo}{\ensuremath{H_2}}
\newcommand{\hthree}{\ensuremath{H_3}}

\newcommand{\cone}{\ensuremath{c_{\alpha_1}}}
\newcommand{\ctwo}{\ensuremath{c_{\alpha_2}}}
\newcommand{\cthree}{\ensuremath{c_{\alpha_3}}}
\newcommand{\sone}{\ensuremath{s_{\alpha_1}}}
\newcommand{\stwo}{\ensuremath{s_{\alpha_2}}}
\newcommand{\sthree}{\ensuremath{s_{\alpha_3}}}
\newcommand{\done}{\ensuremath{\delta_{\alpha_1}}}
\newcommand{\dtwo}{\ensuremath{\delta_{\alpha_2}}}
\newcommand{\dthree}{\ensuremath{\delta_{\alpha_3}}}
\renewcommand{\sb}{\ensuremath{s_{\beta}}}
\newcommand{\cb}{\ensuremath{c_{\beta}}}
\newcommand{\sbd}{\ensuremath{s^2_{\beta}}}
\newcommand{\cbd}{\ensuremath{c^2_{\beta}}}

\newcommand{\tb}{\ensuremath{t_{\beta}}}

\newcommand{\lag}{\ensuremath{\mathcal{L}}}

\newcommand{\mzd}{\ensuremath{m_Z^2}}
\newcommand{\mw}{\ensuremath{m_W}}

\newcommand{\swd}{\ensuremath{s^2_W}}

\newcommand{\aone}{\ensuremath{\alpha_1}}
\newcommand{\atwo}{\ensuremath{\alpha_2}}
\newcommand{\athree}{\ensuremath{\alpha_3}}

\newcommand{\pstaronetwo}{\ensuremath{p^{2}_{\star,12}}}
\newcommand{\pstaronethree}{\ensuremath{p^{2}_{\star,13}}}
\newcommand{\pstartwothree}{\ensuremath{p^{2}_{\star,23}}}

\newcommand{\msbar}{\ensuremath{\overline{\text{MS}}}}

\begin{document}

\title{
\vspace*{-3cm}
\phantom{h} \hfill\mbox{\small KA-TP-27-2017}
\\[1cm]
\textbf{Gauge-independent Renormalization of the \\
 N2HDM \\[4mm]}}

\date{}
\author{
Marcel Krause$^{1\,}$\footnote{E-mail:
  \texttt{marcel.krause@kit.edu}} , 
David L\'opez-Val$^{1\,}$\footnote{E-mail:
  \texttt{david.val@kit.edu}},
Margarete M\"{u}hlleitner$^{1\,}$\footnote{E-mail:
  \texttt{margarete.muehlleitner@kit.edu}} ,
Rui Santos$^{2\,, \, 3\,, \, 4\,}$\footnote{E-mail:
  \texttt{rasantos@fc.ul.pt}} 
\\[9mm]
{\small\it
$^1$Institute for Theoretical Physics, Karlsruhe Institute of Technology,} \\
{\small\it 76128 Karlsruhe, Germany}\\[3mm]
{\small\it
$^2$ISEL - Instituto Superior de Engenharia de Lisboa,} \\
{\small \it  Instituto Polit\'ecnico de Lisboa,  1959-007 Lisboa, Portugal}\\[3mm]
{\small\it
$^3$Centro de F\'{\i}sica Te\'{o}rica e Computacional, Faculdade de Ci\^{e}ncias,} \\
{\small \it    Universidade de Lisboa, Campo Grande, Edif\'{\i}cio C8
   1749-016 Lisboa, Portugal} \\[3mm]
{\small \it
$^4$LIP, Departamento de F\'{\i}sica, Universidade do Minho, 4710-057 Braga, Portugal}
}
\maketitle

\begin{abstract}
The Next-to-Minimal 2-Higgs-Doublet Model (N2HDM) is an interesting
benchmark model for a Higgs sector consisting of two complex doublet
and one real singlet fields. Like the Next-to-Minimal Supersymmetric
extension (NMSSM) it features light Higgs bosons that could have
escaped discovery due to their singlet admixture. Thereby, the model
allows for various different Higgs-to-Higgs decay modes. Contrary to
the NMSSM, however, the model is not subject to supersymmetric relations
restraining its allowed parameter space and its phenomenology. For
the correct determination of the allowed parameter space, the correct
interpretation of the LHC Higgs data and the possible distinction of
beyond-the-Standard Model Higgs sectors higher order corrections to
the Higgs boson observables are crucial. This requires not only their
computation but also the development of a suitable renormalization
scheme. In this paper we have worked out the renormalization of the
complete N2HDM and provide a scheme for the gauge-independent renormalization
of the mixing angles. We discuss the renormalization of the $\mathbb{Z}_2$ soft
breaking parameter $m_{12}^2$ and the singlet vacuum expectation
value $v_S$. Both enter the Higgs self-couplings relevant for
Higgs-to-Higgs decays. We apply our renormalization scheme to 
different sample processes such as Higgs decays into $Z$ bosons and
decays into a lighter Higgs pair. Our results show that the
corrections may be sizeable and have to be taken into account for reliable
predictions. 
\end{abstract}
\thispagestyle{empty}
\vfill
\newpage
\setcounter{page}{1}


\tableofcontents

\section{Introduction}
Even after the discovery of the Higgs boson by the LHC
experiments ATLAS \cite{Aad:2012tfa} and CMS
\cite{Chatrchyan:2012ufa} there remain many open 
questions that cannot be solved within the Standard Model (SM). This calls
for New Physics (NP) extensions, which feature predominantly extended Higgs
sectors. The precise investigation of the Higgs sector has become an
important tool in the search for NP, in particular since its direct
manifestation through the discovery of new non-SM particles remains 
elusive. Among the beyond-the-SM (BSM) Higgs sectors those with
singlet and doublet extensions are particularly attractive as they are
at the same time rather simple and compatible with custodial symmetry. 
The 2-Higgs-doublet model (2HDM)
\cite{Gunion:1989we,Lee:1973iz,Branco:2011iw}  is interesting due to
its relation to supersymmetry and has been extensively studied and
considered as a possible benchmark model in experimental analyses. It
features 5 physical Higgs bosons, 2 CP-even and 1 CP-odd neutral
states and a charged Higgs pair. The next-to-minimal 2HDM (N2HDM) is
obtained upon extension of the 2HDM by a real singlet field with a $\mathbb{Z}_2$
parity symmetry. It contains in its symmetric phase a viable Dark
Matter (DM) candidate. The N2HDM has been the subject of numerous
investigations, both in its symmetric
\cite{He:2008qm,Grzadkowski:2009iz,Logan:2010nw,Boucenna:2011hy,He:2011gc,Bai:2012nv,He:2013suk,Cai:2013zga,Guo:2014bha,Wang:2014elb,Drozd:2014yla,Campbell:2015fra,Drozd:2015gda,vonBuddenbrock:2016rmr}
and in its broken phase
\cite{Chen:2013jvg,Muhlleitner:2016mzt, Muhlleitner:2017dkd}. The
Higgs sector of the latter consists after electroweak  
symmetry breaking (EWSB) of 3 neutral CP-even scalars, 1 pseudoscalar
and a charged Higgs pair. With the Higgs mass eigenstates being
superpositions of the singlet and doublet fields the N2HDM entails an
interesting phenomenology, namely the possibility of a light Higgs
boson, which is not in conflict with the experimental Higgs data in
case of a sufficiently large singlet admixture so that its couplings
to SM particles are suppressed. The enlarged Higgs sector together
with the possibility of light Higgs states allows for cascade Higgs-to-Higgs
decays that provide alternative production channels for the heavier
Higgs bosons and also give access to the trilinear Higgs
self-couplings. Their measurement provides important insights in the
understanding of the Higgs mechanism
\cite{Djouadi:1999gv,Djouadi:1999rca,Muhlleitner:2000jj}. \s

Obviously, any NP extension has to comply with the relevant
theoretical and experimental constraints. Thus, also the N2HDM has to
provide at least one Higgs boson with a mass of 125~GeV compatible with
the LHC data on the discovered Higgs resonance
\cite{{Aad:2015zhl}}. The additional Higgs 
bosons must not violate the LHC exclusion limits. The compatibility
with the electroweak (EW) precision data has to be guaranteed as well
as the compatibility with $B$-physics and low-energy constraints. The
symmetric N2HDM furthermore has to provide a DM candidate that
complies with the DM observables. From the theoretical point of view,
the N2HDM Higgs potential has to be bounded from below, its vacuum has
to be the global minimum and perturbative unitarity has to be
respected. In \cite{Muhlleitner:2016mzt}, part of our group
investigated the N2HDM in great detail with respect to these
constraints. The allowed parameter  space was determined and the
phenomenological implications were investigated. In the 
course of this work the model was implemented in {\tt HDECAY}
\cite{Djouadi:1997yw,Butterworth:2010ym}. The generated code, {\tt
  N2HDECAY} \cite{n2hdecay}, computes the N2HDM Higgs decay widths 
and branching ratios including the state-of-the-art higher order QCD
corrections and off-shell decays. The model was furthermore included in {\tt ScannerS} 
\cite{Coimbra:2013qq,scanners} along with the theoretical conditions and the available
experimental constraints, which then allowed to perform extensive
parameter scans for the model. In \cite{Muhlleitner:2017dkd}, the 
work was extended and we compared the N2HDM to other NP extensions
with the aim to work out observables that can be used to distinguish between various
well-motivated BSM Higgs sectors by using collider data. \s

Since the discovered Higgs bosons behaves very SM-like
\cite{Khachatryan:2014kca,Aad:2015mxa,Khachatryan:2014jba,Aad:2015gba},
the search for NP in the Higgs sector requires on the theoretical side
precise predictions for parameters and observables including
higher-order (HO) corrections. In the framework of the 2HDM, some of
the authors of this work provided an important basis for the 
computation of HO corrections in the 2HDM by working out a manifestly
gauge-independent renormalization of the two 2HDM mixing angles
$\alpha$ and $\beta$, which is also numerically stable and process
independent \cite{Krause:2016oke}. These angles, which diagonalise the
neutral CP-even and the neutral CP-odd or charged Higgs sectors, respectively, enter all
Higgs couplings so that they are relevant for Higgs boson phenomenology.  We
completed the renormalization of the 2HDM Higgs sector in \cite{Krause:2016xku} by
investigating Higgs-to-Higgs decays at EW next-to-leading order
(NLO). Subsequent works 
\cite{Denner:2016etu, Altenkamp:2017ldc, Kanemura:2017wtm} on the 2HDM renormalization applied different approaches and
renormalization conditions, confirming our findings where they
overlapped.\footnote{For the renormalization of
non-minimal Higgs sectors, see also \cite{Fox:2017hbw}.} The
renormalization of the N2HDM is more involved due to 
the additional mixing angles and the additional vacuum expectation value
related to the singlet field in the broken phase. One of our authors
worked on the renormalization of the SM extended by a real
singlet field, {\it cf.}~\cite{Bojarski:2015kra}. In this paper, we
combine our expertise gained in the 
renormalization of the 2HDM and the singlet-extended SM and provide
the complete renormalization of the N2HDM. The renormalization of the
mixing angles $\alpha_i$ ($i=1,2,3$) of the neutral sector and the
angle $\beta$ of the CP-odd/charged sector is manifestly
gauge independent as well as process independent. Where not
parametrically enhanced, it is furthermore numerically stable with
respect to missing higher order corrections. We will demonstrate this in the
numerical analysis where we explicitly compute the NLO EW corrections
to sample Higgs decays. We also use the occasion and clarify in this
paper the notion of the alternative tadpole approach with regard to the
renormalization framework applied to achieve a 
manifestly gauge-independent renormalization of the mixing
angles. With this paper we provide another important step in the
program of precise predictions for BSM Higgs sector parameters and
observables including higher order corrections, an indispensable
requisite for the correct interpretation of the experimental
results. \s

The paper is organised as follows: In Sec.~\ref{sec:model} we
introduce our model, set our notation and provide the relevant
couplings. Starting with Sec.~\ref{sec:renorm}, we describe the 
renormalization of the model. In Sec.~\ref{sec:tadpoles} we explain the way we
treat the tadpoles in our renormalization procedure, before we give
in Sec.~\ref{sec:renconditions} the renormalization
conditions. Section \ref{sec:oneloopdec} is dedicated to the computation of the
one-loop EW sample decay widths. In Sec.~\ref{sec:numerical} we
present our numerical analysis before we conclude in
Sec.~\ref{sec:concl}. The paper is accompanied by an extensive
appendix presenting the details of the computation of the pinched
self-energies in the N2HDM.

\section{Model setup \label{sec:model}}

\setcounter{equation}{0}
The N2HDM is obtained from the CP-conserving (or real) 2HDM with a
softly broken $\mathbb{Z}_2$ symmetry upon extension by a real singlet
field $\Phi_S$ with a discrete symmetry, under which $\Phi_S \to -
\Phi_S$. The kinetic term of the two $SU(2)_L$ Higgs
doublets $\Phi_1$ and $\Phi_2$ and the singlet field $\Phi_S$ is given by
\beq
{\cal L}_{\text{kin}} = (D_\mu \Phi_1)^\dagger (D^\mu \Phi_1) + 
(D_\mu \Phi_2)^\dagger (D^\mu \Phi_2) + \cfrac{1}{2}\,(\partial_\mu
\Phi_S)^2 \;,
\eeq
in terms of the covariant derivative
\beq
D_\mu = \partial_\mu + \frac{i}{2} g \sum_{a=1}^3 \tau^a W_\mu^a +
\frac{i}{2} g' B_\mu \;, \label{eq:covdiv}
\eeq
where $\tau^a$ denote the Pauli matrices, $W_\mu^a$ and $B_\mu$ the
$SU(2)_L$ and $U(1)_Y$ gauge bosons, respectively, and $g$ and $g'$ the
corresponding gauge couplings.
The scalar potential built from the
two $SU(2)_L$ Higgs doublets and the scalar singlet can be written as
\beq
V &=& m_{11}^2 |\Phi_1|^2 + m_{22}^2 |\Phi_2|^2 - m_{12}^2 (\Phi_1^\dagger
\Phi_2 + h.c.) + \frac{\lambda_1}{2} (\Phi_1^\dagger \Phi_1)^2 +
\frac{\lambda_2}{2} (\Phi_2^\dagger \Phi_2)^2 \nonumber \\
&& + \lambda_3
(\Phi_1^\dagger \Phi_1) (\Phi_2^\dagger \Phi_2) + \lambda_4
(\Phi_1^\dagger \Phi_2) (\Phi_2^\dagger \Phi_1) + \frac{\lambda_5}{2}
[(\Phi_1^\dagger \Phi_2)^2 + h.c.] \nonumber \\
&& + \frac{1}{2} m_S^2 \Phi_S^2 + \frac{\lambda_6}{8} \Phi_S^4 +
\frac{\lambda_7}{2} (\Phi_1^\dagger \Phi_1) \Phi_S^2 +
\frac{\lambda_8}{2} (\Phi_2^\dagger \Phi_2) \Phi_S^2 \;.
\label{eq:n2hdmpot}
\eeq
The first two lines correspond to the 2HDM part of the N2HDM, and the
last line contains the contribution of the singlet field $\Phi_S$. 
The potential is based on two $\mathbb{Z}_2$ symmetries, where the
first one is the trivial generalization of the usual 2HDM
$\mathbb{Z}_2$ symmetry to the N2HDM, 
\begin{align}
  \Phi_1 \to \Phi_1\;, \quad \Phi_2 \to - \Phi_2\;, \quad \Phi_S \to
  \Phi_S  \;. \label{eq:2HDMZ2} 
\end{align}
It is softly broken by the term involving $m_{12}^2$. Its extension to
the Yukawa sector ensures the absence of tree-level Flavour
Changing Neutral Currents (FCNC). 
The second $\mathbb{Z}_2$ symmetry on the other hand, under which
\begin{align}
  \Phi_1 \to \Phi_1\;, \quad \Phi_2 \to \Phi_2\;, \quad \Phi_S \to
  -\Phi_S \;, \label{eq:singZ2}
\end{align}
is not explicitly broken. 
After EWSB the neutral components of the Higgs fields develop
vacuum expectation values (VEVs), which are real in the CP-conserving
case. Expanding the elementary field excitations 
around the doublet VEVs $v_1$ and $v_2$ and the singlet VEV $v_S$, we
may write 
\beq
\Phi_1 = \left( \begin{array}{c} \phi_1^+ \\ \frac{1}{\sqrt{2}} (v_1 +
    \rho_1 + i \eta_1) \end{array} \right) \;, \quad
\Phi_2 = \left( \begin{array}{c} \phi_2^+ \\ \frac{1}{\sqrt{2}} (v_2 +
    \rho_2 + i \eta_2) \end{array} \right) \;, \quad
\Phi_S = v_S + \rho_S \;, \label{eq:n2hdmfields}
\eeq
where the field content of the model is parametrised in terms of the
charged complex fields $\phi_i^+$ ($i=1,2$), the real neutral
CP-even fields $\rho_1, \rho_2, \rho_3 \equiv \rho_S$ and the CP-odd
fields $\eta_i$. 
The minimisation conditions of the Higgs potential,
\beq
\left\langle\frac{\partial V}{\partial \Phi_1}\right\rangle = 
\left\langle\frac{\partial V}{\partial \Phi_2}\right\rangle =
\left\langle\frac{\partial V}{\partial \Phi_S}\right\rangle = 0 \;,
\label{eq:tadcond} 
\eeq
where the brackets denote the vacuum state, require the terms linear
in the Higgs fields, the tree-level Higgs tadpole parameters $T_i$
($i=1,2,3$), to vanish in the vacuum. Equation (\ref{eq:tadcond})
leads to the three minimum conditions 
\beq
\left\langle\frac{\partial V}{\partial \Phi_1}\right\rangle \equiv
\cfrac{T_1}{v_1}  &=& -\frac{v_2}{v_1} m_{12}^2 + m_{11}^2 + 
\frac{1}{2} (v_1^2 \lambda_1 + v_2^2 \lambda_{345} + v_S^2 \lambda_7)
= 0
 \label{eq:n2hdmmin1} \\ 
\left\langle\frac{\partial V}{\partial \Phi_2}\right\rangle \equiv
\cfrac{T_2}{v_2} &=& -\frac{v_1}{v_2} m_{12}^2 + m_{22}^2  + 
\frac{1}{2} (v_1^2 \lambda_{345} + v_2^2 \lambda_2 + v_S^2
\lambda_8) = 0 \label{eq:n2hdmmin2} \\ 
\left\langle\frac{\partial V}{\partial \Phi_S}\right\rangle \equiv
\cfrac{T_3}{v_S} &=&  m_S^2  + \frac{1}{2} (v_1^2 \lambda_7 + v_2^2 
\lambda_8 + v_S^2 \lambda_6) = 0\;, \label{eq:n2hdmmin3}
\eeq
with
\beq
\lambda_{345} \equiv \lambda_3 + \lambda_4 + \lambda_5 \;.
\label{eq:l345}
\eeq
At lowest order, the three tadpole conditions can be used to trade the
mass terms $m^2_{11}, \, m^2_{22}$ and $m_S^2$ in favor of the 
other parameters of the
potential. However, non-vanishing tadpole  contributions are relevant
at higher orders and must be included in the renormalization 
procedure, this being the reason why we shall retain them in our notation. 
The mass matrices of the Higgs fields in the gauge basis are obtained
from the second derivatives with respect to
these fields after replacing the doublet and singlet fields in the Higgs
potential by 
the parametrisations (\ref{eq:n2hdmfields}). Due to charge and CP
conservation the $7\times 7$ mass matrix decomposes into three blocks.
These are given by $2\times 2$ matrices for the charged and the CP-odd
fields, respectively, and a $3\times 3$ matrix for the CP-even states.
The former two are identical to the 2HDM case and read
\beq
M_\eta ^2 &=& \left( \frac{m_{12}^2}{v_1v_2} - \lambda _5
        \right) \begin{pmatrix} v_2^2 && -v_1v_2 \\ -v_1v_2 &&
          v_1^2 \end{pmatrix} + \begin{pmatrix} \frac{T_1}{v_1} && 0
          \\ 0 && \frac{T_2}{v_2} \end{pmatrix}  
\label{eq:etamatrix} \\
	M_{\phi^\pm} ^2 &=& \left( \frac{m_{12}^2}{v_1v_2} - \frac{\lambda
            _4 + \lambda _5}{2} \right) \begin{pmatrix} v_2^2 &&
          -v_1v_2 \\ -v_1v_2 && v_1^2 \end{pmatrix} + \begin{pmatrix}
          \frac{T_1}{v_1} && 0 \\ 0 && \frac{T_2}{v_2} \end{pmatrix}
        \;,
\label{eq:charmatrix}
\eeq
where we have kept explicitly the dependence on the tadpole
parameters. They can be diagonalised as 
\beq 
D_\eta^2 &=& R(\beta) M_\eta^2 R^T(\beta) \label{eq:deta} \\
D_{\phi^\pm}^2 &=& R (\beta) M_{\phi^\pm}^2 R^T(\beta) \label{eq:dphi} \;,
\eeq
\noindent with the rotation matrix
\beq
R(\beta) = \left( \begin{array}{cc} c_\beta & s_\beta \\ -s_\beta &
    c_\beta \end{array} \right) \;,
\eeq
where we have introduced the abbreviations $\sin x \equiv s_x$ and
$\cos x \equiv c_x$. 
This yields the neutral CP-odd mass eigenstates, $G^0$
and $A$, and the charged mass eigenstates, $G^\pm$ and $H^\pm$,
respectively. The would-be Goldstone bosons $G^0$ and $G^\pm$ 
are massless. 
Due to the additional real singlet field, the CP-even neutral sector
differs from the 2HDM, now featuring a $3\times 3$ mass matrix. In
the basis $(\rho_1, \rho_2, \rho_3)$ it can be cast into the form
\begin{equation}
M_\rho^2 = \left( \begin{array}{ccc} \lambda_1 c_\beta^2 v^2 + t_\beta
    m_{12}^2 & \lambda_{345} c_\beta s_\beta v^2 - m_{12}^2 &
    \lambda_7 c_\beta v v_S \\ \lambda_{345} c_\beta s_\beta v^2 - m_{12}^2 &
    \lambda_2 s_\beta^2 v^2 + m_{12}^2/t_\beta & \lambda_8 s_\beta v
    v_S \\ \lambda_7 c_\beta v v_S & \lambda_8 s_\beta v v_S &
    \lambda_6 v_S^2 \end{array} \right)
    +  \left( \begin{array}{ccc}  \frac{T_1}{v_1} & 0 & 0 \\ 0 & \frac{T_2}{v_2} & 0 \\ 
               0 & 0 & \frac{T_3}{v_S}
              \end{array} \right)
  \; ,
\label{eq:neutralmassmatrix} 
\end{equation}
\noindent 
where $t_\beta$ stands for the ratio
\beq
t_\beta = \frac{v_2}{v_1}
\eeq
and $v$ is defined as
\beq
v^2 = v_1^2 + v_2^2 \;,
\eeq
with $v \approx 246$~GeV denoting the SM VEV. We have furthermore used 
Eqs.~(\ref{eq:n2hdmmin1})-(\ref{eq:n2hdmmin3}) to
trade the mass parameters $m_{11}^2$, $m_{22}^2$ and $m_S^2$ for $v$,
$t_\beta$ and $v_S$. It is diagonalised by the rotation matrix $R(\alpha_i)$,
which can be parametrised in terms of three mixing angles $\alpha_1$
to $\alpha_3$ as 
\beq
R(\alpha_i) &= 
\left( \begin{array}{ccc} 
c_{\alpha_1} c_{\alpha_2} & s_{\alpha_1} c_{\alpha_2} & s_{\alpha_2}\\
-(c_{\alpha_1} s_{\alpha_2} s_{\alpha_3} + s_{\alpha_1} c_{\alpha_3})
& c_{\alpha_1} c_{\alpha_3} - s_{\alpha_1} s_{\alpha_2} s_{\alpha_3}
& c_{\alpha_2} s_{\alpha_3} \\
- c_{\alpha_1} s_{\alpha_2} c_{\alpha_3} + s_{\alpha_1} s_{\alpha_3} &
-(c_{\alpha_1} s_{\alpha_3} + s_{\alpha_1} s_{\alpha_2} c_{\alpha_3})
& c_{\alpha_2}  c_{\alpha_3}
\end{array} \right) \;.
\label{eq:mixingmatrix}
\eeq
Without loss of generality the angles can be chosen in the range
\beq
- \frac{\pi}{2} \le \alpha_{1,2,3} < \frac{\pi}{2} \;.
\eeq
The mass eigenstates $H_1$, $H_2$ and $H_3$ are obtained from the
gauge basis $(\rho_1, \rho_2, \rho_3)$ as  
\beq
\left( \begin{array}{c} H_1 \\ H_2 \\ H_3 \end{array} \right) = R
\left( \begin{array}{c} \rho_1 \\ \rho_2 \\ \rho_3 \end{array} \right) \;,
\eeq
and the diagonal mass matrix $D^2_\rho$ is given by
\beq
D_\rho^2 = R(\alpha_i) M_\rho^2 R^T(\alpha_i) \equiv 
\mbox{diag}(m_{H_1}^2,m_{H_2}^2,m_{H_3}^2) \;.
\eeq
We use the convention where the mass eigenstates are ordered by ascending mass as
\beq
m_{H_1} < m_{H_2} < m_{H_3} \;.
\eeq
The full set of the N2HDM parameters is given by the parameters of the
N2HDM potential Eq.~(\ref{eq:n2hdmpot}), the VEVs and the free
parameters of the SM. We hence have the following set of free parameters in the
gauge basis of the N2HDM
\beq
\lambda_1, ..., \lambda_8 \;, \quad m_{11}^2 \;, \quad m_{22}^2 \;,
\quad m_S^2 \;, \quad m_{12}^2 \;, \quad v_1 \;, \quad v_2 \;, \quad v_S \;,
\quad g \;, \quad g' \;, \quad y_\Psi \;,
\eeq
where $y_\Psi$ denotes the Yukawa couplings. For the renormalization of
the model it is convenient to relate as many parameters as possible
to physical parameters, like for example masses and the electric
charge. This allows then to apply physical conditions in the
renormalization of the respective parameters. Furthermore, the minimum
conditions can be used to trade $m_{11}^2$, $m_{22}^2$ and $m_S^2$ for
the tadpole parameters $T_{1,2,3}$. Denoting by $m_\Psi$ the fermion
masses, by $m_W$ and $m_Z$ the $W$ and $Z$ boson masses, respectively, 
and by $e$ the electric charge, the 'physical' set of N2HDM parameters is given by
\beq
m_{H_{1,2,3}} \;, \; m_A \;, \; m_{H^\pm} \;, \; \alpha_1 \;
, \; \alpha_2 \; , \; \alpha_3 \; , \; T_1 \;, \; T_2 \;, \; T_3 \;,
\; m_{12}^2 \;, \; v_S \;, \; t_\beta \;, \; e \;, \; m_W, \; m_Z,
\; m_\Psi\;. \label{eq:n2hdminputpars}
\eeq
We will specify in the following sections how these parameters get
renormalized in our way of treating the tadpoles. 
Note also that later in our renormalization procedure we will express 
$v_S$ through a physical quantity that depends on it, given by a
Higgs-to-Higgs decay width.  \s

For the computation of the electroweak corrections to the Higgs
decays we need the Higgs couplings, which we briefly summarise here. 
\begin{table}[b!]
\begin{center}
 \begin{tabular}{cc}
\toprule
\multicolumn{2}{c}{$\kappa_{H_i VV}$} \\
\midrule
$H_1$ & $c_{\alpha_2} c_{\beta-\alpha_1}$ \\
$H_2$ & $-c_{\beta-\alpha_1} s_{\alpha_2} s_{\alpha_3} + c_{\alpha_3}
s_{\beta-\alpha_1}$ \\
$H_3$ & $-c_{\alpha_3} c_{\beta-\alpha_1} s_{\alpha_2} - s_{\alpha_3}
s_{\beta-\alpha_1}$ \\
\bottomrule
\end{tabular}
 \caption{Neutral CP-even Higgs $H_i$ couplings
   to the massive gauge bosons $V=W,Z$. \label{tab:gaugecoupn2hdm}}
\end{center}
\end{table}
Since the singlet field $\rho_3$ does not couple directly to the SM
particles, any change in the tree-level Higgs couplings with respect to the
2HDM is due to the mixing of the three neutral fields
$\rho_i$ ($i=1,2,3$). This means that any coupling not involving 
the CP-even neutral Higgs bosons remains unchanged compared to the
2HDM and can be found {\it 
  e.g.}~in \cite{Branco:2011iw}. Introducing the Feynman rules for the coupling of the Higgs fields $H_i$ to the massive gauge bosons $V\equiv W,Z$ via
\beq
i \, g_{\mu\nu} \, \kappa_{H_i VV} \, g_{H^{\text{SM}}VV}\,
H_i\,V^\mu\,V^\nu \;, \label{eq:gaugecoupdef} 
\eeq
where $g_{H^{\text SM} VV}$ denotes the SM Higgs coupling factor,
we obtain the effective couplings 
\beq
\kappa_{H_i VV} = c_\beta R_{i1} + s_\beta
R_{i2} \label{eq:n2hdmgaugecoup} \;.
\eeq
The SM coupling in terms of the gauge boson masses $m_W$ and $m_Z$,
the $SU(2)_L$ gauge coupling $g$ and the Weinberg angle
$\theta_W$, is given by
\beq
g^{\text SM}_{HVV} = \left\{ \begin{array}{ll} g m_W & \quad
    \mbox{for } V=W \\
 g m_Z/\cos\theta_W & \quad \mbox{for } V=Z \end{array} \right. \;.
\eeq
In Table~\ref{tab:gaugecoupn2hdm} we list the effective couplings
after replacing the $R_{ij}$ by their parametrisation in terms of the mixing angles. \s

\begin{table}[t!]
\begin{center}
\begin{tabular}{rccc} \toprule
& $u$-type & $d$-type & leptons \\ \midrule
type I & $\Phi_2$ & $\Phi_2$ & $\Phi_2$ \\
type II & $\Phi_2$ & $\Phi_1$ & $\Phi_1$ \\
lepton-specific & $\Phi_2$ & $\Phi_2$ & $\Phi_1$ \\
flipped & $\Phi_2$ & $\Phi_1$ & $\Phi_2$ \\ \bottomrule
\end{tabular}
\caption{The four Yukawa types of the $\mathbb{Z}_2$-symmetric 2HDM
  defined by the Higgs doublet that couples to each kind of fermion. \label{tab:types}}
\end{center}
\end{table}
In the Yukawa sector there exist four types of coupling structures
after extending the $\mathbb{Z}_2$ symmetry (\ref{eq:2HDMZ2}) to the
Yukawa sector to avoid tree-level FCNCs. They are the same as in the
2HDM and summarised in Table \ref{tab:types}.
\begin{table}[b!]
\begin{center}
  \begin{tabular}{rccc} \toprule
& $u$-type & $d$-type & leptons \\ \midrule
type I & $\frac{R_{i2}}{s_\beta}$
& $\frac{R_{i2}}{s_\beta}$ &
$\frac{R_{i2}}{s_\beta}$ \\
type II & $\frac{R_{i2}}{s_\beta} $
& $\frac{R_{i1}}{c_\beta} $ &
$\frac{R_{i1}}{c_\beta} $ \\
lepton-specific & $\frac{R_{i2}}{s_\beta}$
& $\frac{R_{i2}}{s_\beta}$ &
$\frac{R_{i1}}{c_\beta}$ \\
flipped & $\frac{R_{i2}}{s_\beta}$
& $\frac{R_{i1}}{c_\beta}$ &
$\frac{R_{i2}}{s_\beta}$ \\ \bottomrule
\end{tabular}
\caption{Coupling coefficients $\kappa_{H_i ff}$ of the Yukawa couplings of
   the N2HDM Higgs bosons $H_i$ as defined in
   Eq.~(\ref{eq:lyukn2hdm}). \label{tab:yukcoup}}
\end{center}
\end{table}
The CP-even $H_i$ Yukawa couplings can be derived from the N2HDM
Yukawa Lagrangian  
\beq
{\cal L}_Y = - \sum_{i=1}^3 \frac{m_f}{v} \kappa_{H_i ff} 
\bar{\psi}_f \psi_f H_i \:.
\label{eq:lyukn2hdm}
\eeq
The effective coupling factors $\kappa_{H_i ff}$ in
terms of the mixing matrix elements $R_{ij}$ and the mixing angle $\beta$
are provided in Table~\ref{tab:yukcoup}.
Replacing the $R_{ij}$ by their parametrisation in terms
of the $\alpha_i$ results in the effective coupling expressions given for
type I and II in Table~\ref{tab:rescaling-yukawas}. \s
\begin{table}[t!]
\begin{center}
  \begin{tabular}{cccc}
\multicolumn{4}{c}{Type I} \\ \toprule
$\kappa_{H_i ff}$ & $u$ & $d$ & $l$ \\ \midrule
$H_1$ & $(c_{\alpha_2} s_{\alpha_1} )/s_\beta$ & $(c_{\alpha_2}
s_{\alpha_1}) / s_\beta$ & $(c_{\alpha_2} s_{\alpha_1})/s_\beta$ \\
$H_2$ & $(c_{\alpha_1} c_{\alpha_3} - s_{\alpha_1} s_{\alpha_2}
s_{\alpha_3})/s_\beta$ & $(c_{\alpha_1} c_{\alpha_3}- s_{\alpha_1}
s_{\alpha_2} s_{\alpha_3})/s_\beta$ & $(c_{\alpha_1} c_{\alpha_3}-
s_{\alpha_1} s_{\alpha_2} s_{\alpha_3})/s_\beta$ \\
$H_3$ & $-(c_{\alpha_1} s_{\alpha_3} + c_{\alpha_3} s_{\alpha_1}
s_{\alpha_2} )/s_\beta$ & $-(c_{\alpha_1} s_{\alpha_3} + c_{\alpha_3}
s_{\alpha_1} s_{\alpha_2} )/s_\beta$ & $-(c_{\alpha_1} s_{\alpha_3} +
c_{\alpha_3} s_{\alpha_1} s_{\alpha_2}) /s_\beta$ \\ \bottomrule\\
\multicolumn{4}{c}{Type II} \\ \toprule
$\kappa_{H_i ff}$ & $u$ & $d$ & $l$ \\ \midrule
$H_1$ & $(c_{\alpha_2} s_{\alpha_1} )/s_\beta$ & $(c_{\alpha_1}
c_{\alpha_2}) / c_\beta$ & $(c_{\alpha_1} c_{\alpha_2})/c_\beta$ \\
$H_2$ & $(c_{\alpha_1} c_{\alpha_3} - s_{\alpha_1} s_{\alpha_2}
s_{\alpha_3})/s_\beta$ & $-(c_{\alpha_3} s_{\alpha_1}+ c_{\alpha_1}
s_{\alpha_2} s_{\alpha_3})/c_\beta$ & $-(c_{\alpha_3} s_{\alpha_1}+
c_{\alpha_1} s_{\alpha_2} s_{\alpha_3})/c_\beta$ \\
$H_3$ & $-(c_{\alpha_1} s_{\alpha_3} + c_{\alpha_3} s_{\alpha_1}
s_{\alpha_2} )/s_\beta$ & $(s_{\alpha_1} s_{\alpha_3} - c_{\alpha_1}
c_{\alpha_3} s_{\alpha_2} )/c_\beta$ & $(s_{\alpha_1} s_{\alpha_3} -
c_{\alpha_1} c_{\alpha_3} s_{\alpha_2}) /c_\beta$ \\ \bottomrule
\end{tabular}
\caption{The effective Yukawa couplings $\kappa_{H_i ff}$ of the N2HDM Higgs
   bosons $H_i$, as defined in Eq.~(\ref{eq:lyukn2hdm}), in type I and
   type II. \label{tab:rescaling-yukawas}}
\end{center}
\end{table}

For the $H_i$ couplings to the $Z$ boson and the pseudoscalar $A$ or
the Goldstone $G^0$ the Feynman rules read 
\begin{align}
\lambda_\mu (H_i ZA) &= \frac{\sqrt{g^2 + g'^2}}{2} (p_{H_i} -
p_A)_\mu \tilde{\kappa}_{H_i V H}
\;, \label{eq:couphiaz} \\
\lambda_\mu (H_i ZG^0) &= \frac{\sqrt{g^2 + g'^2}}{2} (p_{H_i} -
p_{G^0})_\mu \kappa_{H_i V V}
\;, \label{eq:couphigz} 
\end{align}
where $p_A$, $p_{G^0}$ and $p_{H_i}$ are the incoming four-momenta
of the pseudoscalar, the Goldstone boson and the $H_i$, respectively. The 
tilde over the coupling factor for the pseudoscalar indicates that it is not
an effective coupling in the sense introduced above, as it is not normalized to a
corresponding SM coupling, since there is no SM counterpart. The
Feynman rules for the $H_i$ couplings to the
charged pairs $W^\mp$ and $H^\pm$ or $G^\pm$ read
\begin{align}
\lambda_\mu (H_i W^\mp H^\pm) &= \pm  \frac{ig}{2} (p_{H_i} -
p_{H^\pm})_\mu \tilde{\kappa}_{H_i V H}\, \;, \label{eq:couphihw} \\
\lambda_\mu (H_i W^\mp G^\pm) &= \pm  \frac{ig}{2} (p_{H_i} -
p_{G^\pm})_\mu \kappa_{H_i V V}\, \;, \label{eq:couphigw}
\end{align}
where $p_{H^\pm}$ and $p_{G^\pm}$ denote the four-momenta of $H^\pm$
and $G^\pm$ and again all momenta are taken as incoming. The coupling
factors $\tilde{\kappa}_{H_i VH}$ are listed in Table~\ref{tab:couphivh}. \s
\begin{table}[b!]
\begin{center}
\begin{tabular}{cc}
\toprule
 & $\tilde{\kappa}_{H_i VH}$ \\ \midrule
$H_1$ & $- c_{\alpha_2} s_{\beta-\alpha_1}$ \\
$H_2$ &  $s_{\beta-\alpha_1} s_{\alpha_2} s_{\alpha_3} + c_{\alpha_3}
c_{\beta-\alpha_1}$ \\
$H_3$ & $c_{\alpha_3} s_{\beta-\alpha_1} s_{\alpha_2} - s_{\alpha_3}
c_{\beta-\alpha_1}$ \\ \bottomrule
\end{tabular}
\caption{The coupling factors $\tilde{\kappa}_{H_i VH}$ as defined in the
 Feynman rules Eqs.~(\ref{eq:couphiaz}) and (\ref{eq:couphihw})
 for the $H_i$ couplings to a pair of Higgs and gauge
 bosons. \label{tab:couphivh}}
\end{center} 
\end{table}

The trilinear Higgs self-couplings needed for the Higgs decays into a
pair of lighter Higgs bosons are quite lengthy. For their explicit
form, we refer the reader to the appendix of
Ref.~\cite{Muhlleitner:2016mzt}. \s

Note finally, that by letting $\alpha_1 \to \alpha + \pi/2$ and $\alpha_{2,3}
\to 0$, we obtain the limit of a 2HDM with an additional decoupled
singlet. By the shift $\pi/2$ the usual 2HDM convention is matched,
and $\alpha$ diagonalises the $2\times 2$ mass matrix in the CP-even
Higgs sector yielding the two CP-even mass eigenstates $h$ and $H$,
respectively, with $m_h \le m_H$ by convention. Hence,
\beq
\mbox{N2HDM } \rightarrow \mbox{ 2HDM } \; \Longleftrightarrow \;
\left\{ \begin{array}{lcl}
\alpha_1 &\to& \alpha +\frac{\pi}{2} \\
\alpha_2 &\to& 0 \\
\alpha_3 &\to& 0
\end{array} \right. \;. \label{eq:2hdmlimit}
\eeq

\section{Renormalization \label{sec:renorm}}
\setcounter{equation}{0}
The computation of the EW corrections to the Higgs decays involves
ultraviolet (UV) divergences. Decays with external charged
particles additionally induce infrared (IR) divergences. The UV divergences are
canceled by the renormalization of the parameters and wave functions
involved in the process. In the following we will present the
renormalization of the N2HDM Higgs sector. For the purpose of this
work we must deal with the renormalization of the electroweak and the
Higgs sectors. With the main focus being on the renormalization of the N2HDM
Higgs sector, in the sample decays presented in the numerical analysis
we do not include processes that require the treatment
of IR divergences or the renormalization of the fermion sector. 
Note also that we do not need to renormalize the gauge-fixing
Lagrangian since we choose to 
write it already in terms of renormalized fields and 
parameters~\cite{Ross:1973fp,Baulieu:1983tg,Santos:1996vt}.
In the renormalization of the N2HDM Higgs sector we closely
follow the procedure applied in the 2HDM renormalization of
Refs.~\cite{Krause:2016oke,Krause:2016xku}. 
There, for the first time, a gauge-independent 
renormalization has been worked out for the 2HDM mixing angles by
applying the treatment of the tadpoles of Ref.~\cite{Fleischer:1980ub},
which we call the  alternative tadpole scheme, in combination with the
pinch technique. The pinch technique allows to unambiguously extract the
gauge-parameter independent parts of the decay amplitude and in
particular of the angular counterterms. The N2HDM
encounters four mixing angles instead of only two in the 2HDM. This
leads to more complicated renormalization conditions compared
to the 2HDM, as will be shown below. Additionally, the pinched
self-energies needed in this renormalization program have to be
worked out explicitly for the N2HDM. This has been done here for the
first time. Since the formulae are quite lengthy, we defer them to 
App.~\ref{app:pinchself}, which is part of
App.~\ref{app:pinchtech} that is dedicated to the detailed
presentation of the pinch technique in the N2HDM. We hope our
results to be useful for further works on this subject in the future. \s

For the renormalization we replace the bare parameters $p_0$, that are
involved in 
the process and participate in the EW interactions, by the
renormalized ones, $p$, and the corresponding counterterms $\delta p$, 
\beq
p_0 = p + \delta p \;. \label{eq:barrenct}
\eeq
Denoting generically scalar and vector fields by $\Psi$, the fields
are renormalized through their field renormalization constants $Z_\Psi$
as
\beq
\Psi_0 = \sqrt{Z_\Psi} \Psi \;.
\eeq
Note that in case the different field components mix $Z_\Psi$ is a matrix. \s

\noindent
\underline{\it Gauge sector:} The counterterms to be introduced in the
gauge sector are independent of the Higgs sector under
investigation. For convenience of the reader and to set our notation,
we still repeat the necessary replacements here. The massive gauge
boson masses and the electric charge are replaced by\footnote{The
  quantities on the left-hand side are the bare ones, where for
  convenience we dropped the index '0'. The ones on the right-hand
  side are the renormalized ones plus the corresponding counterterms.}
\beq
m_W^2 &\to& m_W^2 + \delta m_W^2 \\
m_Z^2 &\to& m_Z^2 + \delta m_Z^2 \\
e &\to& (1+\delta Z_e) \, e \;.
\eeq
The gauge boson fields are renormalized by their field
renormalization constants $\delta Z$,
\beq
W^\pm &\to& \left( 1 + \frac{1}{2} \delta Z_{WW} \right) W^\pm \\
\left( \begin{array}{c} Z \\ \gamma \end{array} \right) &\to& 
\left( \begin{array}{cc} 1 + \frac{1}{2} \delta Z_{ZZ} & \frac{1}{2}
    \delta Z_{Z\gamma} \\ \frac{1}{2} \delta Z_{\gamma Z} & 1 +
    \frac{1}{2} \delta Z_{\gamma\gamma} \end{array} \right) 
\left( \begin{array}{c} Z \\ \gamma \end{array} \right) \;.
\eeq

\noindent
\underline{\it Fermion sector:} Although not needed in the
computation  
of our sample decay widths in the numerical analysis, for completeness
we also include the renormalization of the fermion sector. The
counterterms of the fermion masses $m_f$ are defined through
\beq
m_f \to m_f + \delta m_f \;.
\eeq
And the bare left- and right-handed fermion fields
\beq
f_{L/R} \equiv P_{L/R} f \quad \;, \quad \mbox{ with } \quad P_{L/R} =
(1\mp \gamma_5)/2 \;, \label{eq:projectors}
\eeq
are replaced by their corresponding
renormalized fields according to
\beq
f_{L/R} \to \left( 1 + \frac{1}{2} \delta Z^{L/R}_f \right) f_{L/R} \;.
\eeq

\noindent
\underline{\it Higgs sector:} The renormalization is performed in the
mass basis and the mass counterterms are defined through
\beq
m_{\Phi}^2 \to  m_\Phi^2 + \delta m_\Phi^2 \;.
\eeq
The field $\Phi$ stands generically for the N2HDM
Higgs mass eigenstates, $\Phi \equiv H_1, H_2, H_3, A, H^\pm$. The
replacement of the fields by the renormalized ones and their
counterterms differs from the 2HDM case only by the fact that the wave
function counterterm matrix in the CP-even neutral Higgs sector is now
a $3 \times 3$ instead of a $2 \times 2$ matrix. Hence,
\beq
\left( \begin{array}{c}  H_1 \\ H_2 \\ H_3 \end{array} \right) &\to&
\left( \begin{array}{ccc} 1 + \frac{1}{2} \delta Z_{H_1H_1} & \frac{1}{2}
    \delta Z_{H_1H_2} & \frac{1}{2}
    \delta Z_{H_1H_3} \\ \frac{1}{2} \delta Z_{H_2H_1} & 1 + \frac{1}{2}
    \delta Z_{H_2H_2} & \frac{1}{2}\delta Z_{H_2H_3} \\  \frac{1}{2}\,\delta Z_{H_3H_1} &
     \frac{1}{2}\,\delta Z_{H_3H_2} & 1 + \frac{1}{2}\,\delta Z_{H_3H_3}
    \end{array} \right) \left( \begin{array}{c}  H_1 \\ H_2 \\ H_3 \end{array} \right) \label{eq:renconst1} 
    \\[0.2cm]
\left( \begin{array}{c}  G^0 \\ A \end{array} \right) &\to&
\left( \begin{array}{cc} 1 + \frac{1}{2} \delta Z_{G^0 G^0} & \frac{1}{2}
    \delta Z_{G^0 A} \\ \frac{1}{2} \delta Z_{AG^0} & 1 + \frac{1}{2}
    \delta Z_{AA} \end{array} \right) \left( \begin{array}{c} G^0 \\
    A \end{array} \right) \label{eq:renconst2} \\[0.2cm]
\left( \begin{array}{c}  G^\pm \\ H^\pm \end{array} \right) &\to&
\left( \begin{array}{cc} 1 + \frac{1}{2} \delta Z_{G^\pm G^\pm} & \frac{1}{2}
    \delta Z_{G^\pm H^\pm} \\ \frac{1}{2} \delta Z_{H^\pm G^\pm} & 1 + \frac{1}{2}
    \delta Z_{H^\pm H^\pm} \end{array} \right) \left( \begin{array}{c} G^\pm \\
    H^\pm \end{array} \right)  \; .\label{eq:renconst3}
\eeq
And for the mixing angles we make the replacements
\beq
\alpha_{i} &\to& \alpha_i + \delta \alpha_i\:, \quad i = 1,2,3 \\
\beta &\to& \beta + \delta \beta \;.
\eeq
For the soft $\mathbb{Z}_2$-breaking mass parameter $m_{12}^2$, finally,
we replace
\beq
m_{12}^2 \to m_{12}^2 + \delta m_{12}^2 \;.
\eeq
The tadpoles vanish at leading order, but the terms linear in the
Higgs fields get loop contributions at higher orders. 
It must therefore be ensured that the correct
vacuum is reproduced also at higher orders. As outlined in the
following, there are two different approaches, depending on whether
one chooses the tadpoles or the VEVs to be renormalized. 
The tadpole parameters $T_i$ ($i=1,2,3$) and the VEVs
$v_{1,2,S}$ are correspondingly replaced by
\beq
T_i \to T_i + \delta T_i \;, \label{eq:tadpshift}
\eeq
or alternatively by
\beq
v_{1,2,S} \to v_{1,2,S} + \delta v_{1,2,S} \;.
\eeq

\section{Treatment of the tadpoles \label{sec:tadpoles}}
The renormalization conditions fix the finite parts of the
counterterms. Throughout this paper we will fix the renormalization constants for
the masses and fields through on-shell (OS) conditions. Using an OS scheme 
provides an unambiguous interpretation
of the bare parameters in the classical Lagrangian in terms of
physically measurable quantities. In Ref.~\cite{Krause:2016oke} it has
been shown that the renormalization of the 2HDM 
mixing angles requires special care. Schemes used in the
literature before, which are based on the definition of the
counterterms through off-diagonal wave function renormalization
constants and a naive treatment of the tadpoles, were shown to lead to
gauge-dependent quantities. In order to cure this problem, in
\cite{Krause:2016oke} for the first time a 
renormalization scheme has been worked out in which the angular
counterterms are explicitly gauge independent. This guarantees the
gauge independence of the decay amplitudes also in case the angular
counterterms are not defined via a physical scheme as given {\it
  e.g.}~by the renormalization through a physical process. The
renormalization scheme developed in \cite{Krause:2016oke}  is based
on the 
combination of the alternative tadpole scheme with the pinch
technique. The pinch technique allows for the extraction of the 
truly gauge-independent parts of the angular counterterms and requires
the use of the alternative tadpole scheme. \s

As alluded to above, we treat the tadpoles in the alternative tadpole
scheme in order to be able to define the angular (and also mass)
counterterm in a gauge-independent way. While this procedure has been
introduced in \cite{Krause:2016oke}, we take here the occasion to
explicitly pin down the differences between the standard and the
alternative tadpole scheme. This, in particular, also reveals how
these differences reflect in the renormalization of the singlet VEV. \s

\begin{figure}[tb]
\vspace*{-0.7cm}
\hspace*{-2.5cm}
\includegraphics[width=1\linewidth , trim = -10mm 2mm 2mm 10mm, clip]{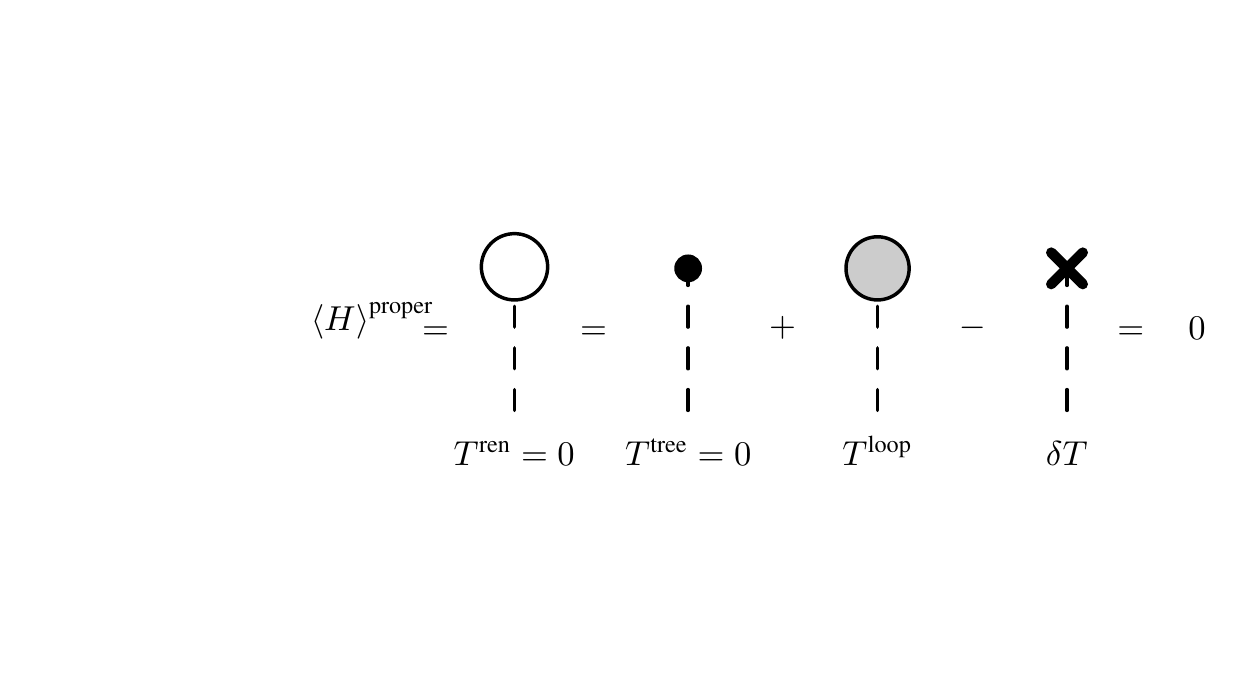}
\vspace*{-1.8cm}
\caption{Renormalization condition in the standard tadpole
  scheme: The requirement of the renormalized tadpole
  graph (white blob) to vanish together with the tree-level tadpole
  being zero fixes the tadpole counterterm.}
\label{fig:vevcondstand}
\end{figure}

The basic difference between the two schemes is the fact that in the
alternative scheme as introduced by Fleischer and Jegerlehner in
\cite{Fleischer:1980ub}, also referred to by 'FJ' in the following,
the VEVs are renormalized, while in the standard scheme the tadpole
parameters are renormalized. We call the {\it proper} VEV the
all-order Higgs vacuum expectation value $\langle \Phi \rangle = v
/ \sqrt{2}$. It represents the true ground state of the
theory and is connected to the particle masses and electroweak
couplings. At tree level the proper VEV and the bare VEV coincide
while at arbitrary loop orders the proper VEV corresponds to the renormalized VEV. 
In the alternative tadpole scheme the proper VEV coincides with the
tree-level VEV and hence is gauge-parameter independent. In this
scheme one renormalizes the VEV explicitly and its counterterm $\delta v$ is fixed
by ensuring the proper VEV to be $v/\sqrt{2}= v^{\text{tree}}/\sqrt{2}$ to all orders. This
renormalization condition yields $\delta v = T^{\text{loop}}/m_H^2$,
where $T^{\text{loop}}$ denotes the tadpole parameter at loop 
level. The condition generalises to multi-Higgs sectors, and we will
show below in the example of the N2HDM, how the renormalization
condition for the VEV counterterm is obtained. In practice, this
scheme is equivalent to inserting tadpole graphs explicitly in the calculations.
\begin{figure}[t!]
\vspace*{-0.7cm}
\hspace*{-2.5cm}
\includegraphics[width=1\linewidth , trim = -10mm 2mm 2mm 10mm,
clip]{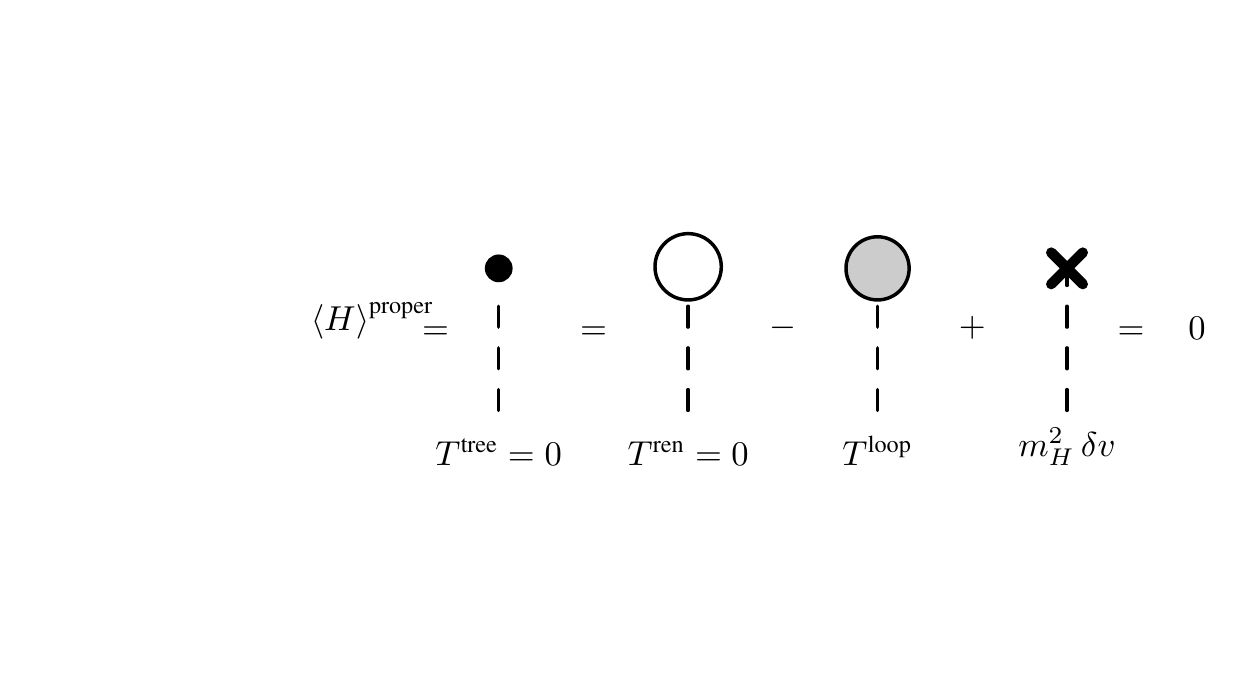}
\vspace*{-1.8cm}
\caption{Renormalization condition in the alternative tadpole
  scheme: With the neutral component $\Phi^0$ of the Higgs doublet
  $\Phi$ defined as $\Phi^0 = (v+H)/\sqrt{2}$, the requirement for the VEV to
  represent the true minimum of the Higgs potential translates into
  $\langle H \rangle^{\text{proper}} = 0$ or, equivalently, the renormalized tadpole
  graph (white blob) to vanish. The proper VEV coincides with the
  tree-level VEV (fixed by the condition $T^{\text{tree}}=0$). Together with the
  renormalization condition $T^{\text{ren}} = 0$, this relates the
  tadpole loop diagram (grey blob) at a given loop order to the VEV counterterm.}
\label{fig:vevcondalt}
\end{figure} 
Since at loop level the proper VEV is given by the renormalized one, and in
the FJ scheme coincides with the tree-level VEV, we have
\beq
\left.v^{\text{ren}}\right| _{\text{FJ}}  = v^\text{tree} = \left. \frac{2m_W}{g}\right|^{\text{tree}} \;.
\eeq
When a given $v$-dependent Lagrangian is used at
higher orders these tree-level parameters $\{ g,m_W \}^{\text{tree}}$
still have to be renormalized, and they are then replaced by their
corresponding renormalized parameters as 
\beq
\left. \frac{2m_W}{g}\right|^{\text{tree}} \to
\left. \frac{2m_W}{g}\right|^{\text{ren}}_{\text{FJ}} +
\underbrace{\left. \frac{2m_W}{g}\left( \frac{\delta m_W^2}{2m_W^2} - \frac{\delta
        g}{g} \right)\right|_{\text{FJ}}}_{\equiv \Delta v} \;.  \label{eq:doubletvevshift}
\eeq
It is important to note that $\Delta v$ is a mere label and not a
VEV counterterm as such. This makes obvious that $\delta v$ and $\Delta v$
are completely unrelated. In particular, they feature
a totally different divergence structure.  
Figure~\ref{fig:vevcondalt} depicts the renormalization
condition for the alternative tadpole scheme.
In the standard scheme, on the other hand, the proper VEV is obtained
from the minimisation of the gauge-dependent loop-corrected potential
and hence is in principle gauge dependent.
An equivalent condition to the FJ scheme requires the renormalized
tadpole to vanish. Together with the requirement of the tree-level
tadpole to be zero, this fixes the tadpole counterterm, which features
here explicitly, as the tadpole is an input parameter in the standard
scheme, {\it cf.}~Fig.~\ref{fig:vevcondstand}. \s 

For the singlet VEV $v_S$ a similar distinction, {\it i.e.}~$\delta
v_S$ versus $\Delta v_S$ has to be made. 
When $v_S$ is related to measurable parameters the NLO VEV shift $\Delta v_s$
denotes the corresponding combination of parameter counterterms,
similarly to ~\eqref{eq:doubletvevshift}. 
In
Ref.~\cite{Sperling:2013eva} it was shown that, in an 
$R_\xi$ gauge, a divergent part for $\Delta v_S$
in the standard scheme is precluded at one loop if 
the scalar field obeys a rigid invariance. This is the case for typical
singlet-extended Higgs sectors, {\it e.g.} the real singlet
model~\cite{Bojarski:2015kra}, and thereby  the N2HDM singlet
scalar. In all these cases the singlet field is disconnected from the
gauge sector and hence invariant under global gauge transformations. The
conclusion of Ref.~\cite{Sperling:2013eva} relies on the use of the
standard scheme, where the renormalized VEV coincides with the
loop-corrected one as the renormalized tadpoles are set to
zero\footnote{Let us also notice that 
  Ref.~\cite{Sperling:2013eva} 
  distinguishes two (equivalent) parametrisations for the
  renormalization transformation of a generic scalar field VEV,
  $\langle \Phi \rangle = \frac{v}{\sqrt{2}}$:\begin{alignat}{5} v \to v +
    \delta v = \sqrt{Z}_{\Phi}(v+\delta \overline{v}) \;,\end{alignat}
  where $\sqrt{Z_{\Phi}}$ is the field renormalization constant of the
  respective scalar field, whereas $\delta \overline{v}$ quantifies
  how the VEV itself is shifted differently by higher-order
  contributions with respect to the field. In our current conventions, 
$\delta v\to \Delta v$ and $\delta \overline{v}\to \Delta \overline{v}$.
The results of Ref.~\cite{Sperling:2013eva}, together with
\cite{Bojarski:2015kra}, show that for a gauge-singlet scalar
the quantity $\Delta v_s$ in the standard scheme is UV finite at one
loop order.}. However, this no longer applies
if the VEVs are renormalized in the alternative tadpole scheme.
In this case $\Delta v^{\text{FJ}}_S$ becomes indeed a UV-divergent quantity. 
We can prove it to cancel part of the UV poles that genuinely appear 
if one-loop amplitudes are computed in the FJ-scheme, when the corresponding
tree-level amplitudes are directly sensitive to the singlet VEV
$v_S$. Salient examples are the Higgs-to-Higgs decays,
which we discuss in detail in Section~\ref{sec:oneloopdec}. 

\subsection{Alternative tadpole scheme for the N2HDM \label{sec:altern}}
In the following, we elaborate in detail the implications of the
alternative tadpole scheme. We derive the necessary relations for the N2HDM,
highlighting the differences 
with respect to the 2HDM case, derived in \cite{Krause:2016oke}. At
tree level the minimum conditions of the N2HDM potential lead to the three relations
Eqs.~(\ref{eq:n2hdmmin1})-(\ref{eq:n2hdmmin3}) for the tadpole
parameters, or alternatively
\beq
T_{1}^{\text{tree}} = 0 \; , \quad  T_{2}^{\text{tree}} = 0 \quad
\mbox{and} \quad T_{3}^{\text{tree}} = 0 \;.
\label{eq:treecondt}
\eeq
These can be used to replace the parameters $m_{11}^2$,
$m_{22}^2$ and $m_S^2$ by the VEVs $v_1$, $v_2$ and $v_S$. Note,
however, that at arbitrary loop order, this may only be done
\textit{after} the \textit{proper VEVs} are taken into account in the
Higgs potential. More precisely, at NLO the
VEVs are modified in order to take into account the NLO effects,
as 
\beq
v_i^{\text{bare}} = v_i^{\text{ren}} + \delta v_i
\stackrel{\text{FJ}}{=} v_i^{\text{tree}} + \delta v_i
 \, , \quad i=1,2,S \;.
\label{eq:vevshift}
\eeq 
In the alternative tadpole scheme,  $\delta v_1$, $\delta v_2$ and $\delta v_S$
correspond to the proper doublet and singlet VEV counterterms in the gauge basis. In
turn, $v_{i}^{\text{ren}}$ are the \textit{proper} VEVs,
{\it i.e.}~in the FJ scheme the renormalized VEVs (coinciding with
the tree-level VEVs), and hence the VEVs
that generate the necessary mass relations for the gauge bosons,
fermions and the scalars. The VEVs are called the \textit{proper} VEVs
if the gauge-invariant relations presented in
Fig.\,\ref{fig:vevcondalt} (for the SM case) are
fulfilled at all orders, which means that the VEVs represent the true vacuum
state of the theory at all orders in perturbation theory. At NLO, we 
insert the relations Eq.~(\ref{eq:vevshift}) into the tadpole relations
Eqs.~(\ref{eq:n2hdmmin1})-(\ref{eq:n2hdmmin3}). At NLO, the left-hand side of
the equations is given by
\beq
T_{i}^{\text{bare}} = \underbrace{T_i^{\text{tree}}}_{=0} + T_i^{\text{loop}} =
T_i^{\text{loop}} \;, \quad
i\equiv 1,2,3\;.
\eeq
We then get the NLO expressions for
Eqs.~(\ref{eq:n2hdmmin1})-(\ref{eq:n2hdmmin3}), 
\beq
T_1^{\text{loop}} &=& T_{1}^{\text{tree}} + \bigg( m_{12}^2
\frac{v_{2}^{\text{tree}}}{v_{1}^{\text{tree}}} + \lambda_1
(v_{1}^{\text{tree}})^2 \bigg) \delta v_{1} + \bigg( - m_{12}^2 +
\lambda _{345} v_{1}^{\text{tree}} v_{2}^{\text{tree}} \bigg) \delta
v_{2} + \lambda_7 v_{1}^{\text{tree}} v_{S}^{\text{tree}}\,\delta
v_{S} \notag \\ \label{eq:delt1shift} \\  
T_2^{\text{loop}} &=& T_{2}^{\text{tree}} + \bigg( - m_{12}^2 +
\lambda _{345} v_{1}^{\text{tree}} v_{2}^{\text{tree}} \bigg) \delta
v_{1} + \bigg( m_{12}^2
\frac{v_{1}^{\text{tree}}}{v_{2}^{\text{tree}}} + \lambda _2
(v_{2}^{\text{tree}})^2 \bigg) \delta v_{2} + \lambda_8
v_{2}^{\text{tree}} v_{S}^{\text{tree}}\,\delta v_{S}  \notag \\ ~ \label{eq:delt2shift} \\
T_3^{\text{loop}}  &=& T_{3}^{\text{tree}} +  \lambda_6
(v_{S}^{\text{tree}})^2 \delta v_{S} + \lambda_7 v_{1}^{\text{tree}}
v_{S}^{\text{tree}} \delta v_{1}  + \lambda_8 v_{2}^{\text{tree}} v_{S}^{\text{tree}} \delta
v_{2}  
~. \label{eq:delt3shift}
\eeq
Since the NLO effects for the VEVs have been taken into account in
form of the counterterms in Eq.~(\ref{eq:vevshift}), the
FJ-renormalized VEVs $v_{i}^{\text{tree}}= v_i^{\text{ren}}$ now represent
the true ground states of the theory, namely those for which
$\langle \rho_i \rangle = 0$. The tree-level relations in
Eq.~(\ref{eq:treecondt}) can therefore be applied, and, in so doing,
the VEV counterterms $\delta v_1,$ $\delta v_2$ and $\delta v_S$ are
given in terms of the tadpole loops $T_1^{\text{loop}}$, $T_2^{\text{loop}}$ and
$T_3^{\text{loop}}$. 
By comparing with the squared mass matrix $M_\rho^2$ of
Eq.~(\ref{eq:neutralmassmatrix}) we find analytically
\beq
\left( 
\begin{array}{c} T_1^{\text{loop}} \\ T_2^{\text{loop}} \\ T_3^{\text{loop}}
\end{array} 
\right) \hspace*{-0.35cm}&= &\hspace*{-0.35cm}
\left( 
\begin{array}{ccc} m_{12}^2 \tb + \lambda_1\,(v^{\text{tree}})^2 \,
  \cb^2  & -m_{12}^2 +
  \lambda_{345}\,(v^{\text{tree}})^2\,\sb\cb
  &\lambda_7\cb\,v^{\text{tree}}\,v^{\text{tree}}_S   
\\
 -m_{12}^2 + \lambda_{345}\,(v^{\text{tree}})^2\,\sb\cb & m_{12}^2/\tb +
 \lambda_2\,(v^{\text{tree}})^2 \,\sb^2
  &\lambda_8\sb\,v^{\text{tree}}\,v^{\text{tree}}_S   
\\
\lambda_7\cb\,v^{\text{tree}}\, v^{\text{tree}}_S &  \lambda_8\sb\,v^{\text{tree}}\,
v^{\text{tree}}_S &  \lambda_6\,(v^{\text{tree}}_S)^2 
\end{array} \right) \hspace*{-0.2cm}\left( 
\begin{array}{c} \delta v_1 \\  \delta v_2 \\  \delta v_S  
\end{array} 
\right) \notag \\
\hspace*{-0.35cm}&=&\hspace*{-0.25cm} \mathcal{M}^2_{\rho}\Big{\lvert}_{T_i = 0}\, \left( 
\begin{array}{c} \delta v_1 \\  \delta v_2 \\  \delta v_S  
\end{array} 
\right)
\label{eq:vevshift-tadpole}\, .
\eeq
Rotation to the mass basis yields
\begin{equation}
 \left( 
\begin{array}{c} \delta v_{H_1} \\  \delta v_{H_2} \\  \delta v_{H_3}  
\end{array} 
\right)\, =
 \left( 
\begin{array}{c} \frac{T_{H_1}^{\text{loop}}}{m^2_{H_1}} \\  \frac{
  T_{H_2}^{\text{loop}}}{m^2_{H_2}} \\ \frac{T_{H_3}^{\text{loop}}}{m^2_{H_3}}
\end{array} \right) \;,
\label{eq:relmassbasis}
\end{equation}
where $T_{H_i}^{\text{loop}} = R(\alpha_i) T_i^{\text{loop}}$, and hence
\begin{equation}
 \left( 
\begin{array}{c} \delta v_1 \\  \delta v_2 \\  \delta v_S  
\end{array} 
\right) = R(\alpha_i)^T\,  \left( 
\begin{array}{c} \frac{T_{H_1}^{\text{loop}}}{m^2_{H_1}} \\
  \frac{T_{H_2}^{\text{loop}}}{m^2_{H_2}} \\ \frac{T_{H_3}^{\text{loop}}}{m^2_{H_3}}
\end{array} 
\right) 
\label{eq:vevshift-tadpole-mass}\, .
\end{equation}
\noindent The latter identity is helpful in practice, as the calculation of
the tadpole diagrams is usually performed in the mass basis, but the
VEV shifts are introduced most conveniently in the gauge basis.
Rewriting Eq.~(\ref{eq:relmassbasis}), the quantities $\delta v_{H_i}$ can
be interpreted as connected tadpole diagrams, containing the Higgs
tadpole and its propagator at zero momentum transfer,
\begin{equation}
\delta v_{H_i} = \frac{-i}{m_{H_i}^2} i T_{H_i}^{\text{loop}} =
\frac{-i}{m_{H_i}^2} ~ \mathord{
  \left(\rule{0cm}{30px}\right. \vcenter{
    \hbox{ \includegraphics[height=57px , trim = 18mm 12mm 17mm 10mm,
      clip]{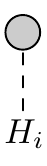} }
  } \left.\rule{0cm}{30px}\right) =
  \left(\rule{0cm}{30px}\right. \vcenter{
    \hbox{ \includegraphics[height=57px , trim = 17.5mm 12mm 14.6mm
      10mm,
      clip]{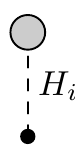}
    } } \left.\rule{0cm}{30px}\right) } ~.
\label{eq:vevshiftinterpret}
\end{equation}
\\
\noindent    
We want to emphasize again that in the alternative tadpole scheme 
Eq.~(\ref{eq:vevshiftinterpret}) defines the counterterms of the
vacuum expectation values. In contrast to the standard scheme, no
tadpole counterterms are introduced. Tadpole graphs appear through the
gauge-invariant condition in Fig.\,\ref{fig:vevcondalt}. \s

Once the leading-order VEVs are promoted to higher orders, namely by
inserting Eq.~(\ref{eq:vevshift}) into a generic VEV-dependent Lagrangian
${\cal L} (v_1,v_2,v_S)$, the contribution of the VEV counterterms
$\delta v_1$, $\delta v_2$ and $\delta v_S$, as given by
Eq.~(\ref{eq:vevshift-tadpole-mass}), is equivalent to introducing
explicit tadpole graphs in all loop amplitudes. Moreover, all
tree-level relations between the VEVs and the weak sector parameters
(masses, coupling constants) hold again. In particular, for the
doublet VEVs this means with ($v_1^2 + v_2^2 = v^2$)
\begin{alignat}{5}
\left. v^\text{ren} \right| _\text{FJ} = v^{\text{tree}} = \cfrac{2m_W}{g}\Big{\lvert}^{\text{tree}} 
 \label{eq:vevRelation}\, 
\end{alignat}
then
\beq
\left. v_1^\text{ren} \right| _\text{FJ} = v_1^{\text{tree}} = \cfrac{2m_W
  c_\beta}{g}\Big{\lvert}^{\text{tree}} \qquad
\mbox{and} \qquad
\left. v_2^\text{ren} \right| _\text{FJ} = v_2^{\text{tree}} = \cfrac{2m_W
  s_\beta}{g}\Big{\lvert}^{\text{tree}} \;.
\label{eq:vev12Relation}
\eeq
By applying the renormalization conditions for the VEVs, the
tree-level VEVs ensure the true ground state of the potential. Since
they are not directly related to a physical observable, we express the
FJ-renormalized doublet VEVs in terms of physical tree-level
parameters, here $m_W$, $g$ and the mixing angle $\beta$.
In higher order calculations, these parameters
  are then renormalized by choosing physical renormalization
  conditions.\footnote{We call the mixing angles \textit{physical} in the sense
  that they appear in the Higgs couplings and hence enter physical observables.}
To better illustrate the implications of the alternative tadpole
scheme, we consider the scalar-vector-vector vertex between the 
physical $H_1$ and a $W$ boson pair. We first define the Feynman
rules, needed in the following, by
\beq
H_1 W^\mu W^\nu &:& i g_{H_1 WW} \, g^{\mu\nu} \\
H_1 H_j W^\mu W^\nu &:& i g_{H_1 H_j WW} \, g^{\mu\nu}\:, \quad
j=1,2,3 \;.
\eeq 
The coupling constants for the triple vertex in terms of the mixing angles and the VEVs $v_1$ and $v_2$ are
\beq
g_{H_1 WW} &\equiv& g_{HWW}^{\text{SM}} \, \kappa_{H_1 WW} \nonumber \\
 &=&  g m_W
  c_{\alpha_2} c_{\beta 
    -\alpha_1} = \frac{g^2 v
  c_{\alpha_2} c_{\beta 
    -\alpha_1}}{2} = \frac{g^2 c_{\alpha_2}}{2} (c_{\alpha_1} v_1 + s_{\alpha_1} v_2) \;,
\label{eq:triple}
\eeq
and for the quartic vertices 
\beq
g_{H_1 H_1 WW} &\equiv& \kappa_{H_1 H_1 ZZ} \,g_{HWW}^{\text{SM}}=
\frac{g^2\,\ctwo^2}{2} \nonumber \\
g_{H_1 H_2 WW} &\equiv& \kappa_{H_1 H_2 ZZ} \,g_{HWW}^{\text{SM}}=
-\frac{g^2\,\ctwo\stwo\sthree}{2} \nonumber \\
g_{H_1 H_3 WW} &\equiv& \kappa_{H_1 H_3 ZZ} \,g_{HWW}^{\text{SM}}=
-\frac{g^2\,\ctwo\stwo\cthree}{2} \; . \label{eq:quartic}
\eeq
When expressing the couplings in terms of the VEVs, care has to be
taken to differentiate between the angle $\beta$ in the sense of a mixing angle
and $\beta$ in the sense of the ratio of the VEVs. Only the latter is
to be replaced by the VEVs that are to be renormalized. The same
distinction must be applied for the $\alpha_i$. Note that in all couplings but the trilinear
  and quartic Higgs self-couplings the angles $\alpha_i$ have the roles
  of mixing angles. Only in the Higgs self-couplings, the $\alpha_i$ partly appear in
  the sense of the ratio of N2HDM potential parameters.
Bearing these considerations in mind, we see that the quartic couplings do not
receive any $\delta v_i$, whereas $g_{H_1 WW}$ contains $\beta$ as
ratio of the VEVs. Instead, the angles $\aone$ and $\alpha_2$ are mixing angles
here. At NLO, we therefore have to make the replacement
\begin{equation}
\begin{split}
i g_{H_1 WW} ~ &= ~ \frac{i g^2 \ctwo}{2} (c_{\alpha_1} v_{1}^{\text{tree}} +
s_{\alpha_1} v_{2}^{\text{tree}}) + \frac{i g^2 \ctwo}{2} (c_{\alpha_1}
\delta v_1 + s_{\alpha_1} \delta v_2) \\
&\stackrel{(\ref{eq:vevshift-tadpole-mass})}{=} i g_{H_1 WW} +
\frac{ig^2 \ctwo}{2 } \left[\ctwo\,\cfrac{
    T_{H_1}^{\text{loop}}}{m^2_{H_1}}  - \stwo\sthree\,\cfrac{ T_{H_2}^{\text{loop}}}{m^2_{H_2}}
- \stwo\cthree\,\cfrac{ T_{H_3}^{\text{loop}}}{m^2_{H_3}}
\right]\\
&= i g_{H_1 WW} \\
&+ i g_{H_1H_1 WW} \, \left( \frac{-i}{m_{H_1}^2} \right) \, i
T_{H_1}^{\text{loop}}  + i g_{H_1H_2 WW} \, \left( \frac{-i}{m_{H_2}^2} \right) \, i
T_{H_2}^{\text{loop}}  
+ i g_{H_1H_3 WW} \, \left( \frac{-i}{m_{H_3}^2} \right) \, i
T_{H_3}^{\text{loop}}  
\\
&= i g_{H_1 WW} +  \mathord{ \left(\rule{0cm}{40px}\right.
 \hspace*{-0.5cm}  \vcenter{
   \hbox{ \includegraphics[height=80px,
      clip]{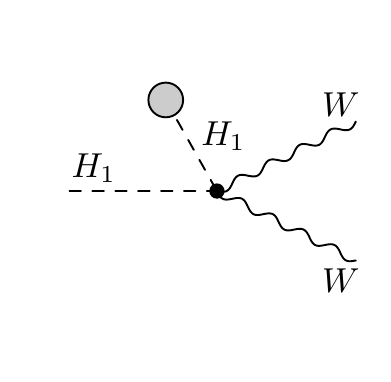} } } 
+
\vcenter{
    \hbox{ \includegraphics[height=80px, clip]{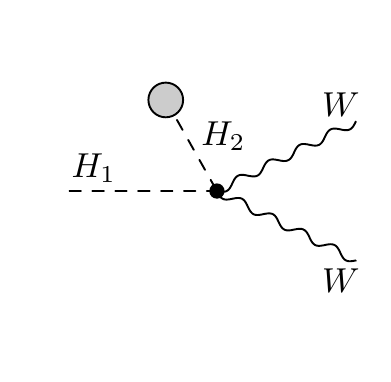} } }
+
\vcenter{
    \hbox{ \includegraphics[height=80px,
      clip]{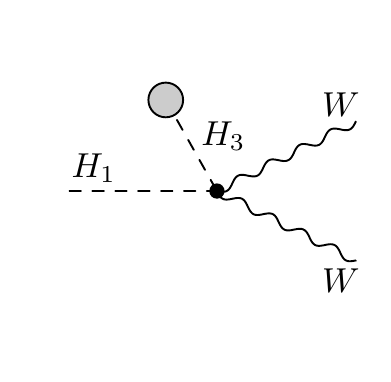} } }
   \left.\rule{0cm}{40px}\right)_\textrm{trunc}   
  }  \\
&\equiv i g_{H_1 WW}^{\textrm{tad}} 
\;.
\end{split}
\label{eq:vertexchange}
\end{equation}
The subscript 'trunc' means that all Lorentz structure of the
vector bosons as well as the Lorentz structure of the coupling has
been suppressed here for simplicity. The second term in the second
line generates, through the VEV counterterms $\delta v_i$, the tadpole
diagrams contributing to the scalar-vector-vector vertex. On the other
hand, as the VEVs in this expression have already been expanded to NLO
through $v_i \to v_{i}^{\text{tree}} + \delta v_i$, we use all tree-level
relations, in particular Eq.~(\ref{eq:vev12Relation}), to fix the
(FJ-renormalized) VEVs $v_{i}^{\text{tree}}$ in terms of the tree-level weak
sector parameters and the angle
  $\beta$.\footnote{Note, that since we use the tree-level
    relations, the angle $\beta$ in the sense of the ratio of the VEVs
    and in the sense of the mixing angle coincide.}
At loop level the EW parameters and
mixing angles that enter the coupling, here $g$, $m_W$,
$\beta$, $\alpha_1$ and $\alpha_2$ have to be renormalized, {\it i.e.}~we
  replace them by their renormalized values plus the corresponding 
  counterterms, {\it cf.}~Eq.~(\ref{eq:barrenct}). We then get for the
  vertex of Eq.~(\ref{eq:vertexchange}) 
\beq
ig_{H_1 WW}^{\text{tad}} + igm_W c_{\alpha_2} c_{\beta - \alpha_1}
\left[ \frac{\delta g}{g} + \frac{\delta m_W^2}{2 m_W^2} -
  t_{\alpha_2} \delta \alpha_2 - t_{\beta-\alpha_1} (\delta \beta -
  \delta \alpha_1) \right] \;.
\eeq
The exact form of these counterterms\footnote{Since the $SU(2)_L$
  coupling is not chosen to be an independent input parameter, it
  will be given in terms of the counterterms for $m_W$, $m_Z$ and $e$.}
depends on the renormalization 
conditions, which will be given in the next section. \s

Our derivation also shows the difference with respect to the 2HDM,
namely the last two terms in Eq.~(\ref{eq:vertexchange}) do not arise 
in the 2HDM. They are due to the additional 
singlet-doublet mixing and have no counterpart in a pure 2HDM structure
({\it cf.}~Eq.~(A.61) of \cite{Krause:2016oke}).  
\s

As a final remark, let us summarise the key differences with respect
to the standard tadpole scheme. In the latter case, VEV counterterms
of the form of Eq.~(\ref{eq:vevshiftinterpret}) are strictly speaking
not introduced. Instead, one introduces renormalized tadpoles and
tadpole counterterms, fulfilling the same condition as in
Fig.\,\ref{fig:vevcondalt} - that is, $T_i^{\text{ren}} = 0$ with
$T_i^{\text{ren}} = T_i^{\text{loop}} - \delta T_i$. In doing so,
the VEVs correspond to the ground state of the loop-corrected scalar
potential, and the corresponding VEV relations to weak sector
parameters hold order-by-order. Due to the fact that in the standard
tadpole scheme one considers the VEVs from the one-loop corrected
potential (in contrast to the alternative scheme, where one considers
the tree-level VEVs), VEV diagrams in the self-energies and vertices
explicitly vanish and thus need not be taken into account, at the
expense of defining mass counterterms which become manifestly
gauge dependent. 
\s

In practice, the rigorous introduction of the VEV counterterms in the
alternative tadpole scheme yields the following rules for its
application in the renormalization of a generic process within the
N2HDM: 

\begin{figure}[tb]
\centering
\includegraphics[width=\linewidth , trim = 19mm 13mm 1mm 7mm,clip]{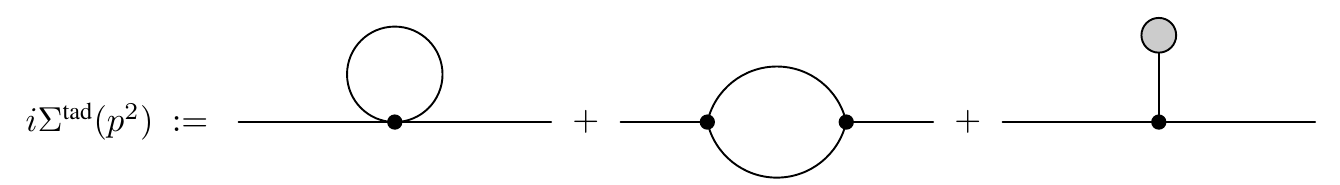} 
\caption{Modified self-energy $i\Sigma ^\textrm{tad} (p^2)$ in the
  alternative tadpole scheme, consisting of all 1 particle-irreducible (PI)
  self-energy diagrams together 
  with the one-loop tadpole diagrams, indicated by a gray blob.
} 
\label{fig:tadpoleselfen}
\end{figure}
\noindent
\begin{enumerate}
 \item Include explicit tadpole contributions in all self-energies
 used to define the (off-diagonal) wave function
renormalization constants\footnote{Diagonal wave function corrections,
  instead, are constructed from derivatives
of the corresponding self-energies with respect to $p^2$, hence the
tadpole-dependent contributions vanish.} and wherever the
self-energies appear in 
the counterterms, such that now $\Sigma^{\text{tad}} (p^2)$ contains
the additional tadpole contributions, {\it cf.} Fig.~\ref{fig:tadpoleselfen}.
\item Include explicit tadpole contributions in the virtual vertex corrections, 
if the tadpole insertions are connected to an existing coupling. This
is applicable \textit{e.g.}~to all triple Higgs self-interactions as
well as to the Higgs couplings to gauge bosons. 
\end{enumerate}

In the alternative tadpole scheme not only the angular counterterms but also the
mass counterterms become gauge independent. This has been shown for
the electroweak sector in \cite{Gambino:1999ai}. All counterterms of
the electroweak sector have exactly the same structure as in the standard 
scheme. Only the self-energies $\Sigma$ 
have to be replaced by the self-energies $\Sigma^{\text{tad}}$
containing the tadpole contributions. Note however, that there are no
tadpole contributions to the transverse photon-$Z$ self-energy
$\Sigma_{\gamma Z}^T$ nor to the transverse photon
  self-energy $\Sigma_{\gamma\gamma}^T$ so that 
\beq
\Sigma_{\gamma Z/\gamma\gamma}^{\text{tad},T} = \Sigma_{\gamma
  Z/\gamma\gamma}^T \;. 
\eeq
Having introduced the tadpole scheme, we now list explicitly the
counterterms needed in the computation of the electroweak
corrections. In particular, we illustrate the
renormalization of the N2HDM Higgs sector. \s

\section{Renormalization conditions \label{sec:renconditions}}
With the previous section we are now able to specify the counterterms
needed in the renorma\-li\-zation of the N2HDM. Those of the EW and Yukawa
sector correspond to the ones of the SM, while differences obviously
arise in the Higgs sector itself. For completeness, all
counterterms of the model will be listed, although not all of them will be
necessary to study the sample processes discussed in Section~\ref{sec:numerical}. 

\subsection{Counterterms of the gauge sector}
The gauge bosons are
renormalized through OS conditions implying the mass counterterms
\beq
\delta m_W^2 = \mbox{Re} \Sigma^{\text{tad},T}_{WW} (m_W^2) \quad
\mbox{and} \quad
\delta m_Z^2 = \mbox{Re} \Sigma^{\text{tad},T}_{ZZ} (m_Z^2) \;,
\eeq
where $T$ denotes the transverse part of the self-energy including the
tadpole contributions. The wave function renormalization constants
that guarantee the correct OS properties are given by
\beq
\delta Z_{WW} &=& - \mbox{Re} \left.\frac{\partial \Sigma^T_{WW}
  (p^2)}{\partial p^2}\right|_{p^2=m_W^2} \label{eq:sigmaww} \\
\left( \begin{array}{cc} \delta Z_{ZZ} & \delta Z_{Z\gamma} \\ \delta Z_{\gamma Z} &
  \delta Z_{\gamma\gamma} \end{array} \right) &=&
\left( \begin{array}{cc} - \mbox{Re} \left.\frac{\partial \Sigma^T_{ZZ}
  (p^2)}{\partial p^2}\right|_{p^2=m_Z^2} & 
2 \frac{\Sigma^T_{Z\gamma} (0)}{m_Z^2} \\
 -2 \mbox{Re} \frac{\Sigma^T_{Z\gamma} (m_Z^2)}{m_Z^2} & 
- \left.\frac{\partial \Sigma^T_{\gamma\gamma} (p^2)}{\partial
    p^2}\right|_{p^2=0} \end{array}\right) \;. \label{eq:sigmagamz}
\eeq
Note that in Eqs.~(\ref{eq:sigmaww}) and (\ref{eq:sigmagamz}) they are
the same in the standard and in the alternative 
tadpole scheme introduced above. The reason is that the tadpoles are
independent of the external momentum so that the derivatives of the
self-energies do not change. Furthermore, $\Sigma_{\gamma Z}^T$ is
identical in both schemes, as alluded to above. 
For better readability we therefore drop the superscript
'tad' here and wherever possible. For the same reasons the counterterm for the electric
charge is invariant with respect to the choice of the tadpole
scheme. The electric charge is renormalized to be the full
electron-positron photon coupling for OS external particles in the
Thomson limit. This implies that all corrections to this vertex vanish OS
and for zero momentum transfer. The counterterm for the electric
charge in terms of the transverse photon-photon and photon-Z self-energies reads 
\cite{Denner:1991kt}
\beq
\delta Z_e^{\alpha(0)} &=& \frac{1}{2} \left.\frac{\partial
  \Sigma^T_{\gamma\gamma} (k^2)}{\partial k^2}\right|_{k^2=0} +
\frac{s_W}{c_W} \frac{\Sigma_{\gamma Z}^T (0)}{m_Z^2} \;.
\label{eq:deltaze0}
\eeq
The sign in the second term of Eq.~(\ref{eq:deltaze0}) differs from
the one in \cite{Denner:1991kt}  because we have adopted different
sign conventions in the covariant derivative of
Eq.~(\ref{eq:covdiv}). In our computation we will use the
fine structure constant at the $Z$ boson mass $\alpha (m_Z^2)$ as
input. This way the results are independent of large logarithms due to
light fermions $f\ne t$. The counterterm $\delta Z_e$ is therefore modified as
\cite{Denner:1991kt} 
\beq
\delta Z_e^{\alpha (m_Z^2)} &=& \delta Z_e^{\alpha(0)} - \frac{1}{2}
\Delta \alpha (m_Z^2) \\
\Delta \alpha (m_Z^2) &=& \left. \frac{\partial
    \Sigma^T_{\gamma\gamma} (k^2)}{\partial k^2}\right|_{k^2=0} -
\frac{\Sigma^{T}_{\gamma\gamma} (m_Z^2)}{m_Z^2}
\;, \label{eq:delalphmz} 
\eeq
where the transverse part of the photon self-energy
$\Sigma^{T}_{\gamma\gamma}$ in
Eq.~(\ref{eq:delalphmz}) includes only the light fermion contributions. 
The calculation of the EW one-loop corrected Higgs decay widths also
requires the renormalization of the weak coupling $g$, which can be
related to $e$ and the gauge boson masses as
\beq
g &=& \frac{em_Z}{\sqrt{m_Z^2- m_W^2}} \;.
\eeq
Its counterterm can therefore be expressed in terms of the electric charge
and gauge boson mass counterterms through 
\beq
\frac{\delta g}{g} = \delta Z_e - \frac{1}{2(1-m_Z^2/m_W^2)}
\left( \frac{\delta m_W^2}{m_W^2} - \frac{\delta m_Z^2}{m_Z^2} \right) \;.
\eeq

\subsection{Counterterms of the fermion sector}
Defining the following structure for the fermion self-energies
\beq
\Sigma_f (p^2) = \slash{\!\!\! p} \Sigma^L_f (p^2) P_L + \slash{\!\!\!
  p} \Sigma^R_f
(p^2) P_R + m_f \Sigma^{Ls}_f (p^2) P_L + m_f \Sigma^{Rs}_f (p^2) P_R 
\eeq
the fermion mass counterterms applying OS conditions are given by
\beq
\frac{\delta m_f}{m_f} = \frac{1}{2} \mbox{Re} \left[ \Sigma^{\text{tad},L}_f
(m_f^2) + \Sigma^{\text{tad},R}_f (m_f^2) + \Sigma^{\text{tad},Ls}_f
  (m_f^2) + \Sigma^{\text{tad},Rs}_f 
(m_f^2) \right] \;.
\eeq
The fermion wave function renormalization constants are determined from
\beq
\delta Z^{L/R}_f &=& -\mbox{Re} \Sigma^{\text{tad},L/R}_f (m_f^2) 
\\
&&- m_f^2
\frac{\partial}{\partial p^2} \mbox{Re} \left.\left(
    \Sigma^{L/R}_f (p^2) + 
  \Sigma^{R/L}_f (p^2) + \Sigma^{L/Rs}_f (p^2) +
  \Sigma^{R/Ls}_f (p^2) 
\right)\right|_{p^2=m_f^2} \,. \nonumber
\eeq

\subsection{Higgs field and mass counterterms}
The OS conditions for the physical Higgs bosons yield the mass
counterterms ($i=1,2,3$) 
\beq
\delta m_{H_i}^2 &=& \mbox{Re} [\Sigma^{\text{tad}}_{H_iH_i}
(m_{H_i}^2)]  \\
\delta m_A^2 &=& \mbox{Re} [\Sigma^{\text{tad}}_{AA} (m_A^2)] \\
\delta m_{H^\pm}^2 &=& \mbox{Re} [\Sigma^{\text{tad}}_{H^\pm H^\pm}
(m_{H^\pm}^2)]  \;. \label{eq:scalarmasstadp}
\eeq
Having absorbed the tadpoles into the self-energies, no tadpole
counterterms appear explicitly in the mass counterterms any more, in
contrast to the corresponding expressions in the standard tadpole scheme. The
OS conditions for the Higgs bosons yield the following wave function
renormalization counterterm $3 \times 3$ matrix for the CP-even
neutral N2HDM scalars,
\beq
&\delta Z_{H_i H_j}
= \left( \begin{array}{ccc} - \mbox{Re} \left.\frac{\partial
        \Sigma_{H_1H_1} (k^2)}{\partial k^2}\right|_{k^2=m_{H_1}^2} &
    2 \frac{\mbox{Re} \left[\Sigma^{\text{tad}}_{H_1H_2} (m_{H_2}^2)\right]}{m_{H_1}^2-m_{H_2}^2} 
   &
    2 \frac{\mbox{Re} \left[\Sigma^{\text{tad}}_{H_1H_3} (m_{H_3}^2)\right]}{m_{H_1}^2-m_{H_3}^2}     
       \notag \\[0.3cm] 
2 \frac{\mbox{Re} \left[\Sigma^{\text{tad}}_{H_2H_1} (m_{H_1}^2)\right]}{m_{H_2}^2-m_{H_1}^2} & 
- \mbox{Re} \left.\frac{\partial \Sigma_{H_2H_2} (k^2)}{\partial k^2}\right|_{k^2=m_{H_2}^2} &  
 2 \frac{\mbox{Re} \left[\Sigma^{\text{tad}}_{H_2H_3} (m_{H_3}^2)\right]}{m_{H_2}^2-m_{H_3}^2} \\ 
2 \frac{\mbox{Re} \left[\Sigma^{\text{tad}}_{H_3H_1} (m_{H_1}^2)\right]}{m_{H_3}^2-m_{H_1}^2} &  2 \frac{\mbox{Re} 
\left[\Sigma^{\text{tad}}_{H_3H_2} (m_{H_2}^2)\right]}{m_{H_3}^2-m_{H_2}^2} & - \mbox{Re} \left.\frac{\partial
        \Sigma_{H_3H_3} (k^2)}{\partial k^2}\right|_{k^2=m_{H_3}^2}    
 \end{array} \right) \label{eq:wavefunc1tad} \\[0.2cm]
 \eeq
And in the CP-odd and charged sector we have the $2\times 2$ matrices
\beq
\left( \begin{array}{cc} \delta Z_{G^0 G^0} & \delta Z_{G^0 A} \\
  \delta Z_{A G^0} & \delta Z_{AA} \end{array} \right)
\hspace*{-0.2cm} &=& \hspace*{-0.2cm} 
\left( \begin{array}{cc} - \mbox{Re} \left.\frac{\partial
        \Sigma_{G^0 G^0} (k^2)}{\partial
        k^2}\right|_{k^2=0} & -2 \frac{\mbox{Re}
      \left[\Sigma^{\text{tad}}_{G^0 A} (m_A^2)\right] }{m_A^2}
  \\[0.3cm] 
2 \frac{\mbox{Re} \left[ \Sigma^{\text{tad}}_{G^0 A} (0) \right]}{m_A^2} & 
- \mbox{Re} \left.\frac{\partial \Sigma_{AA}
    (k^2)}{\partial k^2}\right|_{k^2=m_A^2} 
 \end{array} \right) \label{eq:wavefunc2tadp} \\[0.2cm]
\left( \begin{array}{cc} \delta Z_{G^\pm G^\pm} & \delta Z_{G^\pm H^\pm} \\
  \delta Z_{H^\pm G^\pm} & \delta Z_{H^\pm H^\pm} \end{array} \right) 
\hspace*{-0.2cm} &=& \hspace*{-0.2cm}
\left( \begin{array}{cc} - \mbox{Re} \left.\frac{\partial
        \Sigma_{G^\pm G^\pm} (k^2)}{\partial
        k^2}\right|_{k^2=0} & -2 \frac{\mbox{Re}
      \left[\Sigma^{\text{tad}}_{G^\pm H^\pm} (m_{H^\pm}^2)
      \right]}{m_{H^\pm}^2} \\[0.3cm] 
2 \frac{\mbox{Re} \left[ \Sigma^{\text{tad}}_{G^\pm H^\pm} (0)
  \right]}{m_{H^\pm}^2} & 
- \mbox{Re} \left.\frac{\partial \Sigma_{H^\pm H^\pm} (k^2)}{\partial
    k^2}\right|_{k^2=m_{H^\pm}^2} 
 \end{array} \right) 
\label{eq:wavefunc3tadp}
\eeq

\subsection{Angular counterterms}
As in the 2HDM, we renormalize the mixing angles based on the definition of the
counterterms through off-diagonal wave function renormalization
constants and combine this with the alternative tadpole approach
together with the application of the pinch technique in order to
arrive at an unambiguous gauge-independent definition of the mixing
angle counterterms. Let us note that a process-dependent
renormalization of the mixing 
angles would also lead to a gauge-independent renormalization, as shown
in~\cite{Krause:2016oke} for the 2HDM case. In the N2HDM the
situation becomes more involved as {\it four} different processes need
to be identified to fix all mixing angle counterterms $\delta
\alpha_i$ and $\delta \beta$. Moreover, the construction of such a
process-dependent scheme is complicated by the fact that the different Higgs decay
modes typically rely on more than one mixing angle, implying that
the different angular counterterms appear as linear combinations
in each individual vertex counterterm. It is therefore imperative to
choose a set of processes where the angular counterterm dependences enter as 
a linearly independent combination, such that they can be fixed unambiguously
through linear combinations of the different decay widths. Moreover, all
these processes have to be phenomenologically accessible. The
process-dependent renormalization of the N2HDM mixing angles is hence
rather unpractical from a physical point of view, and we will therefore 
not consider it any further. \s

While the expression for the counterterm in the charged
and CP-odd sector, $\delta \beta$, in terms of the off-diagonal wave
function renormalization constants does not change with respect to the
2HDM, this is not the case for 
the mixing angle counterterms $\delta \alpha_i$ in the CP-even sector. We 
therefore present their derivation here. It is based on the idea of making
the counterterms $\delta \alpha_i$ (and also $\delta \beta$) appear in
the inverse propagator matrix and thereby in the wave function
renormalization constants in a way that is consistent with the
internal relations of the N2HDM.\footnote{The renormalization of the mixing matrix in the scalar sector of a theory with an arbitrary number of scalars
was first discussed in \cite{pilaftsis1}} This can be achieved by performing the 
renormalization in the physical basis $(H_1, H_2, H_3)$, but
temporarily switching to the gauge basis $(\rho_1, \rho_2, \rho_3)$,
and back again. For the CP-even sector of the N2HDM this
means,
\beq
\left( \begin{array}{c} H_1 \\ H_2 \\ H_3 \end{array} \right)_{\text{bare}} 
&=& R(\alpha_i)\,\Bigg{\lvert}_{\text{bare}} \left( \begin{array}{c}
    \rho_1 \\ \rho_2 \\ 
 \rho_3 \end{array}\right)_{\text{bare}}\quad \rightarrow\qquad  
R(\alpha_i + \delta \, \alpha_i)\,\sqrt{Z_{\rho_i}}\,\left(
  \begin{array}{c} \rho_1 \\ 
\rho_2 \\ \rho_3 \end{array}\right) \notag \\ 
&=& \underbrace{R(\delta \alpha_i)\, R(\alpha_i)\,\sqrt{Z_{\rho_i}}\,
R(\alpha_i)^T}_{\sqrt{Z_{H_i}}}\, R(\alpha_i)\,
\left( \begin{array}{c} \rho_1 \\ \rho_2\\ \rho_3 \end{array} \right)  
= \sqrt{Z_{H_i}}\, \left( \begin{array}{c} H_1 \\ H_2 \\ H_3 \end{array}
                        \right) \label{eq:kosyscheme1}\, . \notag \\  
\eeq
The field renormalization matrix in the mass basis can be parametrised as
\beq
\sqrt{Z_{H_i}} &=& R(\delta \alpha_i) \left( \begin{array}{ccc} 1 +\frac{\delta Z_{H_1
H_1}}{2} & \delta C_{12} & \delta C_{13} \\ \delta C_{21} & 1 +\frac{\delta Z_{H_2
H_2}}{2} &  \delta C_{23} \\ \delta C_{31} & \delta C_{32} & 1 +\frac{\delta Z_{H_3 
H_3}}{2}\end{array}\right) = \nonumber \\
&& \hspace*{-2cm}\left(\begin{array}{ccc}  1+\cfrac{\delta Z_{H_1H_1}}{2} &  
\cone\cthree \delta \alpha_1 + \sthree \delta \alpha_2 + \delta C_{12}
            & \cthree\delta \alpha_2 - \sthree\ctwo\delta \alpha_1 + \delta C_{13} \\ 
 -\ctwo\cthree\delta \alpha_1 - \sthree\delta\alpha_2 + \delta C_{21} &
1+\cfrac{\delta Z_{H_2H_2}}{2} & \delta\alpha_3+ \stwo\delta \alpha_1
 + \delta C_{23}  \\
-\cthree \delta \alpha_2 + \sthree\ctwo\delta \alpha_1 + \delta C_{31} & 
-\delta \alpha_3 - \stwo\delta \alpha_1 + \delta C_{32}  &1+\cfrac{\delta
Z_{H_3H_3}}{2} \end{array}\right) \notag \\
 \label{eq:kosyscheme2}
\eeq
By identifying the off-diagonal elements with the off-diagonal wave
function renormalization constants $\delta Z_{H_i H_j}$ ($i\ne j$),
the three neutral CP-even angular counterterms are obtained as
\begin{alignat}{5}
 \delta \alpha_1 &= \cfrac{\cthree}{4\,\ctwo}\,(\delta Z_{H_1H_2} - \delta Z_{H_2H_1})
 - \cfrac{\sthree\,}{4\,\ctwo}\,(\delta Z_{H_1H_3} - \delta Z_{H_3H_1})
 \nonumber \\
 \delta \alpha_2 &= \cfrac{\cthree}{4}\,(\delta Z_{H_1H_3} - \delta Z_{H_3H_1})
 + \cfrac{\sthree\,}{4}\,(\delta Z_{H_1H_2} - \delta Z_{H_2H_1})
 \label{eq:neutralmixingCT} \\
 \delta \alpha_3 &= \cfrac{1}{4}\,(\delta Z_{H_2H_3} - \delta Z_{H_3H_2})
 + \cfrac{\stwo}{4\ctwo}\,\left[\sthree \left(\delta Z_{H_1H_3}-\delta Z_{H_3H_1}\right) - 
 \cthree \left(\delta Z_{H_1H_2}-\delta Z_{H_2H_1}\right)\right] 
 \nonumber \,,
\end{alignat}
while the auxiliary counterterms $\delta C_{ij}$ do not play a role in the
remainder of the discussion. \s

The definition of the counterterm $\delta \beta$ can be taken over
from the 2HDM. It is derived ana\-lo\-gously to the $\delta \alpha_i$,
but from the charged and CP-odd Higgs sectors. In this case, there are
altogether four off-diagonal wave function constants, while
only three free parameters to be fixed. For details, we refer
to~Ref.~\cite{Krause:2016oke}. There we proposed two different possible counterterm
choices for $\beta$, one based on the charged and the other on the
CP-odd sector. Also here we will apply these two possible choices,
given by 
\beq
\delta \beta^{(1)} = \frac{1}{4} (\delta Z_{G^\pm H^\pm} -  \delta
Z_{H^\pm G^\pm}) \label{eq:betact1}
\eeq
and
\beq
\delta \beta^{(2)} &=& \frac{1}{4} (\delta Z_{G^0 A} -  \delta Z_{A
  G^0}) \label{eq:betact2} \;.
\eeq
All wave function renormalization constants appearing in the
counterterms Eqs.~(\ref{eq:neutralmixingCT}), (\ref{eq:betact1}) and
(\ref{eq:betact2}) are renormalized in the OS scheme and given by the
corresponding entries in the wave function counterterm matrices
Eqs.~(\ref{eq:wavefunc1tad}), (\ref{eq:wavefunc2tadp}) and
(\ref{eq:wavefunc3tadp}). While the use of the alternative tadpole scheme ensures
that the angular counterterms can be expressed in a gauge-independent
way, at this stage they still contain a dependence on the gauge-fixing
parameter. We therefore combine 
the virtues of the alternative tadpole scheme with the pinch technique
\cite{Binosi:2004qe,Binosi:2009qm,Cornwall:1989gv,Papavassiliou:1989zd,Degrassi:1992ue,Papavassiliou:1994pr,Watson:1994tn,Papavassiliou:1995fq}. The 
pinch technique allows us to extract the truly gauge-independent parts
of the angular coun\-ter\-terms.

\subsubsection{Gauge-independent pinch technique-based angular
  counterterm schemes}
By the application of the pinch technique it possible to define
\textit{pinched} self-energies $\overline{\Sigma}$ which are truly
gauge independent. They are built up by the tadpole self-energies
evaluated in the Feynman gauge and extra pinched
components $\Sigma^{\text{add}}$, {\it i.e.}
\beq
\overline{\Sigma} (p^2) = \left.\Sigma^{\text{tad}}
  (p^2)\right|_{\xi_V=1} + \Sigma^{\text{add}} (p^2) \;,
\label{eq:sigadddef}
\eeq
where $\xi_V$ stands for the gauge fixing parameters $\xi_Z$, $\xi_W$ and
$\xi_\gamma$ of the $R_\xi$ gauge. By $\Sigma^{\text{add}}$ we dub the 
additional (explicitly $\xi_V$-independent) self-energy contributions
obtained via the pinch technique. 
It is important to notice that, in order to apply the pinch technique,
it is necessary to  explicitly include all tadpole topologies, {\it
  i.e.}~to use the alternative tadpole scheme. In
App.~\ref{app:pinchtech} we present the basic idea of the pinch
technique (see also~Refs.~\cite{Binosi:2004qe,Binosi:2009qm,Cornwall:1989gv,Papavassiliou:1989zd,Degrassi:1992ue,Papavassiliou:1994pr,Watson:1994tn,Papavassiliou:1995fq} for a detailed
exposition). We exemplarily show, for the
CP-even sector, how to proceed in the derivation of the pinched
self-energy. Additionally, we give useful formulae on the gauge dependences
of the scalar self-energies and for the application of the pinch
technique in the N2HDM. 

\paragraph{On-shell tadpole-pinched scheme}

The self-energy $\Sigma^{\text{add}}$  in Eq.~(\ref{eq:sigadddef}) is 
explicitly independent of the gauge fixing parameter 
$\xi_V$. By replacing the wave function renormalization constants in the
counterterms Eqs.~(\ref{eq:neutralmixingCT}), (\ref{eq:betact1}) and
(\ref{eq:betact2}) with their OS renormalization definitions given by
the corresponding entries in the wave function counterterm matrices
Eqs.~(\ref{eq:wavefunc1tad}), (\ref{eq:wavefunc2tadp}) and
(\ref{eq:wavefunc3tadp}) we arrive, upon expressing these in terms of
the pinched self-energies, at the following expressions for the angular
counterterms $\delta \alpha_i$,
\begin{alignat}{5}
\done &= \cfrac{\cthree}{2\ctwo}\,\cfrac{\mbox{Re}\,\left(\left[\Sigma^{\text{tad}}_{H_1H_2}(m^2_{H_2})+\Sigma^{\text{tad}}_{H_2H_1}(m^2_{H_1})\right]_{\xi_V=1} + 
\Sigma^{\text{add}}_{H_1H_2}(m^2_{H_2})+\Sigma^{\text{add}}_{H_2H_1}(m^2_{H_1})
\right)}{m^2_{H_1}-m^2_{H_2}} \notag \\
&- \cfrac{\sthree}{2\ctwo}\,\cfrac{\mbox{Re}\,\left(\left[\Sigma^{\text{tad}}_{H_1H_3}(m^2_{H_3})+\Sigma^{\text{tad}}_{H_3H_1}(m^2_{H_1})\right]_{\xi_V=1}
+\Sigma^{\text{add}}_{H_1H_3}(m^2_{H_3})+\Sigma^{\text{add}}_{H_3H_1}(m^2_{H_1})
\right)}{m^2_{H_1}-m^2_{H_3}} \notag \\
\dtwo &= \cfrac{\cthree}{2}\,\cfrac{\mbox{Re}\,\left(\left[\Sigma^{\text{tad}}_{H_1H_3}(m^2_{H_3})+\Sigma^{\text{tad}}_{H_3H_1}(m^2_{H_1})\right]_{\xi_V=1}
+\Sigma^{\text{add}}_{H_1H_3}(m^2_{H_3})+\Sigma^{\text{add}}_{H_3H_1}(m^2_{H_1})\right)}{m^2_{H_1}-m^2_{H_3}} \notag \\
&+ \cfrac{\sthree}{2}\,\cfrac{\mbox{Re}\,\left(\left[\Sigma^{\text{tad}}_{H_1H_2}(m^2_{H_2})+\Sigma^{\text{tad}}_{H_2H_1}(m^2_{H_1})\right]_{\xi_V=1} +
\Sigma^{\text{add}}_{H_1H_2}(m^2_{H_2})+\Sigma^{\text{add}}_{H_2H_1}(m^2_{H_1})
\right)}{m^2_{H_1}-m^2_{H_2}} \notag 
\end{alignat}
\begin{alignat}{5}
\dthree &= \cfrac{1}{2}\,\cfrac{\mbox{Re}\,\left[\Sigma^{\text{tad}}_{H_2H_3}(m^2_{H_3})+\Sigma^{\text{tad}}_{H_3H_2}(m^2_{H_2})\right]_{\xi_V=1} + 
\Sigma^{\text{add}}_{H_2H_3}(m^2_{H_3})+\Sigma^{\text{add}}_{H_3H_2}(m^2_{H_2})
}{m^2_{H_2}-m^2_{H_3}}
\notag \\ &+ \cfrac{\stwo}{2\ctwo}\,\Bigg{\{}
\cfrac{\sthree\, \mbox{Re}\,\left(\left[\Sigma^{\text{tad}}_{H_1H_3}(m^2_{H_3})+\Sigma^{\text{tad}}_{H_3H_1}(m^2_{H_1})\right]_{\xi_V=1} 
+ \Sigma^{\text{add}}_{H_1H_3}(m^2_{H_3})+\Sigma^{\text{add}}_{H_3H_1}(m^2_{H_1})\right)
}{m^2_{H_1}-m^2_{H_3}} \nonumber \\
&- \cfrac{\cthree\,\mbox{Re}\,\left(\left[\Sigma^{\text{tad}}_{H_1H_2}(m^2_{H_2})+\Sigma^{\text{tad}}_{H_2H_1}(m^2_{H_1})\right]_{\xi_V=1} + 
\Sigma^{\text{add}}_{H_1H_2}(m^2_{H_2})+\Sigma^{\text{add}}_{H_2H_1}(m^2_{H_1})\right)}{m^2_{H_1}-m^2_{H_2}}
\Bigg{\}} \;.
\label{eq:posalpha} 
\end{alignat}
And for the two chosen renormalization prescriptions of $\delta \beta$
we get
\beq
\delta \beta^{(1)} &=& - \frac{\mbox{Re}\left(\left[\Sigma_{G^\pm
        H^\pm}^{\text{tad}} (0) + 
  \Sigma^{\text{tad}}_{G^\pm H^\pm} (m_{H^\pm}^2) \right]_{\xi_V=1} +
  \Sigma^{\text{add}}_{G^\pm H^\pm} (0)
  + \Sigma^{\text{add}}_{G^\pm H^\pm} (m_{H^\pm}^2)\right)}{2 m_{H^\pm}^2} 
\label{eq:posbeta1} \\
\delta \beta^{(2)} &=& - \frac{\mbox{Re}\left(\left[\Sigma_{G^0 A}^{\text{tad}} (0) +
  \Sigma^{\text{tad}}_{G^0 A} (m_A^2) \right]_{\xi_V=1} +
  \Sigma^{\text{add}}_{G^0 A} (0)
  + \Sigma^{\text{add}}_{G^0 A} (m_A^2)\right)}{2 m_A^2} 
\;. \label{eq:posbeta2}
\eeq
With this procedure we have now obtained angular counterterms that are
explicitly gauge independent. \s

The additional contribution $\Sigma^{\text{add}}_{Hh}$ has been given
for the MSSM in \cite{Espinosa:2002cd}, and the ones for the 2HDM in 
\cite{Krause:2016oke,Kanemura:2017wtm,MKrause2016}. We
have derived the 
contributions necessary in the N2HDM, given here for the first time ($i,j=1,2,3$),
\beq
\Sigma^{\text{add}}_{H_iH_j} (p^2) &=& -\frac{g^2}{32 \pi^2 c_W^2}
\left( p^2 - \frac{m_{H_i}^2 + 
    m_{H_j}^2}{2} \right) \Big\{ \mathcal{O}^{(1)}_{H_iH_j}\, B_0
(p^2; m_Z^2, m_A^2) + \mathcal{O}^{(2)}_{H_iH_j}\, B_0 (p^2; m_Z^2, 
m_Z^2) \nonumber
\\
&& + 2 c_W^2 \left[\mathcal{O}^{(1)}_{H_iH_j}\, B_0 (p^2; m_W^2,
  m_{H^\pm}^2) +\mathcal{O}^{(2)}_{H_iH_j}\,B_0 (p^2; m_W^2, 
  m_W^2) \right] \Big\}
\label{eq:sigaddhh} \\
\Sigma^{\text{add}}_{G^0 A} (p^2) &=& \frac{-g^2}{32 \pi^2 c_W^2}
\left( p^2 - \frac{m_A^2}{2} 
\right) \,\sum_{i=1}^3 \,\mathcal{O}^{(3)}_{H_i H_i}\, B_0 (p^2;
m_Z^2,m_{H_i}^2) \label{eq:sigaddga} \\ 
\Sigma^{\text{add}}_{G^\pm H^\pm} (p^2) &=& \frac{-g^2}{16 \pi^2}
\left( p^2 - \frac{m_{H^\pm}^2}{2} 
\right) \,\sum_{i=1}^3 \,\mathcal{O}^{(3)}_{H_i H_i}\, B_0 (p^2;
m_W^2,m_{H_i}^2)  \;, \label{eq:sigaddghpm}  
\eeq 
where $B_0$ is the scalar two-point function
\cite{tHooft:1978xw,Passarino:1978jh}, while
the shorthand notation $\mathcal{O}^{(x)}_{H_iH_j}$ ($x=1,...,4$) stands
for different coupling combinations in the Higgs-gauge sector,
\begin{alignat}{5}
\mathcal{O}^{(1)}_{H_iH_j} &= \tilde{\kappa} _{H_iVH}\times
\tilde{\kappa} _{H_jVH}  \notag \\ 
\mathcal{O}^{(2)}_{H_iH_j} &= \kappa_{H_i VV}\times \kappa_{H_j VV}\, \notag \\
\mathcal{O}^{(3)}_{H_iH_j} &= \kappa_{H_i VV} \times \tilde{\kappa} _{H_jVH} \notag \\
\mathcal{O}^{(4)}_{H_iH_j} &= R_{i1} R_{j1} + R_{i2} R_{j2}
\label{eq:higgsgaugecomb}\, .
\end{alignat}
We note that in the N2HDM the following sum rules hold,
\begin{align}
\mathcal{O}^{(1)}_{H_iH_j}+\mathcal{O}^{(2)}_{H_iH_j} =
  \mathcal{O}^{(4)}_{H_iH_j} ~, &~~~~
  \sum_{i=1}^3\,\mathcal{O}^{(1)}_{H_iH_i} =
  \sum_{i=1}^3\,\mathcal{O}^{(2)}_{H_iH_i} = 1 ~, \notag \\
  \sum_{i=1}^3\,\mathcal{O}^{(3)}_{H_iH_i} = 0 ~,&~~~~ \sum_{i=1}^3 \kappa _{H_i VV} \kappa _{H_i ff} = 1 \;.\label{eq:twosumrules} 
\end{align}
Due to the second sum rule, the additional pinched contributions in
Eqs.~(\ref{eq:sigaddga},\ref{eq:sigaddghpm}) are UV-finite in the
N2HDM. In the 2HDM limit ($\alpha_{2,3} = 0$), the combination
$\mathcal{O}^{(4)}_{H_iH_j}$ becomes the Kronecker delta $\delta
_{H_iH_j}$ and hence, for $i\neq j$, the additional pinched
contributions in Eq.~(\ref{eq:sigaddhh}) become UV-finite by
themselves as well. \s

In the general N2HDM case instead,
$\Sigma^{\text{add}}_{H_1H_2},$ $\Sigma^{\text{add}}_{H_2H_1},$ $\Sigma^{\text{add}}_{H_1H_3},$
$\Sigma^{\text{add}}_{H_3H_1},$ $\Sigma^{\text{add}}_{H_2H_3},$
$\Sigma^{\text{add}}_{H_3H_2}$ contain UV-divergent poles, which
nevertheless cancel as they enter the mixing angle
coun\-ter\-terms~\eqref{eq:posalpha} via the additive structure
$\Sigma^{\text{add}}_{H_iH_j}(m_i^2)+
\Sigma^{\text{add}}_{H_jH_i}(m_j^2)$, which is UV-finite. 

\paragraph{$p_\star$ tadpole-pinched scheme}

Along the same lines followed for the 2HDM in
Ref.~\cite{Krause:2016oke}, we now generalise the 
\textit{$p_\star$ tadpole-pinched scheme} to the N2HDM Higgs 
sector. 
Again, we replace the scalar self-energies within the mixing angle 
counterterms 
with the corresponding \textit{pinched} self-energies,
$\overline{\Sigma}$, \eqref{eq:sigadddef}, which we evaluate this time
at the average of the particle momenta squared \cite{Espinosa:2001xu},
\beq
p_{\star,ij}^{2} = \frac{m_{\Phi_i}^2 + m_{\Phi_j}^2}{2} \;,
\eeq
where $(\Phi_i,\Phi_j)=(H_i,H_j)$, $(G^\pm,H^\pm)$ and $(G^0,A)$,
respectively. In this way the additional self-energies
$\Sigma^{\text{add}}$ vanish, and the  pinched self-energies are given by the  
tadpole self-energies $\Sigma^{\text{tad}}$ computed in the Feynman
gauge, {\it i.e.}  
\beq
\overline{\Sigma} (p_\star^2) = \left.\Sigma^{\text{tad}}
  (p_\star^2)\right|_{\xi_V=1} \;.
\eeq
The angular counterterms $\delta \alpha_i$ in
Eq.~(\ref{eq:neutralmixingCT}) then read 
\begin{alignat}{5}
\done  &=  \cfrac{ \cthree
  \,\Sigma^{\text{tad}}_{H_1H_2}(\pstaronetwo)}{\ctwo(m_{H_1}^2-m_{H_2}^2)}  
- \cfrac{\sthree
  \,\Sigma^{\text{tad}}_{H_1H_3}(\pstaronethree)}{\ctwo(m_{H_1}^2-m_{H_3}^2)}
\notag \\ 
\dtwo &=
\cfrac{\cthree\mbox{Re}\,\Sigma^{\text{tad}}_{H_1H_3}(\pstaronethree)}{m^2_{H_1}-m^2_{H_3}} 
+
\cfrac{\sthree\mbox{Re}\,\Sigma^{\text{tad}}_{H_1H_2}(\pstaronetwo)}{m^2_{H_1}-m^2_{H_2}}
\notag \\ 
\dthree &=
\cfrac{\mbox{Re}\,\Sigma^{\text{tad}}_{H_2H_3}(\pstartwothree)}{m^2_{H_3}-m^2_{H_2}} 
+ \cfrac{\stwo}{\ctwo}\,\Bigg{\{}
\cfrac{\sthree\,
  \mbox{Re}\,\Sigma^{\text{tad}}_{H_1H_3}(\pstaronethree)}{m^2_{H_1}-m^2_{H_3}}  
- \cfrac{\cthree\,\mbox{Re}\,\Sigma^{\text{tad}}_{H_1H_2
  }(\pstaronetwo)}{m^2_{H_1}-m^2_{H_2}} \Bigg{\}} \;,
\label{eq:neutralmixingCT-pstar}
\end{alignat}
with the different $p_\star$ scales being
\begin{alignat}{5}
 p_{\star,12}^2 = \cfrac{m^2_{H_1}+m^2_{H_2}}{2}\; , \qquad
 p_{\star,13}^2 = \cfrac{m^2_{H_1}+m^2_{H_3}}{2}\; , \qquad 
 p_{\star,23}^2 = \cfrac{m^2_{H_2}+m^2_{H_3}}{2}\; .
 \label{eq:pstar}
\end{alignat}%
For the counterterm $\delta \beta$ we get
\beq
\delta \beta^{(1)} = - \frac{\mbox{Re}
  \left[\overline{\Sigma}_{G^\pm H^\pm} \left( 
    \frac{m_{H^\pm}^2}{2} \right)\right]}{m_{H^\pm}^2} 
\eeq 
or alternatively
\beq
\delta \beta^{(2)} = - \frac{\mbox{Re} \left[\overline{\Sigma}_{G^0 A}
  \left( \frac{m_A^2}{2} \right) \right]}{m_A^2} \label{eq:delbet2star} \;.
\eeq

\subsection{Renormalization of $m^2_{12}$}
The soft $\mathbb{Z}_2$ breaking parameter $m_{12}^2$ enters the Higgs
self-couplings. For the  computation of higher-order corrections to
Higgs-to-Higgs decays it therefore has to be renormalized as well. We
may consider two different renormalization schemes. \s

\noindent
\underline{\it Modified Minimal Substraction Scheme:}
One possibility is to use a modified $\msbar$ scheme, {\it
  cf.}~\cite{Krause:2016xku},  
where the counterterm $\delta m_{12}^2$ is chosen such that it 
cancels all residual terms of the amplitude that are proportional to  
\beq
\Delta = \frac{1}{\epsilon} - \gamma_E + \ln (4\pi) \;,
\eeq
where $\gamma_E$ denotes the Euler-Mascheroni constant. 
These terms obviously contain the remaining UV divergences given as
poles in $\epsilon$ together with additional finite constants that appear
universally in all loop integrals. The
renormalization of $\delta m_{12}^2$ in this scheme is thereby given by
\beq
\delta m_{12}^2 = \delta m_{12}^2 (\Delta)|_{\overline{\text{MS}}} \;.
\eeq
The right-hand side of the equation symbolically denotes all
terms proportional to $\Delta$ that are necessary to cancel the
$\Delta$ dependence of the remainder of the amplitude. \s

\noindent
\underline{\it Process-dependent renormalization:}
Alternatively, one could resort to a process-dependent sche\-me, in which 
case the divergent parts of $\delta m_{12}^2$, along with additional
finite remainders, 
are related to a physical on-shell Higgs-to-Higgs decay. While this method
provides a physical definition for the counterterm, it relies on
having at least one kinematically accessible on-shell Higgs-to-Higgs
decay. For a generic Higgs-to-Higgs decay process $H_i \to H_j 
H_k$, where the final state pair $H_j H_k$ can also be a pair of
pseudoscalars, if kinematically allowed, the counterterm $\delta
m_{12}^2$ is then fixed by imposing as renormalization condition
\beq
\Gamma^{\text{LO}} (H_i \to H_j H_k) \stackrel{!}{=}
\Gamma^{\text{NLO}} (H_i\to H_j H_k) \;.
\eeq

Note that $\delta m_{12}^2$ is gauge independent in either of the proposed schemes,
and also independently on how the tadpole topologies are treated. 
The key reason is that $m^2_{12}$ is indeed a genuine parameter of
the original N2HDM Higgs potential before EWSB, and hence 
unlinked to the VEV, this being the source for the
potential gauge-parameter dependences that arise at higher orders in
certain schemes. In this paper we will apply the
  $\msbar$ renormalization scheme.

\section{One-Loop EW Corrected Decay Widths \label{sec:oneloopdec}}
Having elaborated in detail the renormalization scheme for the N2HDM,
we compute the NLO EW corrections to a selected set of decay
widths, in order to illustrate their impact. The chosen decays widths
are 
\beq
H_{2/3} &\to& ZZ \label{eq:htozz}\\
H_{2/3} &\to& AA \label{eq:htoaa} \\
H_3 &\to& H_2 H_2 \quad \mbox{and} \qquad H_2 \to H_1 H_1 \label{eq:htohh} \;.
\eeq
All processes require the renormalization of the mixing
angles. The Higgs-to-Higgs decays demand in addition the
renormalization of $m_{12}^2$. And the Higgs decays into CP-even
pairs, Eq.~(\ref{eq:htohh}), additionally involve the renormalization
of $v_S$. The chosen processes are structurally different and involve
the various mixing angles in 
different more or less complicated combinations, allowing us to 
study the impact of our renormalization scheme in different
situations, and enabling us to study the renormalization of the
Higgs potential parameter $m_{12}^2$ as well as of the singlet VEV
$v_S$. Note finally that all these decays only involve 
electrically neutral particles, so that we do not encounter any IR
divergences in the EW corrections. \s

\subsection{The NLO EW corrected decay $H_i  \to ZZ$}
\begin{figure}[t!]
\begin{center}
\includegraphics[width=0.9\linewidth , trim = 0mm 6mm 0mm 6mm, clip]{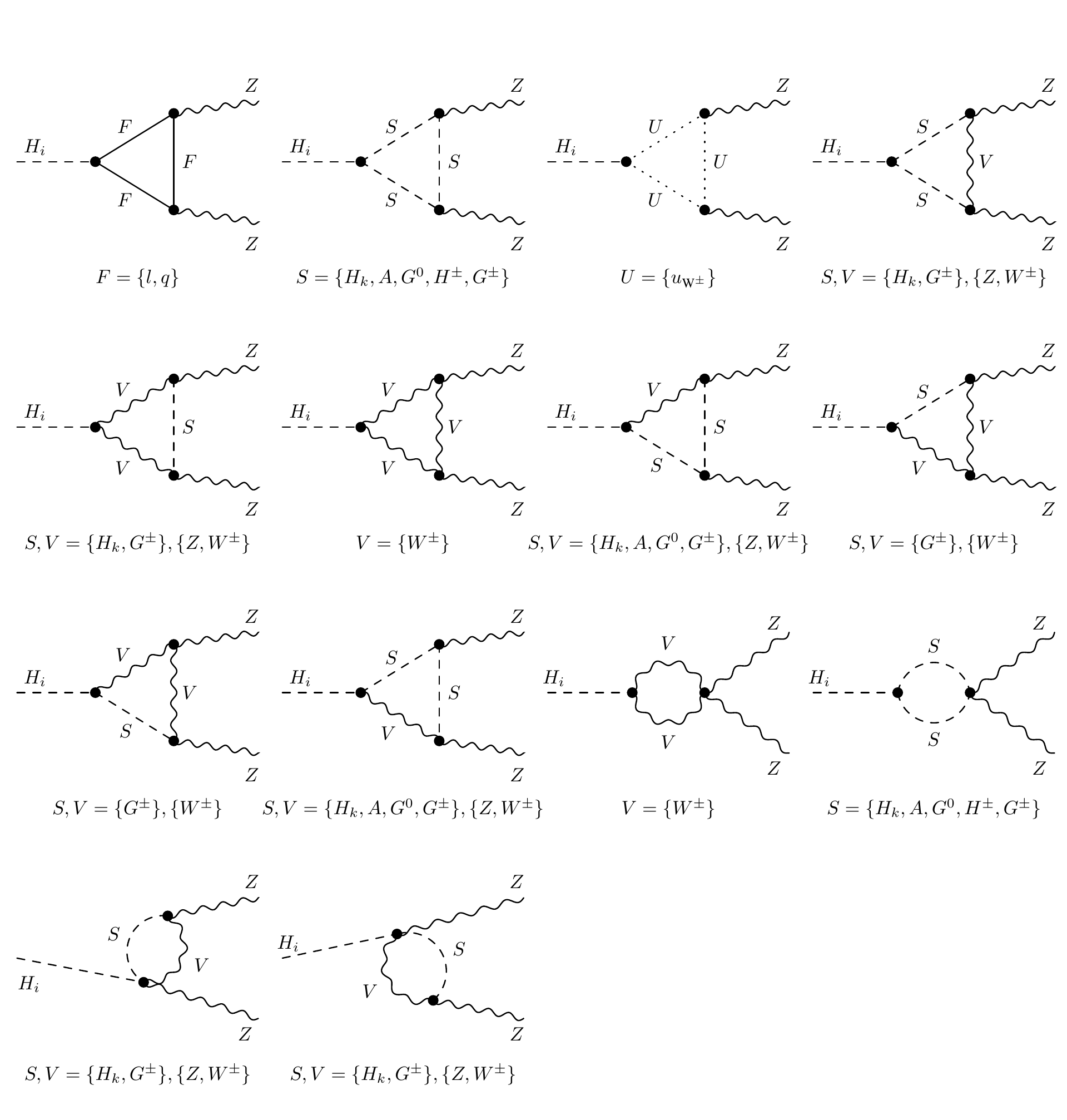}
\end{center}
\caption{Generic diagrams contributing to the vertex
  corrections in $H_i \to ZZ$ with fermions $F$, scalar bosons $S$, 
      gauge bosons $V$ and ghost particles $U$ in the
      loops.}
\label{fig:detailsvirthitozz}
\end{figure}

The LO decay width for the decay of a CP-even Higgs boson
$H_i$ into a pair of $Z$ bosons, 
\begin{alignat}{5}
H_i \to ZZ \;, \label{eq:hzz-process}
\end{alignat}
is given by 
\beq
\Gamma^{\text{LO}} (H_i\to ZZ) = \frac{\alpha \kappa_{H_iVV}^2}{32
  \swd m_W^2 m_{H_i}} (m_{H_i}^4-4 m_{H_i}^2 m_Z^2 + 12 m_Z^4) \sqrt{1-\frac{4
    m_Z^2}{m_{H_i}^2}}   
\eeq
and depends on the mixing angles through the coupling factors 
\begin{alignat}{5}
 \kappa_{H_1 VV} &= R_{11}\,\cb + R_{12}\,\sb = \ctwo c_{\beta-\alpha_1} \notag \\
  \kappa_{H_2 VV} &= R_{21}\,\cb + R_{22}\,\sb = -c_{\beta-\alpha_1} s_{\alpha_2}s_{\alpha_3} + c_{\alpha_3} s_{\beta-\alpha_1} \notag \\
  \kappa_{H_3 VV} &= R_{31}\,\cb + R_{32}\,\sb = -c_{\alpha_3}\,c_{\beta-\alpha_1}\,s_{\alpha_2} - s_{\alpha_3}\,s_{\beta-\alpha_1}\, .
\end{alignat}
\begin{figure}[t]
\begin{center}
\includegraphics[width=0.9\linewidth , trim = 0mm 11mm 0mm 16mm, clip]{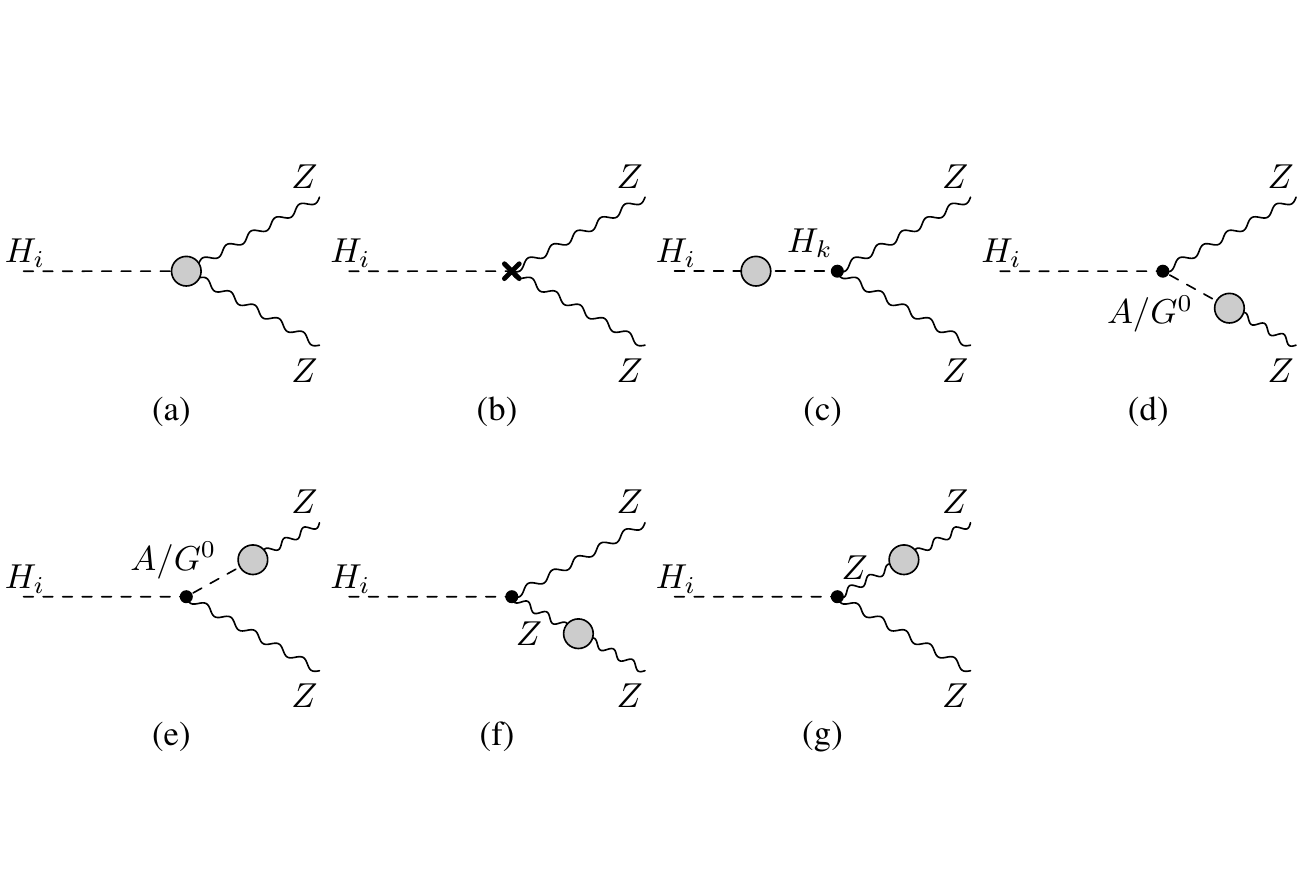}
\end{center}
\caption{Generic diagrams contributing to the virtual corrections of
 the decay $H_i \to ZZ$: vertex corrections (a) and corrections to the
 external legs (c)-(g), where $k=1,2,3$. Diagram (b)
 displays the vertex counterterm.}
\label{fig:genericnlohizz}
\end{figure}
The generic diagrams describing the virtual corrections contributing to the
NLO decay width together with the counterterm diagram introduced to cancel the UV
divergences are displayed in Fig.~\ref{fig:genericnlohizz}. 
With the decay width involving only neutral particles
there are neither IR divergences nor real corrections. The corrections
to the external legs in Fig.~\ref{fig:genericnlohizz} (c), (f) and (g) vanish due
to the OS renormalization of $H_i$ and $Z$, respectively, and the
mixing contributions (d) 
and (e) are zero because of the Ward identity satisfied by the OS $Z$ boson. 
The one-particle irreducible (1PI) diagrams contributing to the vertex
corrections originate from the triangle diagrams with scalars, fermions,
massive gauge bosons and ghost particles in the loops, depicted in the
first three rows of Fig.~\ref{fig:detailsvirthitozz},  and from the 
diagrams involving four-particle vertices, as given by the last 
four diagrams of Fig.~\ref{fig:detailsvirthitozz}. \s

\begin{figure}[t!]
\begin{center}
\includegraphics[width=14cm]{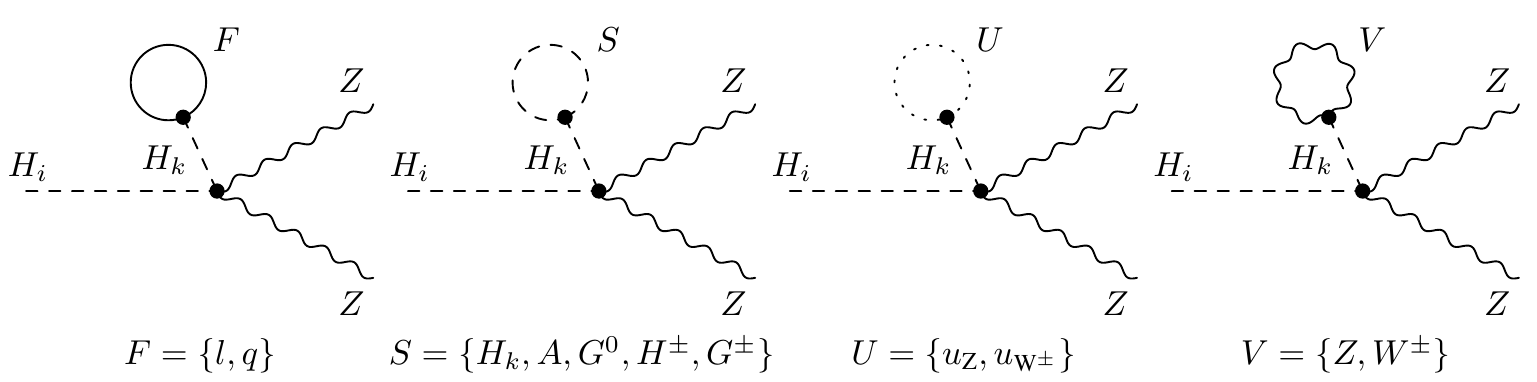}
\end{center}
\caption{Tadpole contributions to the vertex diagrams to be included
  in the decay $H_i \to ZZ$ in the alternative tadpole scheme.}
\label{fig:hizztadpdiags}
\end{figure}

To work out the vertex counterterms, the relations 
\begin{alignat}{5}
 s_{\varphi} \to s_{\varphi} + c_{\varphi}\,\delta\varphi \qquad \mbox{and}
 \qquad  c_{\varphi} \to c_{\varphi} - s_{\varphi}\,\delta\varphi
 \label{eq:trigo-shifts}
\end{alignat}
are helpful for the derivation of the entries in the rotation
matrix counterterm $\delta R$ obtained from~\eqref{eq:mixingmatrix},
\begin{alignat}{5}
\delta R_{11} &=  -\cone\,\stwo\dtwo - \sone\ctwo\done \notag \\
\delta R_{12} &= -\sone\stwo\dtwo + \ctwo\cone\done \notag \\
\delta R_{13} &= \ctwo\dtwo \notag \\
\delta R_{21} &=  -\cone\cthree\done + \sone\sthree\dthree
-\cone(\stwo\cthree\dthree+\sthree\ctwo\dtwo) + \stwo\sthree\sone\done
\notag \\
\delta R_{22} &=
-\cone\sthree\dthree - \sone\cthree\dthree - \sone(\stwo\cthree\dthree + \sthree\ctwo\dtwo) - \cone\stwo\sthree\done \notag \\
\delta R_{23} &= -\stwo\sthree\dtwo + \ctwo\cthree\dthree \notag \\
\delta R_{31} &= \sone\cthree\dthree+\cone\sthree\done-\cone(\ctwo\cthree\dtwo -\stwo\sthree\dthree) + \sone\stwo\cthree\done \notag \\
\delta R_{32} &= \sone\sthree\done-\cone\sthree\dthree+\sone\stwo\sthree\dthree - \cthree(\sone\ctwo\dtwo+\cone\stwo\done) \notag \\
\delta R_{33} &= -\stwo\cthree\dtwo -
\ctwo\sthree\dthree \label{eq:RotMass-shifts}\, .
\end{alignat}
The $H_i ZZ$ vertex counterterm in terms of the different parameter
counterterms and wave function renormalization constants is obtained
from the corresponding counterterm Lagrangian 
\begin{alignat}{5}
 {\cal L}^{\text{ct}}_{H_i ZZ} &= 
\left( \cfrac{g \mzd\,\kappa_{H_i ff}}{m_W}\,\left[ \cfrac{\delta
      \mzd}{\mzd}\, - \left( \frac{\delta
          m_W^2}{2m_W^2} - \frac{\delta g}{g} \right) 
%
%
+ \delta Z_{ZZ} +
    \cfrac{1}{2}\,\delta Z_{H_iH_i} + \cfrac{1}{2}\,\sum_{j\neq
      i}\cfrac{\kappa_{H_j VV}}{\kappa_{H_i VV}}\,\delta Z_{H_jH_i}\,  
\right] \notag \right. \\  
 & \left. \quad + \cfrac{g \mzd}{m_W}\,\left[\delta R_{i1}\,\cb + \delta
   R_{i2}\,\sb - (R_{i1}\,\sb- R_{i2}\,\cb) \delta \beta
 \right] \right) g^{\mu\nu} H_i Z_\mu Z_\nu \label{eq:HZZct}\, , 
\end{alignat}
with the various counterterms given in Section~\ref{sec:renorm}
and the $\delta R_{ij}$ defined in Eq.~(\ref{eq:RotMass-shifts}).
Since we apply the alternative tadpole scheme, tadpole contributions to the $H_i
ZZ$ vertex have to be taken into account explicitly in the computation of the decay
width. They are shown in Fig.~\ref{fig:hizztadpdiags}. 
The formulae for the vertex corrections and counterterms in terms of
the scalar one-, two- and three-point functions are quite lengthy so that we
do not display them explicitly here. 

\subsection{The decay $H_i \to AA$ at NLO EW}
The LO decay width of the CP-even $H_i$ decay into a pair
of CP-odd scalars, 
\beq
H_i \to AA \;,
\eeq
reads 
\beq
\Gamma^{\text{LO}} (H_i \to AA) = \frac{\alpha\,\left| \lambda_{H_iAA} \right| ^2}{8
  \swd\,\,m_{H_i}} \sqrt{1-\frac{4m_A^2}{m_{H_i}^2}} \;.
\eeq
It is governed by the trilinear coupling 
\begin{alignat}{5}
g_{H_iAA} = -i\,\cdot\, \lambda_{H_iAA}\, &= g\,\cfrac{1}{2\mw}\Bigg{\{}
  -M^2\left[\frac{R_{i1}}{\cb}+\frac{R_{i2}}{\sb} \right] + 
  m_{H_i}^2\,\left[\frac{R_{i1}\sb^2}{\cb}+\frac{R_{i2}\,\cb^2}{\sb} \right] \notag \\
  & \qquad + 2\,m_A^2\,\left[R_{i1}\cb+R_{i2}\sb \right]
 \Bigg{\}} \;,
 \label{eq:tripleHAA}
\end{alignat}
where $M^2 \equiv m^2_{12}/(\sb\cb)$. \s

\begin{figure}[t!]
\begin{center}
\includegraphics[width=0.9\linewidth]{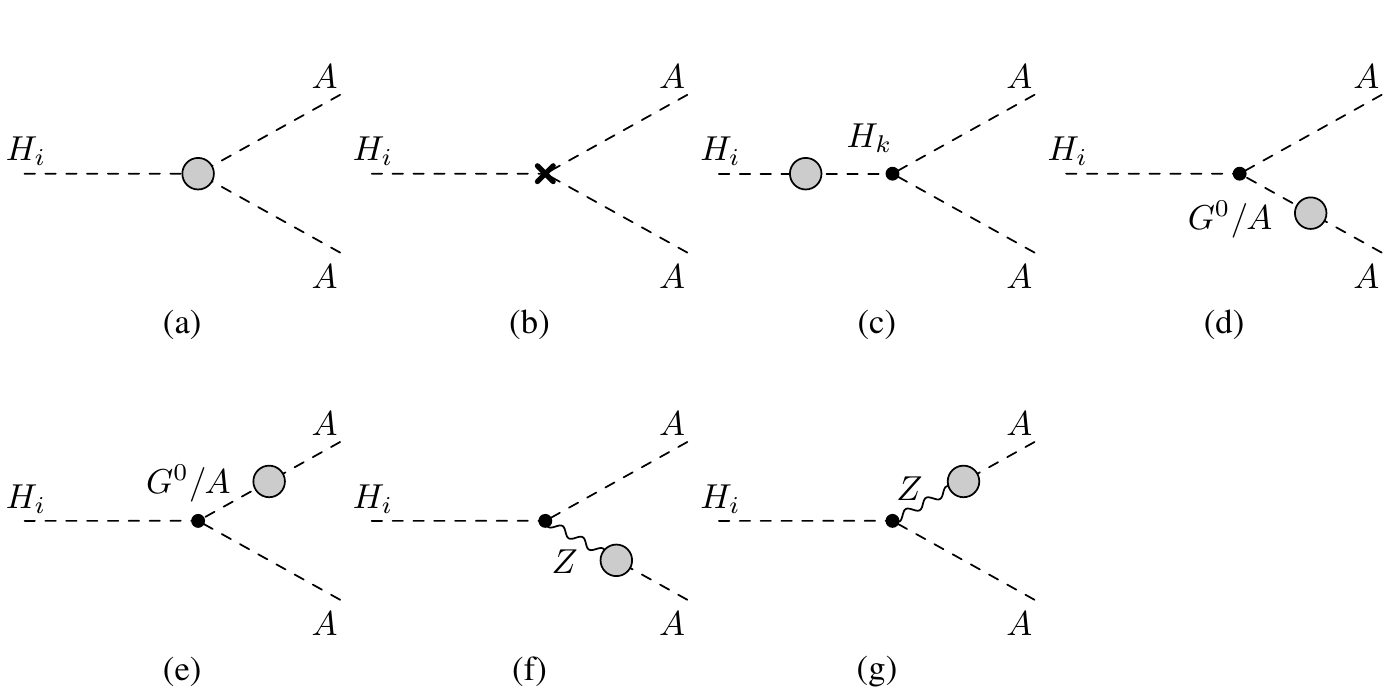}
\end{center}
\caption{Generic diagrams contributing to the virtual corrections of
 the decay $H_i \to AA$: vertex corrections (a) and corrections to the
 external legs (c)-(g). Diagram (b)
 displays the corresponding vertex counterterm.}
\label{fig:virtualHiAA}
\end{figure}
\begin{figure}[t!]
\begin{center}
\includegraphics[width=0.9\linewidth]{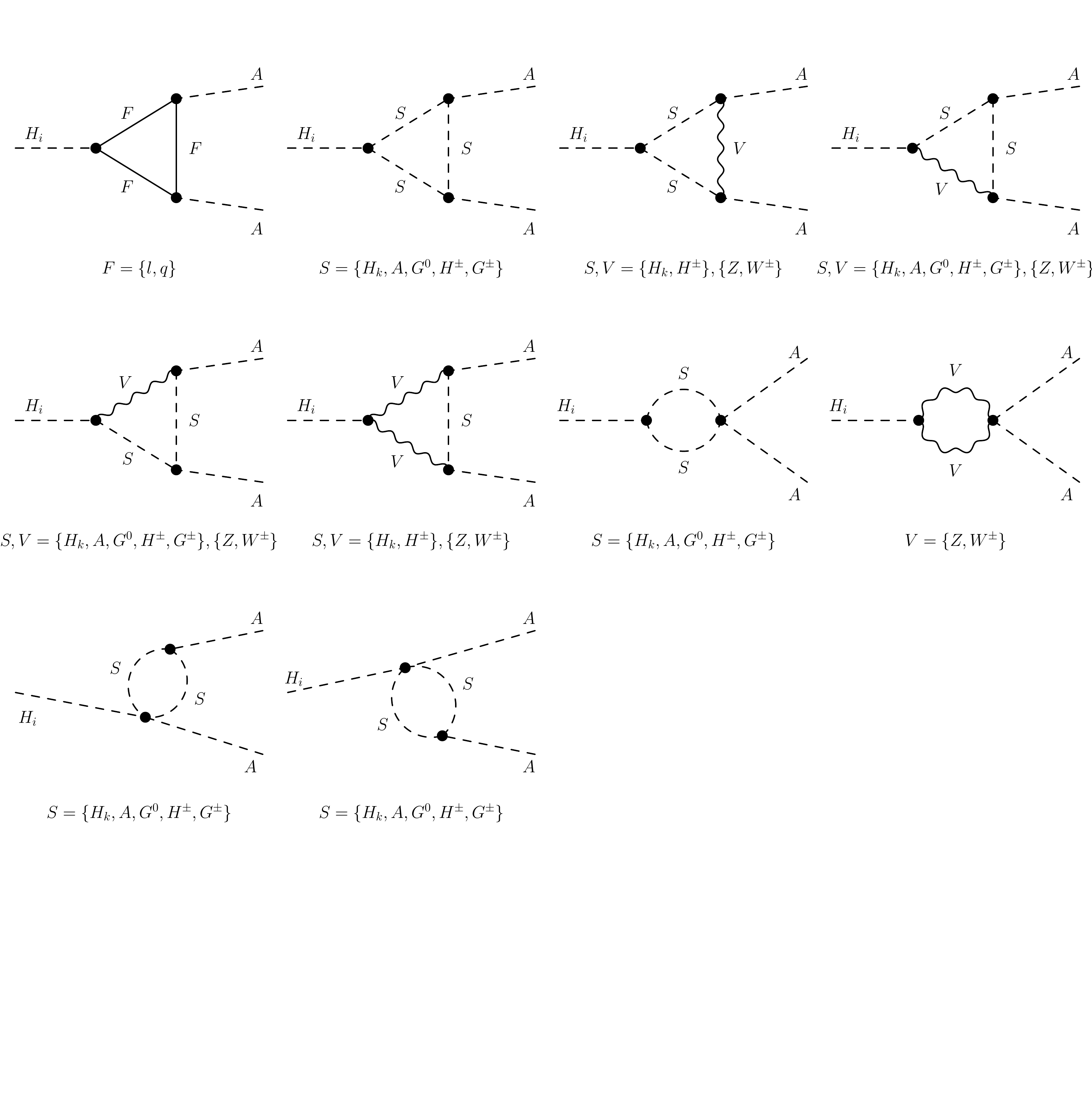}
\end{center}
\vspace*{-3.5cm}
\caption{Generic diagrams contributing to the vertex corrections in $H_i \to AA$.}
\label{fig:vertexcorrsHiAA}
\end{figure}
\begin{figure}[t!]
\begin{center}
\includegraphics[width=14cm]{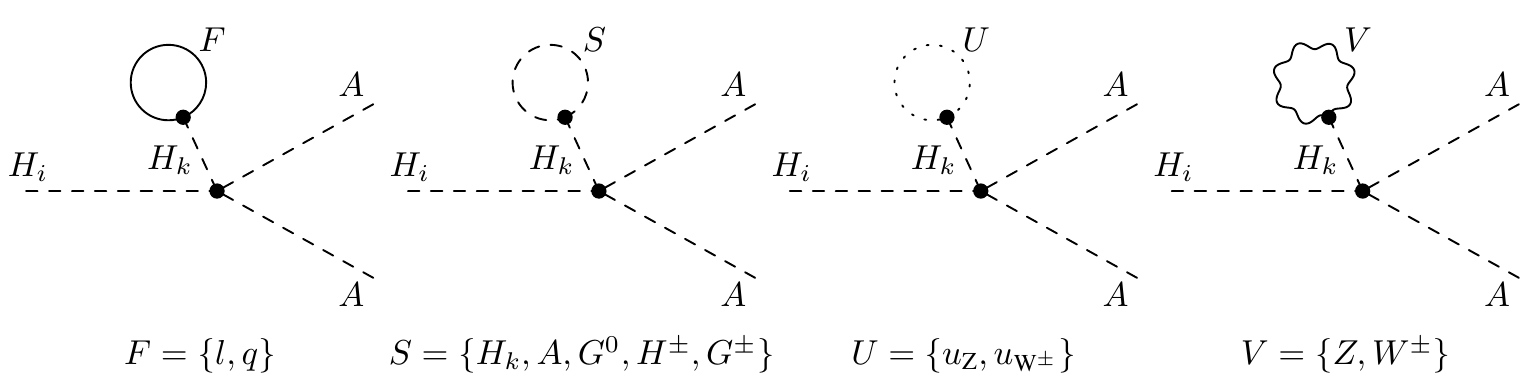}
\end{center}
\caption{Tadpole contributions
  to the vertex diagrams to be included in the decay $H_i\to AA$ in the
  alternative tadpole scheme.}
\label{fig:tadphitoaa}
\end{figure}
The EW one-loop corrections consist of the virtual corrections and the
counterterm contributions ensuring the UV-finiteness of the
decay amplitude. Again we do not have to deal with IR divergences nor
real corrections. The virtual corrections, consisting of the corrections to
the external legs and the pure vertex corrections, are shown in
Fig.~\ref{fig:virtualHiAA}. The corrections to the external legs in
Fig.~\ref{fig:virtualHiAA} (c), (d) and (e) are zero because of the OS
renormalization of the external fields, while diagrams (f) and (g) vanish
due to a Slavnov-Taylor identity \cite{Williams:2011bu}. 
The 1PI diagrams of the vertex corrections are depicted in
Fig.~\ref{fig:vertexcorrsHiAA}. They are given by the 
triangle diagrams with fermions, scalars and gauge bosons in the loops
and by the diagrams containing four-particle vertices. 
The counterterm contributions consist of the genuine vertex
counterterm $\delta g_{H_i AA}^{\text{vertex}}$ and the counterterm insertions on
the external legs $\delta g_{H_i AA}^{\text{field}}$, 
\begin{alignat}{5}
  \delta g_{H_iAA} &= \delta\,g^{\text{field}}_{H_iAA}   + \delta\,g^{\text{vertex}}_{H_iAA}
  \label{eq:tripleHAActone}\, ,
\end{alignat}
with
\begin{alignat}{5}
 \delta\,g^{\text{field}}_{H_iAA} &=
 g_{H_iAA}\, \left[\delta Z_{AA} + \frac{1}{2}\,\delta Z_{H_i H_i} + \frac{1}{2}\,\sum_{i\neq j}\,\cfrac{g_{H_jAA}}{g_{H_iAA}}\,\delta Z_{H_jH_i}
 + \cfrac{g_{H_iAG}}{g_{H_iAA}}\,\delta Z_{G^0 A} \right]
   \label{eq:tripleHAActfield}
\end{alignat}
and 
\begin{alignat}{5}
 \delta\,g^{\text{vertex}}_{H_iAA} =& -g_{H_iAA}\,
 \left( \frac{\delta m_W^2}{2m_W^2} - \frac{\delta g}{g} \right)
+  \cfrac{g}{2 m_W}\Bigg{\{} \left(R_{i1}\cfrac{\sbd}{\cb} +
  R_{i2}\cfrac{\cbd}{\sb} \right)\,\delta m^2_{H_i}  
  -\left(\cfrac{R_{i1}}{\cb} + \cfrac{R_{i2}}{\sb} \right)\,\delta M^2
  \notag \\
 & + 2\,\left[R_{i1}\cb+R_{i2}\sb \right]\,\delta m_A^2
- M^2\, \left(\cfrac{\delta R_{i1}}{\cb} + \cfrac{\delta R_{i2}}{\sb} \right)
 +m^2_{H_i}\, \left(\cfrac{\sbd}{\cb}\,\delta R_{i1} + \cfrac{\cbd}{\sb}\,\delta R_{i2} \right)\,
 \notag \\
 & + 2\,m_A^2\, \left[\cb\delta R_{i1}+\sb\delta R_{i2} \right]
+ M^2\,\left(R_{i1}\cfrac{\delta \cb}{\cb^2} + R_{i2}\cfrac{\delta \sb}{\sb^2} \right) + 
   2\,m_A^2\, \left[R_{i1}\delta \cb +R_{i2}\,\delta\sb \right] \notag \\ 
 & + m_{H_i}^2\,\left[R_{i1}\,\cfrac{\sbd}{\cb}\left(2\cfrac{\delta \sb}{\sb} - \cfrac{\delta\cb}{\cb}\right)
   + R_{i2}\,\cfrac{\cbd}{\sb}\left(2\cfrac{\delta \cb}{\cb} - \cfrac{\delta\sb}{\sb}\right)
   \right]
 \Bigg{\}}
   \label{eq:tripleHAActparam}\, , 
\end{alignat}
with the $\delta R_{ij}$ given in Eq.~(\ref{eq:RotMass-shifts}).
Working in the alternative tadpole scheme, we additionally have to take into account
the vertices dressed with the tadpoles, displayed in
Fig.~\ref{fig:tadphitoaa}. \s

The one-loop correction to the decay is obtained from the interference
of the loop-corrected decay amplitude ${\cal M}_{H_i AA}^{\text{1loop}}$
with the LO amplitude ${\cal M}_{H_i AA}^{\text{LO}}$. The one-loop amplitude 
combines the virtual corrections ${\cal M}_{H_i AA}^{\text{virt}}$,
including external leg and pure vertex corrections, 
and the counterterm amplitude ${\cal M}_{H_i AA}^{\text{ct}} =
\delta g_{H_iAA} + {\cal M}_{H_i AA}^{\text{tad}}$,
with ${\cal M}_{H_i AA}^{\text{tad}}$ denoting the vertices with the tadpoles,
\beq
{\cal M}_{H_i AA}^{\text{1loop}} = {\cal M}_{H_i AA}^{\text{virt}} + {\cal
  M}^{\text{ct}}_{H_iAA} \;.
\eeq
The NLO corrections factorise from the LO amplitude so that the
loop-corrected partial width can be cast into the form
\begin{alignat}{5}
\Gamma^{\text{NLO}} &= 
\Gamma^{\text{LO}} + \frac{m_{H_i} }{32 \pi} \sqrt{1-\frac{4
      m_A^2}{m_{H_i}^2}} \, 2 \, \mbox{Re} \left[ ({\cal 
    M}_{H_i AA}^{\text{LO}})^* {\cal M}_{H_i AA}^{\text{1loop}} \right] \notag \\
    & = \Gamma^{\text{LO}}\,
    [1+ \Delta^{\text{virt}}_{H_i AA} + \Delta^{\text{ct}}_{H_i AA}]\;,
\end{alignat}
with
\beq
\Delta^{\text{virt/ct}}_{H_i AA} \equiv \frac{2 {\cal
    M}^{\text{virt/ct}}_{H_iAA}}{g_{H_iAA}} 
= \frac{2 {\cal
    M}^{\text{virt/ct}}_{H_iAA}}{-i \cdot \lambda_{H_iAA}} 
\;.
\eeq
Again we refrain from giving the explicit expressions for the various
contributions to $\Gamma^{\text{NLO}}$ as they are quite lengthy.

\subsection{Electroweak one-loop corrections to $H_j \to H_i H_i$}
The LO decay width for the decay of a neutral CP-even Higgs boson into
two identical CP-even scalars is given by ($i,j=1,2,3$ )
\beq
\Gamma^{\text{LO}} (H_j \to H_iH_i) = \frac{\alpha\,
  \left| \lambda_{H_i H_i H_j} \right| ^2}{8\,\swd\,m_{H_j}} \sqrt{1-\frac{4m_{H_i}^2}{m_{H_j}^2}} \;,
\eeq
with the trilinear Higgs coupling 
\begin{alignat}{5}
 g_{H_iH_iH_j}\, &= -i \cdot \lambda_{H_iH_iH_j}\, = \cfrac{g}{2m_W}\,\Bigg{\{}
  -\cfrac{1}{2}\,M^2\,\left[\left(\cfrac{R_{i2}}{\sb}-\cfrac{R_{i1}}{\cb} \right)\,\left(6R_{i2}R_{j2}\,\cb^2 -
  6R_{i1}R_{j1}\,\sb^2 \right. \right. \notag \\
& \left. \left. + \sum_k\,\epsilon_{ijk}\,R_{k3}\,s_{2\beta}\right)
\right] + \cfrac{2m^2_{H_i}+m^2_{H_j}}{v_S}\,\left[R^2_{i3}\,R_{j3}\,v
+ R^2_{i2}\,R_{j2}\,\cfrac{v_S}{\sb}+
R^2_{i1}\,R_{j1}\,\cfrac{v_S}{\cb}\right] \Bigg{\}} \;,
 \label{eq:tripleiij}
\end{alignat}
where $\epsilon_{ijk}$ denotes the totally
  antisymmetric tensor in three dimensions with $\epsilon_{123}=1$.
At variance with the processes discussed so far, Higgs-to-Higgs decays
in the CP-even sector are directly sensitive to the singlet VEV $v_S$
at tree level. As discussed in section~\ref{sec:altern}, this explicit
dependence must be handled with care when the NLO calculations are performed in the
alternative tadpole scheme. Here, a non-vanishing UV-divergent singlet VEV shift
$\Delta v_S$ cancels a subset of the UV poles
in the NLO Higgs-to-Higgs decay amplitude which genuinely arise
in this scheme. To fix $\Delta v_S$ we proceed along the same lines
as for the doublet VEV. 
First, we identify the singlet VEV input value in this scheme
with the (would-be) experimental input, to be extracted eventually
through the measurement of an observable 
Higgs-to-Higgs decay width $\Gamma_{H_i\to H_jH_j}$.
When promoted to higher orders,
the tree-level relation $v^{\text{tree}}_S
= f(\Gamma^{\text{tree}}_{H_i\to H_jH_j})$ becomes 
\begin{alignat}{5}
\left. v^{\text{ren}}_S \right| _{\text{FJ}} = v_S^{\text{tree}} = 
f(\Gamma^{\text{tree}}_{H_i\to H_jH_j}) &=
f(\Gamma^{\text{ren}}_{H_i\to H_jH_j} + 
\Gamma^{\text{ct}}_{H_i\to H_jH_j}) = 
 \underbrace{\tilde{f}(\Gamma^{\text{ren}}_{H_i\to
     H_jH_j})}_{v_S^{\text{exp.}}} +
 \underbrace{\delta\tilde{f}(\Gamma^{\text{ct}}_{H_i\to
     H_jH_j})}_{\Delta v_S} 
\label{eq:extraeq-vevsren}\, ,
\end{alignat}
in such a way that the (would-be) experimental value
$v_S^{\text{exp}}$ is properly written in terms of the  
renormalized (physical) width from which it would be extracted.
Notice that the quantity $\Delta v_S$ is simply a shorthand 
for the combination of counterterm contributions contained in
$\Gamma^{\text{ct}}_{H_i\to H_jH_j}$ - the same role that
$\Delta v$ plays in~\eqref{eq:doubletvevshift} for the doublet VEV case.
For our sample processes $H_3 \to H_2 H_2$
and $H_2 \to H_1 H_1$ discussed in the
numerical analysis we assume the $v_S$ input values to be extracted
from the decay $\hthree \to 
\hone\hone$.\footnote{The choice of the process relies on the
  experimental feasibility of measuring it and on its dependence on
  $v_S$ itself. For some scenarios the parameter configurations can be
such that the decay is not measurable or the dependence on $\Delta
v_S$ is almost vanishing, {\it cf.}~also the discussion in \cite{Belanger:2016tqb} on the
renormalization of the NMSSM  where similar issues arise.}
The choice of this process is of course not unique. Therefore, given that the finite parts
included in $\Delta v_S$ are to some degree arbitrary, we could
formally resort to $\msbar$-like conditions to fix $\Delta v_S$  by retaining only
the UV-divergent parts contained in $\Gamma^{\text{ct}}_{H_i\to H_jH_j}$. In
this case the $v_S$ input values could not be extracted directly from
the experimental data. The relation to the to be measured $v^{\text{exp}}_S$
would be given by a scheme-dependent finite shift.
In the process-dependent framework 
$\Delta v_S$ can be fixed  through the requirement 
\begin{alignat}{5}
 \Gamma^{\text{NLO}}_{\hthree \to \hone\hone} & \stackrel{!}{=}
 \Gamma^{\text{LO}}_{\hthree \to \hone\hone} 
 \label{eq:DeltaVS-definition}\,.
\end{alignat}
Factorising the NLO decay width as 
\begin{alignat}{5}
 \Gamma^{\text{NLO}}_{\hthree \to \hone\hone} &= 
 \Gamma^{\text{LO}}_{\hthree \to \hone\hone}\Big[1+ \Delta^{\text{virt}} + 
 \Delta^{\text{ct}}(\Delta v_S = 0) + \Delta^{\text{ct}}(\Delta v_S)\Big] \stackrel{!}{=} \Gamma^{\text{LO}}_{\hthree \to \hone\hone}
  \label{eq:DeltaVS-determinationOne}\,
\end{alignat}
and isolating the $v_S$-dependent part of the corresponding
self-interaction Lagrangian, 
\beq
&&  \lag_{H_1H_1H_3} \supset \cfrac{1}{v_S}\,(2m^2_{H_1} +
m^2_{H_3})\,R^2_{13}\,R_{33} \;, \quad \text{whereby} \nonumber \\
&&
 \delta \lag_{H_1H_1H_3}\supset - \cfrac{1}{v_S}\,(2m^2_{H_1} + m^2_{H_3})\,R^2_{13}\,R_{33}\,\cfrac{\Delta\,v_S}{v_S} 
 \label{eq:VSpart}\, ,
\eeq
the condition~\eqref{eq:DeltaVS-determinationOne} leads to 
\begin{alignat}{5}
 \cfrac{\Delta v_S}{v_S} &= \cfrac{g_{H_iH_iH_j}\,v_S}{2} \Big{[} (2m_{\hone}^2 + m_{\hthree}^2)R^2_{13}\,R_{33}\,\Big{]}^{-1}\,
 \left[\Delta^{\text{virt}} + \Delta^{\text{CT}}(\Delta v_S = 0) \right]
 \label{eq:DeltaVS-determinationTwo} \, .
\end{alignat}

\begin{figure}[t!]
\begin{center}
\includegraphics[width=0.8\linewidth]{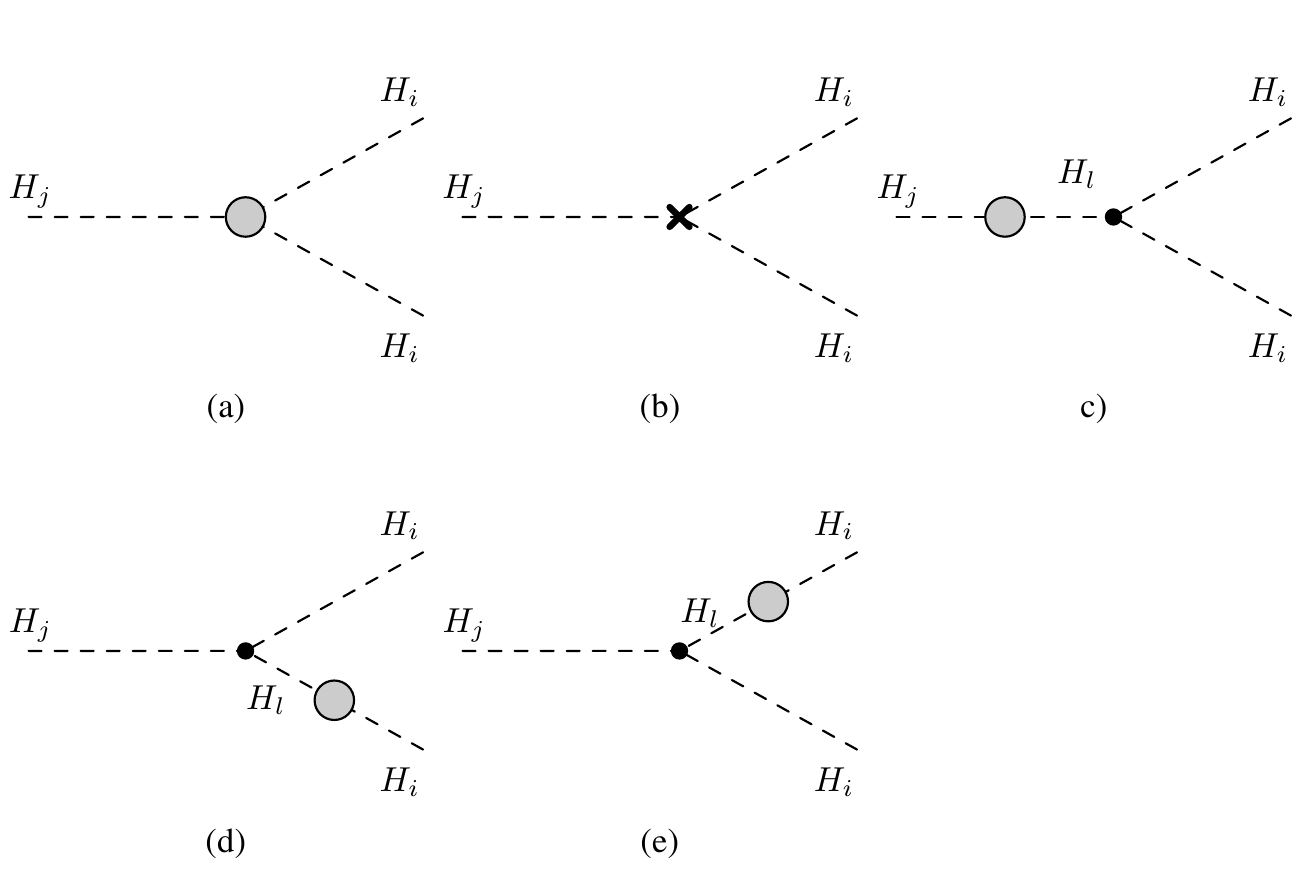}
\end{center}
\caption{Generic diagrams contributing to the virtual corrections of
 the decay $H_j \to H_iH_i$: vertex corrections (a) and corrections to the
 external legs (c)-(e). Diagram (b)
 displays the corresponding vertex counterterm.}
\label{fig:vertexcorrsHiHjHk}
\end{figure}

\begin{figure}[t!]
\begin{center}
\includegraphics[width=0.9\linewidth]{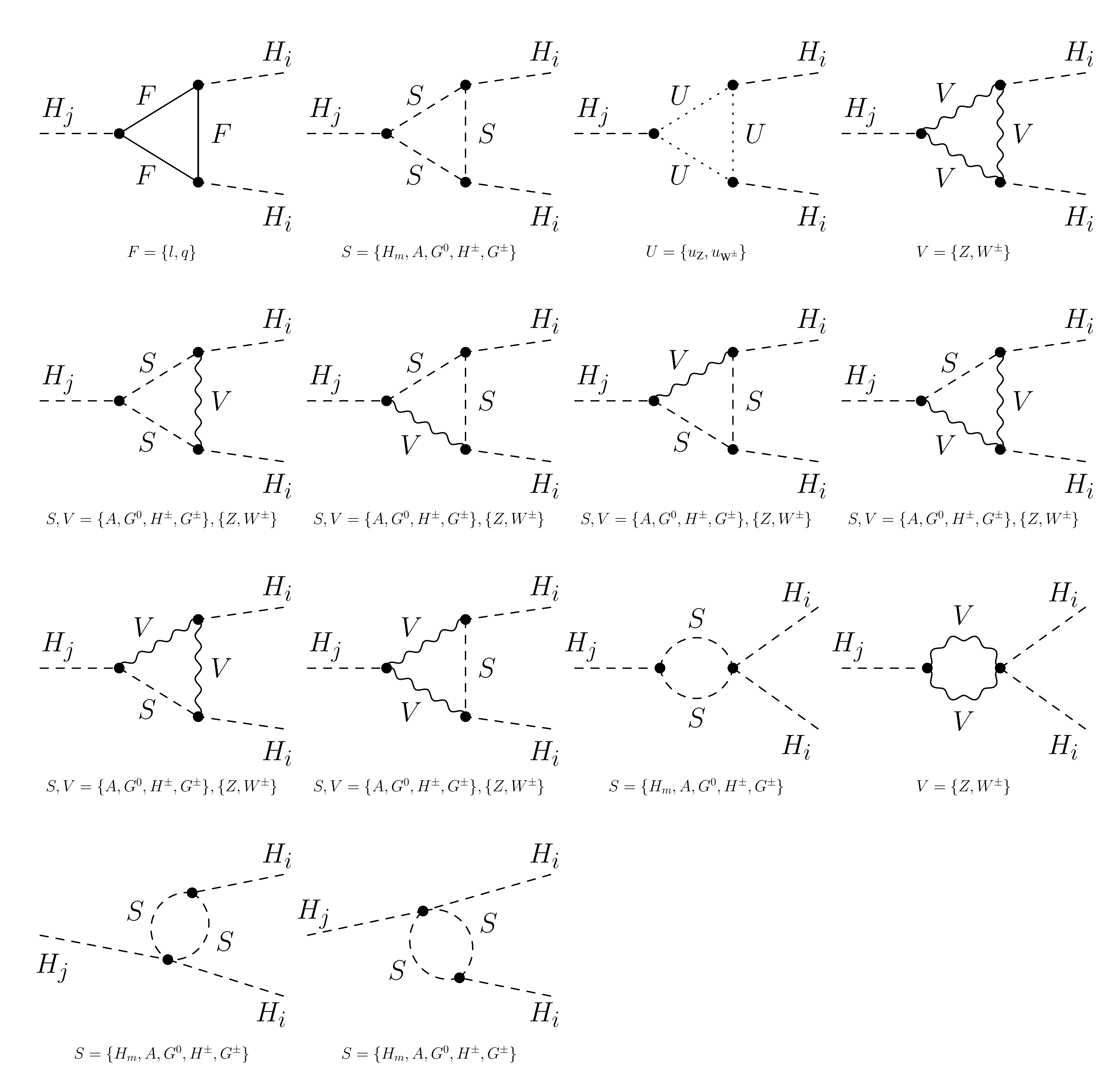}
\end{center}
\caption{Generic diagrams contributing to the vertex corrections
in  $H_j \to H_i H_i$.}
\label{fig:vertexdetails}
\end{figure}

\begin{figure}[t!]
\begin{center}
\includegraphics[width=16cm]{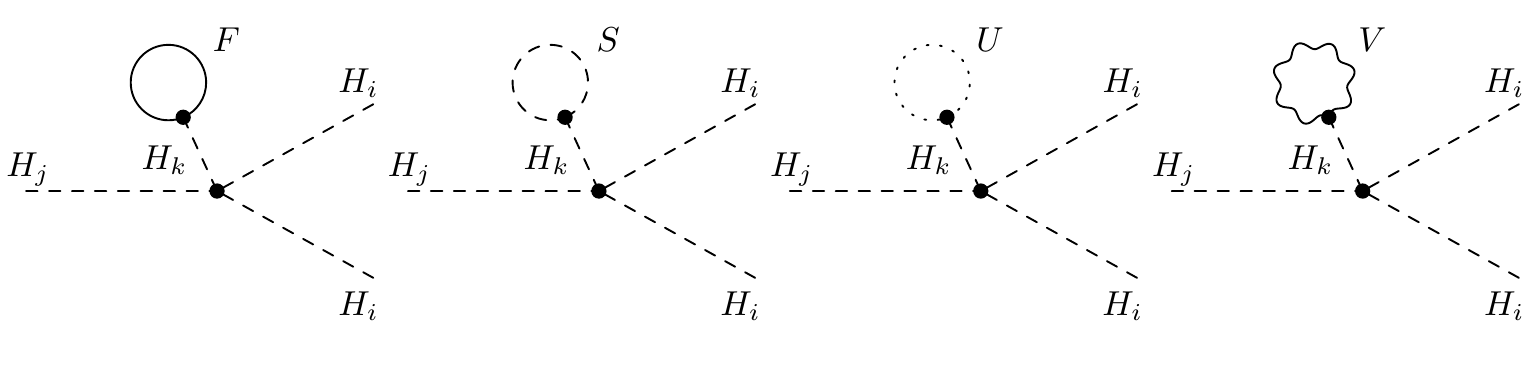}
\end{center}
\vspace*{-0.5cm}
\caption{Tadpole contributions
  to the vertex diagrams to be included in the decay $H_j\to H_i H_i$ in the
  alternative tadpole scheme.}
\label{fig:tadphitophiphi}
\end{figure}

The diagrams contributing to the virtual corrections of our process
$H_j \to H_i H_i$ are shown in 
Fig.~\ref{fig:vertexcorrsHiHjHk}. The 1PI diagrams contributing to 
the vertex corrections are depicted in Fig.~\ref{fig:vertexdetails}
and the tadpole diagrams are shown in
Fig.~\ref{fig:tadphitophiphi}. They have to be included in the 
alternative tadpole scheme. 
The counterterm is given by the genuine vertex counterterm and the counterterm
insertions on the external legs,
\beq
  \delta g_{H_i H_j H_k} = \delta\,g^{\text{field}}_{H_i H_j H_k}   
+ \delta\,g^{\text{vertex}}_{H_i H_j H_k}
  \label{eq:tripleHiHjHkctone}\, ,
\eeq
with 
\beq
 \delta\,g^{\text{field}}_{H_i H_j H_k} &=&
 g_{H_i H_j H_k}\, \left[\frac{1}{2}\,\sum_{l=1}^3 \,\cfrac{g_{H_l H_j
     H_k}}{g_{H_i H_j H_k}}\,\delta Z_{H_lH_i} + \frac{1}{2}\,\sum_{l=1}^3 \,\cfrac{g_{H_l H_i
     H_k}}{g_{H_i H_j H_k}}\,\delta Z_{H_lH_j} \right. \nonumber \\
&& \left. + \frac{1}{2}\,\sum_{l=1}^3 \,\cfrac{g_{H_l H_i
     H_j}}{g_{H_i H_j H_k}}\,\delta Z_{H_lH_k} \right] 
   \label{eq:tripleHiHjHkctfield}\, ,
\eeq
and
\begin{alignat}{5}
\delta g_{H_i H_j H_k}^{\text{vertex}}  
=& - g_{H_iH_iH_j}\,\left(\cfrac{\delta m_W^2}{2 m_W^2} - \cfrac{\delta g}{g} \right) 
 + \frac{1}{v}\,\Bigg{\{}
 -\cfrac{1}{2}\,\delta M^2\,\left[\left(\cfrac{R_{i2}}{\sb}-\cfrac{R_{i1}}{\cb} \right)\, \right. \times \notag \\
  \qquad &\times \left. \left(6R_{i2}R_{j2}\,\cb^2 -
  6R_{i1}R_{j1}\,\sb^2 + \sum_k\,\epsilon_{ijk}\,R_{k3}\,s_{2\beta}\right) \right]
  \notag \\
&  -\cfrac{1}{2}\,M^2\,\left[\left(\cfrac{\delta R_{i2}}{\sb}-\cfrac{\delta R_{i1}}{\cb} \right)\,
\left(6R_{i2}R_{j2}\,\cb^2 -
  6R_{i1}R_{j1}\,\sb^2 + \sum_k\,\epsilon_{ijk}\,R_{k3}\,s_{2\beta}\right) \right] \notag \\
&  -\cfrac{1}{2}\,M^2\,\left[\left(\cfrac{R_{i1}\,\delta \cb}{\cbd}-\cfrac{R_{i2}\,\delta\sb}{\sbd} \right)\,
\left(6R_{i2}R_{j2}\,\cb^2 -
  6R_{i1}R_{j1}\,\sb^2 +
  \sum_k\,\epsilon_{ijk}\,R_{k3}\,s_{2\beta}\right) \right] \notag
\end{alignat}
\begin{alignat}{5}
&  -\cfrac{1}{2}\,M^2\,\left(\cfrac{R_{i2}}{\sb}-\cfrac{R_{i1}}{\cb} \right)\,
\Big{[} 6R_{j2}\cbd\delta R_{i2} + 6R_{i2}\cbd\delta R_{j2} + 12
R_{i2}R_{j2}\cb\delta \cb  -  6R_{i1}\sbd\delta R_{j1} \notag \\
& - 6R_{i1}\sbd\delta R_{j1} - 12 R_{i1}R_{j1}\sb\delta \sb + \sum_k\,\epsilon_{ijk}\,(\sb\,\delta R_{k3} + 2 R_{k3}\,(\cb\delta \sb+\sb\delta \cb) 
\Big{]} \notag \\
&
+ \cfrac{2\delta m^2_{H_i}+\delta m^2_{H_j}}{v_S}\,\left[R^2_{i3}\,R_{j3}\,v
+ R^2_{i2}\,R_{j2}\,\cfrac{v_S}{\sb}+
R^2_{i1}\,R_{j1}\,\cfrac{v_S}{\cb}\right] \notag \\
& - \cfrac{v}{v_S}\,(2m^2_{H_i} + m^2_{H_j})\,R^2_{i3}\,R_{j3}\,\cfrac{\Delta\,v_S}{v_S}
+ \cfrac{2m^2_{H_i}+m^2_{H_j}}{v_S}\,\Big{[}2 R_{i3}\,R_{j3}\,v\,\delta R_{i3} +
R^2_{i3}\,v\,\delta R_{j3} \notag \\
& + R^2_{i3}\,R_{j3}\,\delta v + 2 R_{i2}\,R_{j2}\,\cfrac{v_S}{\sb}
\delta R_{i2}  
+ R^2_{i2}\,\cfrac{v_S}{\sb}\delta R_{j2}
- R^2_{i2}\,R_{j2}\,\cfrac{v_S}{\sbd}\delta \sb
+ 2 R_{i1}\,R_{j1}\,\cfrac{v_S}{\cb} \delta R_{i1} \notag \\
& + R^2_{i1}\,\cfrac{v_S}{\cb}\delta R_{j1}- R^2_{i1}\,R_{j1}\,\cfrac{v_S}{\cbd}\,\delta \cb\Big{]}
\Bigg{\}}
   \label{eq:tripleiijctparam}\, . 
\end{alignat}
The NLO corrections factorise so that the loop-corrected decay
width can be cast into the form
\begin{alignat}{5}
\Gamma^{\text{NLO}} &= \Gamma^{\text{LO}}\,
    [1+ \Delta^{\text{virt}}_{H_i H_j H_k} + \Delta^{\text{ct}}_{H_i
      H_j H_k}]\;,
\end{alignat}
with 
\beq
\Delta^{\text{virt/ct}}_{H_i H_j H_k} \equiv \frac{2 {\cal
    M}^{\text{virt/ct}}_{H_i H_j H_k}}{g_{H_i H_j H_k}} 
= \frac{2 {\cal
    M}^{\text{virt/ct}}_{H_i H_j H_k}}{-i \cdot \lambda_{H_i H_j H_k}} 
\eeq
in terms of the virtual corrections and counterterm amplitude ${\cal
  M}^{\text{virt}}_{H_i H_j H_k}$ and ${\cal M}^{\text{ct}}_{H_i H_j
  H_k}$, respectively, where we have included the vertices with the
tadpoles in ${\cal M}^{\text{ct}}_{H_i H_j H_k}$. Due to rather
lengthy expressions we refrain from giving the explicit expressions of
the various contributions to $\Gamma^{\text{NLO}}$.
%

\section{Numerical Analysis \label{sec:numerical}}
For the computation of the NLO EW corrections to the Higgs decays
presented in the following the tree-level and one-loop decay
amplitudes have been generated with {\tt FeynArts}
\cite{Kublbeck:1990xc,Hahn:2000kx}. The necessary N2HDM Feynman rules have been obtained as UFO \cite{Degrande:2011ua} and FeynArts \cite{Hahn:2000kx}
model files using FeynRules\cite{Alloul:2013bka}, while all
renormalization counterterms have been derived analytically and implemented by hand.
The amplitudes have been analytically processed via {\tt FormCalc}
\cite{Hahn:1998yk}. The dimensionally regularised loop form factors have been 
evaluated in the 't Hooft-Veltman scheme
\cite{'tHooft:1972fi,Bollini:1972ui} and written in terms of  standard
loop integrals. These have been further reduced through 
Passarino-Veltman decomposition and evaluated with the help of {\tt
    LoopTools} \cite{Hahn:1998yk}. \s

In the following we give the input parameters for the
numerical evaluation. As explained in section~\ref{sec:renconditions} we use
the fine structure constant $\alpha$ at the $Z$ boson mass scale,
given by \cite{Agashe:2014kda} 
\beq
\alpha (m_Z^2) = \frac{1}{128.962} \;.
\eeq
The massive gauge bosons are renormalized OS, and their input masses
are chosen as \cite{Agashe:2014kda,Denner:2047636}
\beq
m_W = 80.385 \mbox{ GeV} \qquad \mbox{and} \qquad m_Z = 91.1876 \mbox{
GeV} \;.
\eeq
For the lepton masses we take \cite{Agashe:2014kda,Denner:2047636}
\beq
m_e = 0.510998928 \mbox{ MeV} \;, \quad
m_\mu = 105.6583715 \mbox{ MeV} \;, \quad
m_\tau = 1.77682 \mbox{ GeV} \;.
\eeq
These and the light quark masses, which we set \cite{LHCHXSWG}
\beq
m_u = 100 \mbox{ MeV} \;, \quad m_d = 100 \mbox{ MeV} \;, \quad
m_s = 100 \mbox{ MeV} \;,
\eeq
have only a small impact on our results. Following the recommendation
of the LHC Higgs Cross Section Working Group (HXSWG)
\cite{Denner:2047636,Dittmaier:2011ti}, we use the following
OS value for the top quark mass 
\beq
m_t = 172.5 \mbox{ GeV} \;,
\eeq
which is consistent with the ATLAS and CMS analyses.
The charm and bottom quark OS masses are set to
\beq
m_c = 1.51 \mbox{ GeV} \qquad \mbox{and} \qquad
m_b = 4.92 \mbox{ GeV} \;,
\eeq
as recommended by \cite{Denner:2047636}.
We consider the CKM matrix to be unity. This approximation has
negligible impact on our results. The SM-like Higgs mass value,
denoted by $m_{h}$, has been set to \cite{Aad:2015zhl}
\beq
m_{h} = 125.09 \mbox{ GeV} \;. 
\eeq
Note that, depending on the parameter set, in the N2HDM any of
the three neutral CP-even Higgs bosons can be the SM-like Higgs boson. \s 

In the subsequently presented analysis we only used N2HDM parameter sets
compatible experimental and theoretical constraints. These data sets
have been generated with the tool {\tt ScannerS}
\cite{Coimbra:2013qq,scanners}.\footnote{We thank Marco 
  Sampaio, one of the authors of {\tt ScannerS}, and Jonas Wittbrodt who kindly provided
  us with the necessary data sets.} The applied theoretical
constraints require that the vacuum state found by {\tt ScannerS} is
the global minimum, that the N2HDM potential is bounded from below and
that tree-level unitarity holds. On the experimental side,
compatibility with the EW precision constraints is guaranteed by
requiring the oblique parameters $S$, $T$ and $U$ to be compatible 
with the SM fit \cite{Baak:2014ora} at $2\sigma$, including the full
correlations. The constraints from $B$ physics observables 
\cite{Mahmoudi:2009zx,Deschamps:2009rh,Hermann:2012fc,Misiak:2015xwa,Misiak:2017bgg} and the
measurement of $R_b$ \cite{Deschamps:2009rh,Haber:1999zh} have
been taken into account, as well as the most recent bound of
$m_{H^\pm} \gsim 580$~GeV for the  type II and flipped (N)2HDM
\cite{Misiak:2017bgg}. For the compatibility with the LHC Higgs data
we require one of the 
scalar states, denoted by $h_{125}$, to have a mass of 125.09~GeV
and to match the observed LHC signal rates. Furthermore, the remaining
Higgs bosons have to be consistent with the exclusion bounds from the
collider searches at Tevatron, LEP and LHC. For
further details on these checks and the scan, we refer to
\cite{Muhlleitner:2016mzt,Muhlleitner:2017dkd}. \s

Note that in all scenarios presented in the following we stick to the
N2HDM type I, with the type II scenarios leading to the same overall
results. The only difference between the models comes from the fermion
loops. The Yukawa couplings are, in all Yukawa types, well-behaved
functions of the $\alpha_i$ and $\beta$ because extreme values of
$\beta$ are already disallowed by all the constraints imposed on the
model. Therefore, this is sufficient for our analysis to illustrate
the effects of the EW corrections, without aiming at a full phenomenological
analysis of N2HDM Higgs decays. 

\subsection{Results for $H_{2/3} \to ZZ$}
In this section we investigate the relative size of the NLO EW
corrections as well as the impact of the different renormalization
schemes for the mixing angles on the decay $H_i \to ZZ$. We base our
numerical analysis upon a set of representative N2HDM 
scenarios of phenomenological interest. To this aim we select among
the generated parameter points compatible with the theoretical and
experimental constraints scenarios that either have a large or a small
LO branching fraction (BR) into $ZZ$. Discarding the SM-like decay of the
$H_1$ fixed to be the 125~GeV Higgs boson, we select hence four
scenarios, two for $H_2$ and $H_3$, respectively, which we denote by
'BRH2/3high' and 'BRH2/3low' for high and low branching ratio
scenarios. The corresponding input parameters are listed in
Table~\ref{tab:benchmarkHZZ}. Note that, if not stated otherwise, the
mixing angles are understood to be the angles defined in the OS
tadpole pinched scheme (pOS) with $\delta \beta$ defined via the charged
sector, denoted by the superscript '$c$'.\footnote{While the scheme choice
  is not relevant for the LO 
  width alone, it becomes important when the NLO EW corrections are
  included. The renormalization of the parameters then fixes 
  the scheme of the input parameters at LO.} The
suppressed branching fractions in the BRlow scenarios are 
due to a small tree-level coupling to $ZZ$ of the decaying Higgs
boson. The branching fractions given in this table have been obtained with the Fortran
code {\tt N2HDECAY}.\footnote{\tt N2HDECAY can be obtained from \url{https://www.itp.kit.edu/~maggie/N2HDECAY/}.} 
We insured to consider purely OS decays into massive gauge bosons in {\tt N2HDECAY}, 
as we do not include any gauge boson off-shell effects in the NLO computation.  
\s

\begin{table}[hbt!]
 \begin{center}
  \begin{tabular}{l|ll|ll} \hline & BRH2ZZhigh & BRH3ZZhigh &  BRH2ZZlow & BRH3ZZlow\\ \hline
$m_{H_1}$ & 125.09 & 125.09 & 125.09 & 125.09 \\     
 $m_{H_2}$ & 673.70  & 600.76  & 657.07 & 283.53 \\ 
 $m_{H_3}$ & 692.22 & 713.74 &  658.28 &  751.72\\
 $m_A$ &  669.07 & 743.00  & 543.62 & 763.09   \\ 
 $m_{H^{\pm}}$ & 679.76 & 695.73  &  528.76 & 733.05\\ 
 $t_\beta$ (pOS$^c$) & 6.12  & 8.39 &4.79 &3.53 \\ 
 $\aone$ (pOS) & -1.513  &-1.526 &  -1.489 &1.318 \\
 $\atwo$ (pOS) & 0.098 &-0.308  & 0.225 &0.0362 \\
 $\athree$ (pOS) & -0.495  & -1.421  &-1.001 & 1.504  \\
 $m^2_{12}$ & 74518.4 &60125.0  &87240.8  &143579.0 \\
 $v_s$ & 305.48 &  854.50 & 834.33 & 219.29 \\ \hline
 $\Gamma_H$ & 2.946 & 2.241 &  2.990 & 2.746\\
 BR & 0.327 & 0.329& 0.010 & 0.010\\ \hline
  \end{tabular}
\vspace*{-0.3cm}
 \end{center}
 \caption{Input parameters for the N2HDM benchmark scenarios used in
the numerical analysis of the decay processes $H_{2/3} \to ZZ$. In
round brackets we specify the scheme in which $\alpha$ and $\beta$ are
defined. All masses and $v_S$ are
given in GeV. The LO total width (also given in GeV) and individual
branching fractions in the last two rows correspond to the Higgs
state and decay each benchmark is named after, and have been
generated with {\tt N2HDECAY}.
}
\label{tab:benchmarkHZZ}
\end{table}
\vspace*{0.4cm}
\begin{table}[htb!]
 \begin{center}
  \begin{tabular}{|l|l||l|l|l|l|} \hline
  & & pOS$^c$ & pOS$^o$ & p$_\star^c$ & p$_\star^o$ \\ \hline
\multirow{3}{*}{BRH2ZZhigh}  &$\Gamma^{\text{LO}}(H_2 \to ZZ)$ & 0.989& 0.989 & 1.008& 1.008 \\ 
 &$\Gamma^{\text{NLO}}(H_2 \to ZZ)$ & 1.120 & 1.122 & 1.142  & 1.148  \\
 & $\Delta \Gamma^{H_2 ZZ}$ [\%] & 13.2 & 13.4 & 13.3 & 14.0  \\ \hline
\multirow{3}{*}{BRH3ZZhigh}  &$\Gamma^{\text{LO}}(H_3 \to ZZ)$  & 0.755 & 0.755 & 0.782 & 0.782 \\ 
 &$\Gamma^{\text{NLO}}(H_3 \to ZZ)$ & 0.872 & 0.867  & 0.890   & 0.889  \\
 & $\Delta \Gamma^{H_3 ZZ}$ [\%] & 15.6 & 14.9 & 13.9  & 13.7 \\ \hline \hline
\multirow{3}{*}{BRH2ZZlow}  &$\Gamma^{\text{LO}}(H_2 \to ZZ)$  & 3.130$\times 10^{-2}$ & 3.130$\times 10^{-2}$ & 2.529$\times 10^{-2}$ & 2.533$\times 10^{-2}$  \\ 
 &$\Gamma^{\text{NLO}}(H_2 \to ZZ)$& 3.042$\times 10^{-2}$ &3.040$\times 10^{-2}$ & 2.840$\times 10^{-2}$  &2.745$\times 10^{-2}$\\
 & $\Delta \Gamma^{H_2 ZZ}$ [\%] & -2.8 & -2.9& 12.3& 8.4
\\ \hline 
\multirow{3}{*}{BRH3ZZlow}  &$\Gamma^{\text{LO}}(H_3 \to ZZ)$ & 2.870$\times 10^{-2}$&2.869$\times 10^{-2}$&3.430$\times 10^{-2}$&3.418$\times 10^{-2}$ \\ 
 &$\Gamma^{\text{NLO}}(H_3 \to ZZ)$ & 2.990$\times 10^{-2}$ & 3.011$\times 10^{-2}$ & 3.593$\times 10^{-2}$ & 3.738$\times 10^{-2}$\\
  & $\Delta \Gamma^{H_3 ZZ}$ [\%]  & 4.2 & 5.0& 4.8  & 9.3
\\ \hline
\end{tabular}
 \end{center}
 \caption{Higgs decay widths (in GeV) at LO and NLO EW
   accuracy as well as the relative corrections for the N2HDM benchmarks presented in
   Table~\ref{tab:benchmarkHZZ} and four different renormalization schemes.}
 \label{tab:resultsHZZ}
\end{table}
In Table~\ref{tab:resultsHZZ} we present for all four benchmark
scenarios the results for the LO and the NLO width as well as the
relative corrections $\Delta\Gamma$. They are
given for four different renormalization schemes. These consist of 
the p$_\star$ and the pOS tadpole pinched schemes, which employ two
different renormalization scales, and for these additionally the two
possibilities to renormalize $\beta$, either via the charged sector (denoted
by '$c$') or the CP-odd sector (denoted by '$o$'). The
relative corrections are defined as
\beq
\Delta \Gamma \equiv \frac{\Delta \Gamma^{\text{NLO}}}{\Gamma^{\text{LO}}}
= \frac{\Gamma^{\text{NLO}}-
  \Gamma^{\text{LO}}}{\Gamma^{\text{LO}}} \;.
\label{eq:defdeld}
\eeq

When computing the NLO EW corrected decay width $\Gamma^{\text{NLO}}$
in a different renormalization scheme $b$ than the one of the input
parameters $p$, scheme $a$, these parameters first have to be converted to
the scheme that is applied. We perform this conversion for the mixing
angles $\alpha$ and $\beta$ through ($p=\alpha,\beta$)
\beq
p^b = p^a + \delta p^a - \delta p^b \;,
\eeq
where $\delta p$ denotes the counterterm in either scheme $a$ or
scheme $b$. With the thus obtained input parameters in scheme $b$ we
compute the quantity $\Delta \Gamma^{\text{NLO}}$ and the LO width
$\Gamma^{\text{LO}}$, to which we normalize the relative
correction.\footnote{Note that the LO widths given in
  Table~\ref{tab:resultsHZZ} for the pOS$^c$ scheme slightly differ
  from the values as obtained from the corresponding BRs and total
  widths given in Table~\ref{tab:benchmarkHZZ}, since, in 
  consistency with our NLO computation, we use as
  input parameters $m_W$, $m_Z$ and $\alpha$, while in {\tt N2HDECAY} 
  all decay widths are expressed in terms of the Fermi constant $G_F$ as input value.
  Including in our LO results the SM correction
  $\Delta r^{\text{SM}}$
  \cite{Kennedy:1988sn,Hollik:1988ii,Awramik:2003rn}, which relates $m_W$ to $G_F$, would bring the derived Fermi
  constant numerically very close to the PDG value $G_F=1.166 \cdot
  10^{-5}$~GeV$^{-1}$ used in {\tt N2HDECAY}.} \s

The relative corrections for the scenarios with relatively large
branching ratios turn out to be of moderate size with values between 13.2
and 15.6\%. The variation due to
different renormalization schemes is at most 1.9\%, indicating a
relatively small theoretical error due to missing higher order
corrections. For the low branching ratio scenarios on the other hand,
the differences between the results for the various renormalization
schemes are substantial. This points towards a large theoretical error
due to missing higher order corrections. A reliable prediction 
in these cases would require the inclusion of
corrections beyond one-loop order. This is to be expected as the
tree-level widths are very small in these scenarios so that the
one-loop correction effectively becomes the leading contribution to
the width. When changing from
the charged to the CP-odd based renormalization of $\beta$, 
the change in the relative corrections is rather mild for
most of the scenarios. This is because the two different scales,
$m_{H^\pm}$ or $m_A$, involved in 
these two renormalization schemes of $\beta$ are close in our scenarios.

\subsection{Results for $H_{2/3} \to AA$}
\begin{table}[b!]
 \begin{center}
  \begin{tabular}{l|ll|ll} \hline & BRH2AAhigh & BRH3AAhigh & BRH2AAlow & BRH3AAlow\\ \hline
 $m_{\hone}$ & 125.09 & 125.09 & 125.09 & 125.09 \\     
 $m_{\htwo}$ & 130.48 & 137.15  & 294.92 & 243.70 \\ 
 $m_{\hthree}$ & 347.65 & 146.22 &  503.44 &  903.07\\
 $m_A$ & 58.14 &  70.27  & 74.28 &429.82   \\ 
 $m_{H^{\pm}}$ & 146.93 & 166.83   & 278.19 &426.18\\ 
 $t_\beta$ (pOS$^c$) & 5.89  &  5.55 & 6.12 & 4.01 \\ 
 $\aone$ (pOS) & -1.535  &1.338 &   -1.457 &1.409\\
 $\atwo$ (pOS) & 0.369 &0.095  &  -0.117 &-0.195 \\
 $\athree$ (pOS) &0.029  & -1.28 & -0.118 & -0.078  \\
 $m^2_{12}$ ($\mu_R=2 m_A$) & 864.2 &982.9  &13036.9  &8300.6  \\
 $v_s$ & 538.37 &  638.95 &1352.51  & 991.00\\ \hline
 $\Gamma_H$ &  2.694 &  2.005 &  4.986 $\cdot 10^{-2}$ & 26.140 $\cdot 10^{-2}$\\
 BR & 0.999 & 0.999 & 0.997 & 0.992\\ \hline
  \end{tabular}
 \end{center}
\caption{Input parameters for the N2HDM benchmarks used in the numerical 
 analysis of the decay process $H_{2/3} \to AA$. All masses and $v_S$ are given
 in GeV. The LO total width (also given in GeV) and
 individual branching fractions in the last two rows correspond to the
 Higgs state and decay each benchmark is named after, and have been generated
 with {\tt N2HDECAY}.} 
\label{tab:benchmarkHAA}
\end{table}
%
Here we study the decay into a pair of pseudoscalars and again 
concentrate on the decays of the heavier Higgs bosons $H_2$ and $H_3$ and choose
scenarios where $H_1$ is the 125~GeV Higgs boson\footnote{We do not
  consider $H_1$ decays into $AA$. They would require $m_A$ to be
  below about 65~GeV and care would have to be taken to keep the decay
$H_1 \to AA$ small enough to still be compatible with the LHC Higgs
data.} and with low and high 
branching ratios for $H_{2/3} \to AA$, respectively. The corresponding
benchmark scenarios are called 'BRH2/3AAhigh' and 'BRH2/3AAlow', with
the input values summarised in Tab.~\ref{tab:benchmarkHAA} together
with the LO total widths and branching ratios computed with {\tt N2HDECAY}. The input
mixing angles are given in the pOS scheme and the $\beta$
renormalization is based on the charged sector. The parameter
$m_{12}^2$ is assumed to be given at the scale $\mu_R= 2
m_A$\footnote{This
  choice was shown to yield the most stable results 
for the 2HDM \cite{Krause:2016xku}.}. 
The suppressed decay widths in the BRlow scenarios are due to a
small trilinear coupling $\lambda_{H_{2/3} AA}$. In the BRhigh
scenarios, the $H_{2/3} \to AA$ decays are maximised because $(i)$ the $H_{2/3} AA$
trilinear couplings are enhanced, $(ii)$ the couplings to fermions are
suppressed and $(iii)$ the decays into massive weak bosons are
kinematically closed. \s 

\begin{table}[htb!]
 \begin{center}
  \begin{tabular}{|l|l||l|l|l|l|} \hline
  & & pOS$^c$ & pOS$^o$ & p$_\star^c$ & p$_\star^o$ \\ \hline
\multirow{3}{*}{BRH2AAhigh}  &$\Gamma^{\text{Born}}(\htwo \to AA)$  & 2.761 & 2.759& 2.761& 2.760   \\ 
 &$\Gamma^{\text{NLO}}(\htwo \to AA)$ & 2.454 & 2.500  & 2.459  & 2.500  \\
 & $\Delta \Gamma^{H_2 AA}$ [\%]  & -11.1 & -9.4 & -10.9 & -9.4  \\ \hline
\multirow{3}{*}{BRH3AAhigh}  &$\Gamma^{\text{Born}}(\hthree \to AA)$& 2.054& 2.053 & 2.042& 2.041  \\ 
 &$\Gamma^{\text{NLO}}(\hthree \to AA)$ & 1.840 & 1.885  & 1.848   & 1.886 \\
 & $\Delta \Gamma^{H_3 AA}$ [\%]  & -10.4 & -8.1 & -9.5  & -7.6 \\ \hline \hline
\multirow{3}{*}{BRH2AAlow}  &$\Gamma^{\text{Born}}(\htwo \to AA)$  & 5.097$\times 10^{-2}$ & 5.266$\times 10^{-2}$ & 5.075$\times 10^{-2}$&5.208$\times 10^{-2}$  \\ 
 &$\Gamma^{\text{NLO}}(\htwo \to AA)$ & 5.408$\times 10^{-2}$ & -1.013$\times 10^{-2}$ &  4.071$\times 10^{-2}$   & -9.986$\times 10^{-3}$ \\
 & $\Delta \Gamma^{H_2 AA}$ [\%] & 6.1 & -119.2& -19.8& -119.2 \\ \hline 
\multirow{3}{*}{BRH3AAlow}  &$\Gamma^{\text{Born}}(\hthree \to AA)$& 0.266&0.266 &0.286&0.286  \\ 
 &$\Gamma^{\text{NLO}}(\hthree \to AA)$ & 0.277 & 0.272 & 0.270 & 0.277 \\
 & $\Delta \Gamma^{H_3 AA}$ [\%]  & 4.4 & 2.1& -5.5  & -3.0  \\ \hline
\end{tabular}
 \end{center}
 \caption{Higgs decay widths (in GeV) at LO and NLO EW accuracy as
   well as the relative corrections for the N2HDM
 benchmarks presented in Table~\ref{tab:benchmarkHAA} and four different
 renormalization schemes. The renormalization scale of $m_{12}^2$ is
 set to $\mu_R = 2 m_A$.}
 \label{tab:resultsHAA}
\end{table}
In Table~\ref{tab:resultsHAA} we display for all four benchmark
scenarios the LO and NLO widths as well as the
relative corrections $\Delta\Gamma$. They are
given for the four different renormalization schemes, p$_\star^{c/o}$,
pOS$^{c/o}$. 
As can be inferred from the table, for the BRhigh scenarios we obtain
moderate corrections  of $\mathcal{O}(10)\%$, \textit{i.e.}~of the same order as
for $H_{2/3} \to ZZ$. The associated theoretical uncertainties are
very mild, as indicated by the the rather small influence of the
renormalization schemes of the mixing angles, which lead to
a change of at most 2.8\%, when considering all four schemes. The
$H_{2/3} \to AA$ decays in the BRlow scenarios, on the contrary, 
are dominated by the loop effects. Here, the small trilinear Higgs coupling
suppresses the tree-level width. At one loop, however, the Higgs decay
is also sensitive to the additional trilinear Higgs couplings, some of
which being very large as a result 
of the heavy Higgs masses and the large $m^2_{12}$ scale - yet in agreement with
the unitarity constraints. This results in very large 
NLO effects, and also induces the strong dependence on the
renormalization scheme and 
the renormalization scale. This reflects the fact  that the $H_{2/3} \to AA$
decays in these benchmarks are effectively loop-induced and higher
order corrections beyond the one-loop level need to be considered to
make reliable predictions. These sizable higher-order effects are particularly 
apparent in the BRH2AAlow scenario, where some of the renormalization
schemes even lead to negative and hence unphysical NLO widths. 
Note, furthermore, that the change when switching from the charged to
the CP-odd based renormalization schemes for $\beta$ is now larger when
compared to the results in Table~\ref{tab:resultsHZZ}, due to
the now wider separation  between the scales given by the charged
and the CP-odd Higgs mass as compared to the 
scenarios studied for the decays into a $Z$ boson pair.

\subsection{Results for $H_{3} \to H_2\,H_2$ and $H_2 \to H_1 H_1$}
\begin{table}[b!]
 \begin{center}
  \begin{tabular}{l|llll} \hline & HHHI & HHHII & HHHIII & HHHIV \\ \hline
 $m_{\hone}$ & 125.09 & 125.09 & 125.09 & 125.09  \\     
 $m_{\htwo}$ & 304.18 & 425.61 & 351.65& 298.42    \\ 
 $m_{\hthree}$ & 630.94 & 857.27   &  717.32 & 743.18 \\
 $m_A$ & 325.07 & 547.48  & 487.07 & 362.40  \\ 
 $m_{H^{\pm}}$ &265.81 & 383.85 & 386.42   & 306.19 \\ 
 $t_\beta$ (pOS$^c$) &6.30 & 5.17  & 4.08& 6.26 \\ 
 $\aone$ (pOS) & -1.559 & 1.495 & 1.453   & 1.315   \\
 $\atwo$ (pOS) & -0.330 &0.082  & 0.353 & -0.148 \\
 $\athree$ (pOS) &-0.077 & -0.101 & 0.340 & -0.098 \\
 $m^2_{12}$ ($\mu _R = 2 m_{H_{\text{final}}}$) & 14312.1 &  32824.5 & 35765.3 & 12707.3 \\
 $v_s$  & 1327.57 &  1098.81  & 630.19& 1425.0  \\ \hline
 $\Gamma_{H_3}$ & 24.160 &25.190 & 43.590 &  18.750  \\
 BR($H_3\to H_1 H_1$) & 0.13 & 0.03 &  0.08 & 0.08\\
 BR($H_3\to H_2 H_2$) & 0.05 & 0.10 & 0.15 & 0.15 \\ \hline
 $\Gamma_{H_2}$ & 0.393 & 0.723 & 1.558 & 0.234  \\
 BR($H_2\to H_1 H_1$) & 0.17 & 0.47 & 0.43 & 0.76   \\ \hline  
 \end{tabular}
 \end{center}
 \caption{Input parameters for the N2HDM benchmarks used in the numerical 
 analysis of the decay processes $H_j\to H_iH_i$. All masses and
 $v_S$ are given in GeV. In the last five rows the total $H_2$ and
 $H_3$ widths are given in GeV as well as the branching fractions
 (generated with {\tt N2HDECAY}) of the Higgs-to-Higgs decays $H_3 \to
 H_1 H_1, H_2 H_2$ and $H_2 \to H_1 H_1$.}
\label{tab:benchmarkHHH}
\end{table}

\begin{table}[htb!]
 \begin{center}
  \begin{tabular}{|l|l||l|l|l|l|} \hline
  &  & pOS$^c$ & pOS$^o$ & p$_\star^c$ & p$_\star^o$ \\ \hline
\multirow{3}{*}{HHHI}  &$\Gamma(\hthree \to \hone\hone)$ & 3.206& 3.206 & 3.197& 3.197 \\ \cline{2-6}
 &$\Gamma^{\text{LO}}(\hthree \to \htwo\htwo)$ & 1.229 & 1.229 & 1.242  & 1.242  \\
 &$\Gamma^{\text{NLO}}(\hthree \to \htwo\htwo)$ & 1.344 & 1.343 & 1.344  & 1.341  \\ 
 & $\Delta\Gamma^{\hthree \to \htwo\htwo}$ [\%] & 9.4 & 9.3 & 8.2 & 8.0 \\ \cline{2-6} 
&$\Gamma^{\text{LO}}(\htwo \to \hone\hone)$ & $6.699\times 10^{-2}$ & $6.699\times 10^{-2}$ & $6.667\times 10^{-2}$  & $6.667\times 10^{-2}$  \\ 
 &$\Gamma^{\text{NLO}}(\htwo \to \hone\hone)$ & $7.433\times 10^{-2}$ & $7.429\times 10^{-2}$ & $7.429\times 10^{-2}$  & $7.409\times 10^{-2}$  \\ 
 & $\Delta\Gamma^{\htwo \to \hone\hone}$ [\%] & 11.0 & 10.9 & 11.4 & 11.1 \\ \hline 
 \multirow{3}{*}{HHHII}  &$\Gamma(\hthree \to \hone\hone)$ & 0.719& 0.719 & 0.753& 0.753 \\ \cline{2-6}
 &$\Gamma^{\text{LO}}(\hthree \to \htwo\htwo)$ & 2.580 & 2.580 & 2.730  & 2.730  \\
 &$\Gamma^{\text{NLO}}(\hthree \to \htwo\htwo)$ & 2.453 & 2.454 & 2.493  & 2.492  \\ 
 & $\Delta\Gamma^{\hthree \to \htwo\htwo}$ [\%] & -4.9 & -4.9 & -8.7 & -8.7 \\ \cline{2-6} 
&$\Gamma^{\text{LO}}(\htwo \to \hone\hone)$ & 0.345 & 0.345 & 0.343  & 0.343  \\ 
 &$\Gamma^{\text{NLO}}(\htwo \to \hone\hone)$ & 0.398 & 0.398 & 0.397 & 0.397  \\ 
 & $\Delta\Gamma^{\htwo \to \hone\hone}$ [\%] & 15.2 & 15.2 & 15.9 & 15.9 \\ \hline 
 \multirow{3}{*}{HHHIII}  &$\Gamma(\hthree \to \hone\hone)$ & 3.561& 3.561 & 3.565& 3.564 \\ \cline{2-6}
 &$\Gamma^{\text{LO}}(\hthree \to \htwo\htwo)$ & 6.662 & 6.661 & 6.469  & 6.466  \\
 &$\Gamma^{\text{NLO}}(\hthree \to \htwo\htwo)$ & 6.071 & 6.094 & 6.208  & 6.264  \\ 
 & $\Delta\Gamma^{\hthree \to \htwo\htwo}$ [\%] & -8.883 & -8.515 & -4.027 & -3.118 \\ \cline{2-6} 
&$\Gamma^{\text{LO}}(\htwo \to \hone\hone)$ & 0.687 & 0.687 & 0.684  & 0.683  \\ 
 &$\Gamma^{\text{NLO}}(\htwo \to \hone\hone)$ & 0.678 & 0.679 & 0.675 & 0.676  \\ 
 & $\Delta\Gamma^{\htwo \to \hone\hone}$ [\%] & -1.3 & -1.2 & -1.3 & -1.1 \\ \hline   
 \multirow{3}{*}{HHHIV}  &$\Gamma(\hthree \to \hone\hone)$ & 1.446& 1.446 & 1.422& 1.422 \\ \cline{2-6}
&$\Gamma^{\text{LO}}(\hthree \to \htwo\htwo)$ & 2.873 & 2.874 & 2.860  & 2.859  \\ 
 &$\Gamma^{\text{NLO}}(\hthree \to \htwo\htwo)$ & 2.793 & 2.780 & 2.799 & 2.820  \\ 
 & $\Delta\Gamma^{\hthree \to \htwo\htwo}$ [\%] & -2.8 & -3.3 & -2.1 & -1.4 \\  \cline{2-6}  
&$\Gamma^{\text{LO}}(\htwo \to \hone\hone)$ & 0.183 & 0.183 & 0.185  & 0.185  \\ 
 &$\Gamma^{\text{NLO}}(\htwo \to \hone\hone)$ & 0.151 & 0.144 & 0.147 & 0.158  \\ 
 & $\Delta\Gamma^{\htwo \to \hone\hone}$ [\%] & -17.4 & -21.3 & -20.6 & -14.3 \\ \hline   
\end{tabular}
 \end{center}
 \caption{Higgs decay width predictions (in GeV) at LO and NLO EW
   accuracy as well as the relative corrections for the N2HDM
   benchmarks presented in Table~\ref{tab:benchmarkHHH} and four
   different renormalization schemes.}
 \label{tab:resultsHHH}
\end{table}
Finally, we consider the decay of a heavy neutral CP-even Higgs boson into 
a pair of lighter CP-even Higgs bosons. We evaluate the NLO EW corrections
for a number of illustrative scenarios, given in
Table~\ref{tab:benchmarkHHH}. The scenarios have been chosen such that
their Higgs mass spectra allow simultaneously for the OS $H_3 \to H_2
H_2$ and $H_2 \to H_1 H_1$ decays. Furthermore, the chosen large
$m^2_{12}$ parameter insures these heavy Higgs mass scenarios to be in
agreement with the unitarity and vacuum stability constraints. All scenarios feature
Higgs-to-Higgs decay branching ratios that are of moderate size. Only
HHHIV features a $H_2$ branching ratio into $H_1 H_1$ that is
dominating. All input mixing angles are assumed to be given
in the pOS scheme, with charged sector-based renormalization
for the angle $\beta$, and $m_{12}^2$ is assumed to be defined at the
renormalization scale given by the total final state mass, $\mu_R = 2 m_{H_i}$. 
The LO total widths and branching ratios in this table have been obtained
from {\tt N2HDECAY}. \s

In Table~\ref{tab:resultsHHH} we summarise the relative NLO
corrections for the various decays. Note, that the decay process $H_3
\to H_1 H_1$ appears only at LO because we use it for the renormalization
of $v_S$, as explained in detail in Section~\ref{sec:oneloopdec}. The
sizeable $m^2_{12}$ and heavy Higgs mass values imply large Higgs 
self-couplings and thereby enhanced contributions from the virtual
Higgs exchanges. On the other hand, these enhancements are partly damped by
the inverse Higgs mass powers in the Higgs-mediated loops. The balance between these
dynamical features governing the Higgs-mediated loops, and how they interplay
with the remaining gauge boson and fermion-mediated one-loop
contributions, determines the overall size of the NLO EW effects. 
For most of the decays, the relative NLO corrections are moderate
and reach at most 21\%. Accordingly, they show a mild renormalization scheme
and scale dependence with changes in the predicted NLO widths
typically at the percent level or below. Some decays, however, exhibit
a stronger renormalization scheme and scale dependence. This implies a
larger theoretical uncertainty and can be explained by the mass
hierarchies and couplings governing these cases, which lead to
loop-dominated decays. 

\section{Conclusions \label{sec:concl}}
In this paper we worked out the renormalization of the N2HDM, which is
an interesting benchmark model for studying extended Higgs sectors
involving Higgs-to-Higgs decays. For the mixing angles, we provided a
renormalization scheme 
that is manifestly gauge independent by applying the alternative
tadpole scheme combined with the pinch technique. We explained in
great detail the notion of the alternative tadpole scheme in our
renormalization framework,  
and for the first time provided the formulae for the pinched
self-energies in the N2HDM. Apart from the additional mixing angles as
compared to the 2HDM, in the N2HDM we encounter a singlet VEV that
needs to be renormalized as well. We elaborated in detail the
implications of the alternative tadpole scheme for the renormalization
of the singlet VEV that we renormalize through a physical quantity,
given by a Higgs-to-Higgs decay width. 
The soft $\mathbb{Z}_2$ breaking parameter $m_{12}^2$, which,
like $v_S$, enters the Higgs self-couplings and hence features in
Higgs-to-Higgs decays, is renormalized in the $\overline{\mbox{MS}}$
scheme. We studied the impact of our renormalization scheme by
computing the EW one-loop corrections to various Higgs decay widths,
including the Higgs decays into a massive $Z$-boson pair and into
lighter Higgs pairs. \s

The computation of the EW corrections to our different sample decay
widths has shown that the corrections can be sizeable and 
have to be taken into account in order to make reliable predictions
for the Higgs observables. For a broad range of phenomenologically representative 
scenarios we find a rather
weak renormalization scale and renormalization scheme dependence, indicative of a
rather small theoretical error due to missing higher order
corrections. In some cases the EW corrections can
be sizeable, in particular if the corresponding LO decay widths are suppressed,
so that the NLO-corrected width effectively becomes the leading order decay width.
Higher order corrections beyond NLO would then be 
necessary in order to reduce the theoretical error. \s

With this paper, we have provided an important contribution to the
renormalization of extended Higgs sectors involving singlet
fields. This is crucial input for the computation of the EW
corrections to the Higgs bosons of such models and therefore indispensable
for the correct prediction and interpretation of Higgs observables at
the LHC.

\subsubsection*{Acknowledgments}
The authors acknowledge financial support from the DFG
project ``Precision Calculations in the Higgs Sector - Paving the Way
to the New Physics Landscape'' (ID: MU 3138/1-1). 
We would like to thank Marco Sampaio, Michael Spira and Jonas
Wittbrodt for useful discussions. 

\section*{Appendix}
\setcounter{equation}{0}
\begin{appendix}

\section{The Pinch Technique in the N2HDM \label{app:pinchtech}}
In this section, we present the explicit gauge dependences appearing
in the scalar-scalar and scalar-vector self-energies in the
N2HDM. Additionally, we present the application of the pinch technique
in the N2HDM for the first time, as well as the cancellation of all
gauge dependences by the generation of pinched self-energies.  

\subsection{Gauge dependence of the self-energies}
We begin by setting the notation used in the explicit expressions of
the gauge dependences. Following the notation of
Ref.~\cite{Espinosa:2002cd}, we define the functions
\begin{align}
	f_{\Phi _i \Phi _j} (p^2) &= p^2 - \frac{m_{\Phi _i}^2 + m_{\Phi _j}^2}{2}  \\
	g_{\Phi _i \Phi _j} (p^2, m^2) &= 2\left( p^2 - m^2\right)\left( p^2 - 
\frac{m_{\Phi_i}^2 + m_{\Phi _j}^2}{2}\right) - \left( p^2 -m_{\Phi_i}^2 \right) \left(
 p^2 - m_{\Phi _j}^2 \right)~, 
\label{eq:YamadaDef}
\end{align}
where $\Phi$ stands for an arbitrary neutral or
  charged scalar particle and $m_{\Phi _{i,j}} = 0$ in case $\Phi _{i,j}$ is a Goldstone
boson. We introduce the one-loop integrals
\begin{align}
	\alpha _V &= \frac{1}{\left( 1- \xi _V \right) m_V^2 } \left[ A_0 \left( m_V^2 \right) - A_0 \left( \xi _V m_V^2 \right) \right] = B_0 \left( 0; m_V^2 , \xi _V m_V^2 \right) \\
	\beta _{V \Phi _i } (p^2) &= \frac{1}{\left( 1- \xi _V \right) m_V^2 } \left[ B_0 \left( p^2; m_V^2 , m_{\Phi _i} ^2 \right) - B_0 \left( p^2; \xi _V m_V^2 , m_{\Phi _i} ^2 \right) \right] \\
	&= C_0 \left( 0,p^2,p^2; m_V^2, \xi _V m_V^2, m_{\Phi _i } ^2 \right) \notag \\
	\beta _{V \xi V } (p^2) &= \frac{1}{\left( 1- \xi _V \right) m_V^2 } \left[ B_0 \left( p^2; m_V^2 , \xi _V m_V ^2 \right) - B_0 \left( p^2; \xi _V m_V^2 , \xi _V m_V ^2 \right) \right]  \\
	&= C_0 \left( 0,p^2,p^2; \xi _V m_V^2, \xi _V m_V^2, m_{\Phi _i } ^2 \right) \notag \\
	C_2 ^{V \Phi _i } (p^2) &= C_2 \left( 0, p^2, p^2; m_V^2 , \xi _V m_V ^2 , m_{\Phi _i}^2 \right) ~,
\label{eq:DefOneLoopIntegrals}
\end{align}
where $A_0$, $B_0$ and $C_0$ denote the usual scalar one-, two-
and three-point integrals and $C_2$ denotes the
coefficient integral of the tensor integral $C_\mu$, which can be
expressed solely through $A_0$ and $B_0$ integrals,
\textit{cf.}~Refs.~\cite{tHooft:1978xw,
  Denner:1991kt}. The
  index $V$ denotes a vector boson $V\in \left\{ W^\pm, Z,
    \gamma \right\}$. \s  

In what follows, we extract the gauge dependences of all self-energies
via the definition 
\beq 
	i\Sigma ^\text{tad} (p^2) = \left. i\Sigma ^\text{tad} (p^2)
        \right| _{\xi _V = 1} + \left. i\Sigma (p^2) 
        \right|_\text{g.d.} \;,
	\label{eq:extractionGD}
\eeq
where $i\Sigma ^\text{tad} (p^2)$ is the fully gauge-dependent
modified self-energy with tadpole contributions included,
\textit{cf.}~Fig.~\ref{fig:tadpoleselfen}, $\left. i\Sigma (p^2)
\right| _\text{g.d.}$ represents the truly gauge-dependent part of the
self-energy and $\left. i \Sigma (p^2) \right| _{\xi _V = 1}$ denotes
the evaluation of the self-energy in the 't-Hooft Feynman gauge. The
inclusion of tadpole contributions for the analysis of the
self-energies with respect to gauge dependence is necessary for a
consistent application of the pinch technique
\cite{Binosi:2009qm}. While the extraction of the gauge dependence via
Eq.~(\ref{eq:extractionGD}) is not unique, we show in the following by
applying the pinch technique that $\left. i\Sigma (p^2) \right|
_\text{g.d.}$ is considered to be the truly gauge-dependent part of
the self-energies, since it is precisely these terms which are
cancelled by the pinch contributions. 

\subsubsection{Gauge dependence of the CP-even scalar self-energies}
\begin{figure}[tb]
\centering
\includegraphics[width=\linewidth , trim = 0mm 0mm 0mm 6mm,
clip]{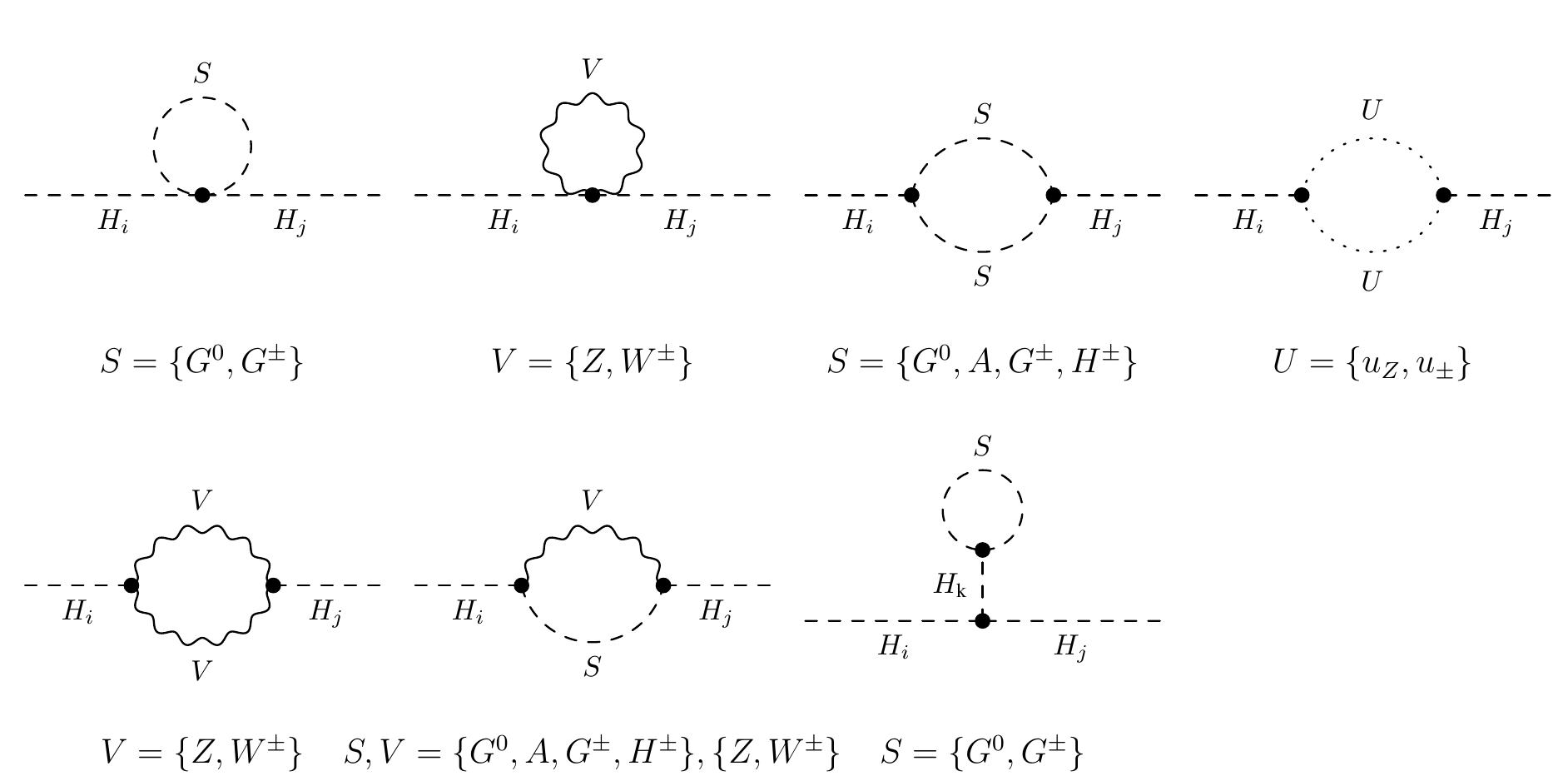} 
\caption{All Feynman diagrams contributing to the gauge dependence of
  the CP-even self-energies $\Sigma ^\text{tad} _{H_i H_j} (p^2)$. For
  the tadpole diagram, a sum over intermediate Higgs states $H_k$
  $(k=1,2,3$) is 
  assumed. Note that the ghost and vector boson contributions in the
  tadpole diagrams precisely cancel against each other, so that these
  are not shown.}
\label{fig:CPevenSelfEnergies}
\end{figure} 
First, we consider the gauge dependence of the CP-even scalar
self-energies, \textit{i.e.}~the self-energies of all possible
combinations of $H_i$ and $H_j$ ($i,j=1,2,3$). All Feynman diagrams
contributing gauge-dependent terms are shown in
Fig.~\ref{fig:CPevenSelfEnergies}. The evaluation of
Eq.~(\ref{eq:extractionGD}) for the CP-even scalars of the N2HDM sector
yields  
\begin{equation}
\begin{split}
	i\Sigma ^\text{tad} _{H_i H_j} (p^2) &= \left. i\Sigma ^\text{tad} _{H_i H_j} (p^2) \right| _{\xi _V = 1} \\
	&+ \frac{ig^2 \left( 1 - \xi _Z \right) }{64 \pi ^2 \cos \Theta _W } \bigg[ g_{H_i H_j} (p^2, m_A^2 ) \mathcal{O}^{(1)}_{H_iH_j} \beta _{ZA} (p^2) - f_{H_i H_j} (p^2) \mathcal{O}^{(4)}_{H_iH_j} \alpha _Z \\
	&\hspace*{2.7cm} + \frac{1}{2} g_{H_i H_j} (p^2, 0) \mathcal{O}^{(2)}_{H_iH_j} \left( \beta _{ZZ} (p^2) + \beta _{Z\xi Z} (p^2) \right) \bigg] \\
	&+ \frac{ig^2 \left( 1 - \xi _W \right) }{32 \pi ^2 } \bigg[ g_{H_i H_j} (p^2, m_{H^\pm}^2 ) \mathcal{O}^{(1)}_{H_iH_j} \beta _{WH^\pm} (p^2) - f_{H_i H_j} (p^2) \mathcal{O}^{(4)}_{H_iH_j} \alpha _W \\
	&\hspace*{2.7cm} + \frac{1}{2} g_{H_i H_j} (p^2, 0) \mathcal{O}^{(2)}_{H_iH_j} \left( \beta _{WW} (p^2) + \beta _{W\xi W} (p^2) \right) \bigg] ~,
\end{split}
\label{eq:gdofhihj}
\end{equation}
where the combinations $\mathcal{O}^{(1)}_{H_iH_j}$,
$\mathcal{O}^{(2)}_{H_iH_j}$ and $\mathcal{O}^{(4)}_{H_iH_j}$ have
been defined in Eq.~(\ref{eq:higgsgaugecomb}). We note that when
evaluating these combinations in the 2HDM limit, \textit{i.e.}~by
applying Eq.~(\ref{eq:2hdmlimit}), where $\mathcal{O}^{(4)}_{H_iH_j}$
reduces to the Kronecker delta $\delta _{H_i H_j}$, the result in
Eq.~(\ref{eq:gdofhihj}) coincides with the results presented in
Refs.~\cite{MKrause2016, Kanemura:2017wtm} for the 2HDM as well as
with the result presented in Ref.~\cite{Espinosa:2002cd} for the MSSM,
since the structure of the gauge-dependence of the CP-even scalar
self-energies does not differ between the MSSM and the 2HDM. 

\subsubsection{Gauge dependence of the charged scalar and vector self-energies}
\begin{figure}[tb]
\centering
\includegraphics[width=\linewidth , trim = 0mm 0mm 0mm 6mm,
clip]{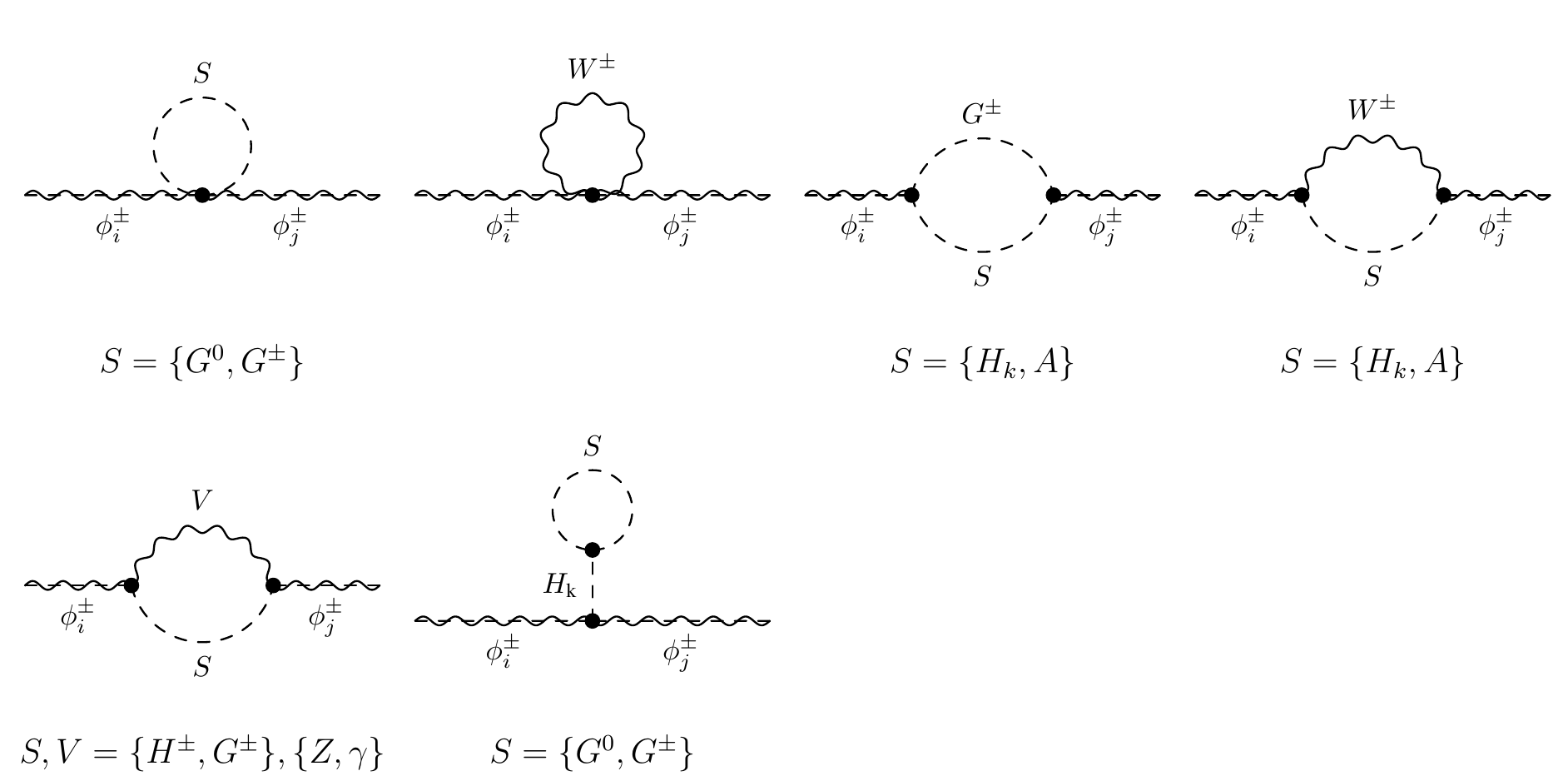} 
\caption{All Feynman diagrams contributing to the gauge-dependence of
  the charged self-energies $\Sigma ^\text{tad} _{\phi ^\pm _i \phi
    ^\pm _j } (p^2)$ where $\phi ^\pm _{i,j} \in \{ W, G^\pm , H^\pm
  \}$. A sum over intermediate Higgs states $H_k$ is assumed wherever
  they appear. Overlapping dashed and twiggled
  lines denote a scalar or a
  gauge boson, respectively, depending on the chosen particles. Note
  that we only consider contributions to the extended scalar sector of
  the N2HDM. Depending on the particles $\phi ^\pm _{i,j}$ chosen,
  some of the diagrams shown may not exist in the
  N2HDM.} 
\label{fig:ChargedSelfEnergies}
\end{figure} 
Next, we consider the charged sector. Due to the mixing of the charged
particles of the N2HDM, we have to consider not only all possible
self-energy combinations of the scalar particles $H^\pm$ and $G^\pm$,
but additionally their mixing with the charged vector bosons
$W^\pm$. In the SM, where only one Higgs boson exists, it was shown
that the Higgs contributions to the gauge dependence of the charged
sector form a gauge-dependent subset which is cancelled by a
corresponding subset of pinch contributions
\cite{Papavassiliou:1994pr}. In the N2HDM we follow the same
approach, \textit{i.e.}~we focus only on gauge-dependent contributions
stemming from the enriched scalar sector
of the N2HDM, which form a 
subset with respect to gauge dependence as well. \s

We first consider the gauge dependence of the self-energies of all
combinations of $W^\pm$ and $G^\pm$. The relevant contributions from
the Higgs sector are given by the Feynman diagrams in
Fig.~\ref{fig:ChargedSelfEnergies} for all possible
self-energies. Note that since we consider only the subset where the
scalars of the N2HDM appear in the loops, only terms containing the
gauge-fixing parameter $\xi _W$ contribute to these
self-energies. They explicitly
read\footnote{Note that in the case of 
  the self-energy $\Sigma ^\text{tad} _{G^\pm G^\pm }$ we subtracted
  an additional term of $f_{G^\pm G^\pm} (p^2) \alpha _W$ with respect
  to the diagrams shown in Fig.~\ref{fig:ChargedSelfEnergies}. This
  term stems from other gauge-dependent subsets of the
  gauge-dependence of the self-energy, which we do not present
  explicitly here. This is in line with \cite{Papavassiliou:1994pr},
  where these additional terms are simply dropped since they cancel
  elsewhere.} 
\begin{align}
	i\Sigma ^\text{tad} _{WW, \mu \nu} (p^2) &= \left. i\Sigma ^\text{tad} _{WW, \mu \nu} (p^2) \right| _{\xi _V = 1} \\
	&- (1-\xi _W ) \frac{ig^2 m_W^2  }{64\pi ^2} p_\mu p_\nu  \sum _{H_i} \mathcal{O}^{(2)}_{H_iH_i} \left\{ \beta _{WH_i} (p^2) + 4 C_2 ^{WH_i} (p^2) \right\}  \notag \\
	i\Sigma ^\text{tad} _{WG^\pm , \mu } (p^2) &= \left. i\Sigma ^\text{tad} _{WG^\pm , \mu } (p^2) \right| _{\xi _V = 1}  \\
	&+ (1-\xi _W ) \frac{ig^2 m_W  }{64\pi ^2} p_\mu \left\{
          \alpha _W + \sum _{H_i} \mathcal{O}^{(2)}_{H_iH_i} \left[
          m_{H_i}^2 \beta _{WH_i} (p^2) + 2 p^2 C_2 ^{WH_i} (p^2)
          \right]  \right\}  \notag 
\end{align}
\begin{align}
	i\Sigma ^\text{tad} _{G^\pm G^\pm } (p^2) &= \left. i\Sigma ^\text{tad} _{G^\pm G^\pm } (p^2) \right| _{\xi _V = 1}  \\
	&+ (1-\xi _W ) \frac{ig^2 }{64\pi ^2} \left\{ -2 f_{G^\pm G^\pm } (p^2) \alpha _W + \sum _{H_i} \mathcal{O}^{(2)}_{H_iH_i} g_{G^\pm G^\pm} (p^2, m_{H_i}^2 ) \beta _{WH_i} (p^2) \right\} \;.\notag
\end{align}
Next, the gauge dependence of the self-energies of all combinations of
$H^\pm$ and $G^\pm$ or $W^\pm$ is given by the relevant contributions
from the Higgs sector as given by the Feynman diagrams in
Fig.~\ref{fig:ChargedSelfEnergies} as well. In the case of the
self-energy for two $H^\pm$ particles, additional dependences on $\xi
_Z$ and $\xi _\gamma$ appear even when focusing on the extended scalar
sector of the N2HDM only, while for the other self-energies only the
dependence on $\xi _W$ is relevant. The self-energies explicitly read 
\begin{align}
	i\Sigma ^\text{tad} _{H^\pm H^\pm } (p^2) &= \left. i\Sigma ^\text{tad} _{H^\pm H^\pm } (p^2) \right| _{\xi _V = 1} \\
	&+ (1-\xi _W ) \frac{ig^2}{64\pi ^2} \left\{ -2 f_{H^\pm H^\pm
          } (p^2) \alpha _W + g_{H^\pm H^\pm } (p^2, m_A^2 ) \beta
          _{WA}  \right. \notag \\
& + \sum _{H_i} \mathcal{O}^{(2)}_{H_iH_i} g_{H^\pm H^\pm } (p^2,
  m_{H_i}^2 ) \beta _{WH_i} \big\} \notag \\
& + (1-\xi _Z ) \frac{ig^2 (\cos ^2 \Theta _W - \sin ^2 \Theta _W ) ^2 }{64\pi ^2 \cos ^2 \Theta _W } \left\{ g_{H^\pm H^\pm } (p^2, m_{H^+}^2 ) \beta _{ZH^\pm } (p^2) - f_{H^\pm H^\pm } (p^2) \alpha _Z \right\} \notag \\
	&+ (1 - \xi _\gamma ) \frac{ie^2}{16\pi ^2} \left\{ - f_{H^\pm H^\pm } (p^2) \alpha _\gamma + g_{H^\pm H^\pm } (p^2 , m_{H^+}^2 ) \beta _{\gamma H^\pm } (p^2) \right\}  \notag \\
	i\Sigma ^\text{tad} _{H^\pm G^\pm } (p^2) &= \left. i\Sigma ^\text{tad} _{H^\pm G^\pm } (p^2) \right| _{\xi _V = 1} \\
	&+ (1-\xi _W ) \frac{ig^2}{64\pi ^2} \sum _{H_i} \mathcal{O}^{(3)}_{H_iH_i} g_{H^\pm G^\pm } (p^2, m_{H_i}^2 ) \beta _{W H_i} (p^2)  \notag \\
	i\Sigma ^\text{tad} _{W H^\pm , \mu } (p^2) &= \left. i\Sigma ^\text{tad} _{WH^\pm , \mu } (p^2) \right| _{\xi _V = 1} \\
	&- (1- \xi _W ) \frac{ig^2 m_W}{64\pi ^2} p_\mu \sum _{H_i}
          \mathcal{O}^{(3)}_{H_iH_i} \left\{ f_{H^\pm H^\pm }
          (m_{H_i}^2) \beta _{WH_i} (p^2) + 2f_{H^\pm H^\pm } (p^2)
          C_2 ^{WH_i} (p^2) \right\}  \notag
\end{align}
Note that when the former two equations are evaluated in the
2HDM-limit, \textit{cf.}~Eq.~(\ref{eq:2hdmlimit}), these reproduce the
formulae given in Ref.~\cite{Kanemura:2017wtm} for the 2HDM. 

\subsubsection{Gauge dependence of the CP-odd scalar and vector self-energies}
\begin{figure}[tb]
\centering
\includegraphics[width=\linewidth , trim = 0mm 0mm 0mm 6mm, clip]{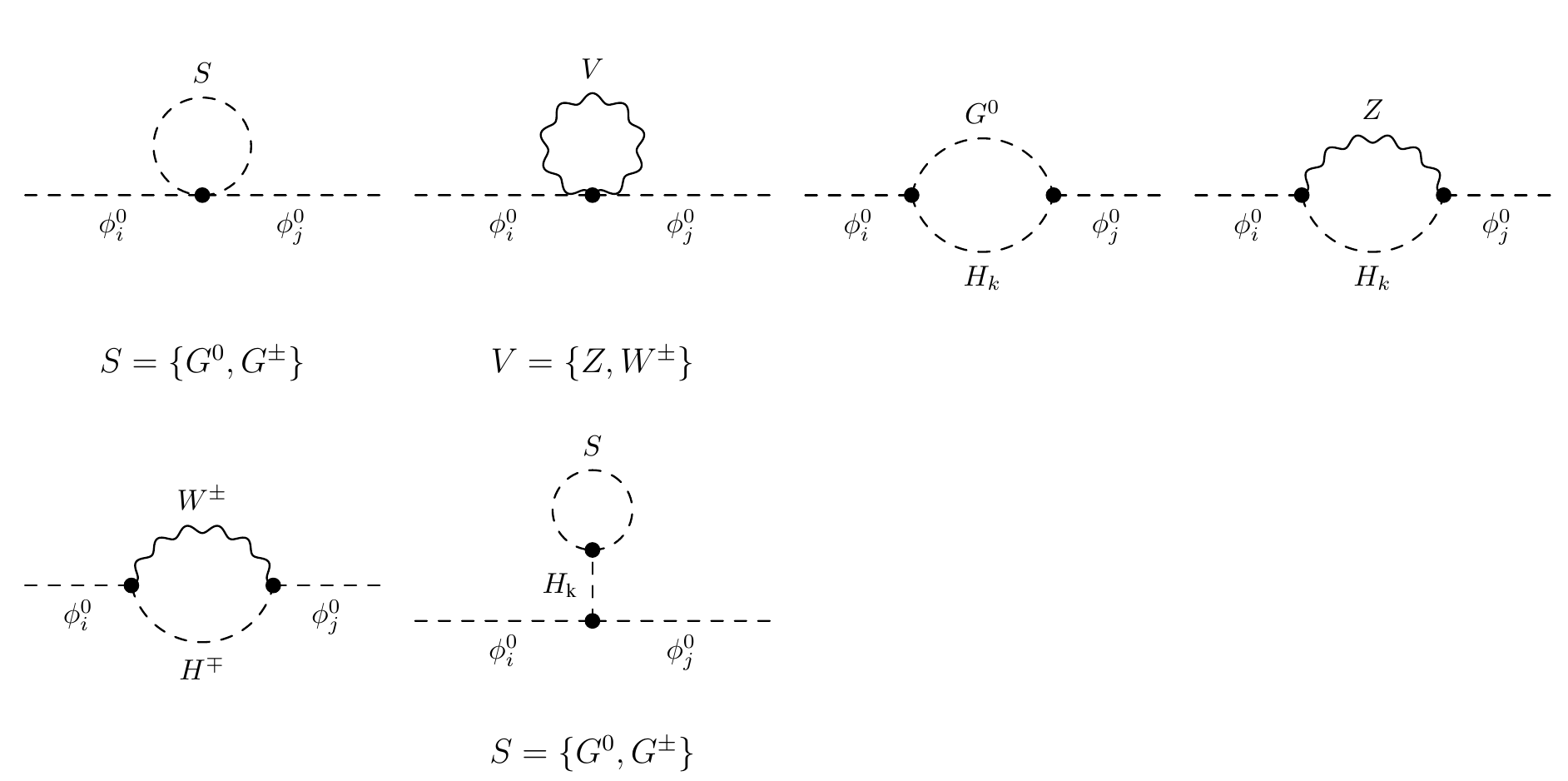}
\caption{All Feynman diagrams contributing to the gauge-dependence of
  the CP-odd self-energies $\Sigma ^\text{tad} _{\phi ^0 _i \phi ^0 _j
  } (p^2)$ where $\phi ^0 _{i,j} \in \{ A , G^0 \}$. A sum over
  intermediate Higgs states $H_k$ is assumed wherever they
  appear. Note that we only consider contributions to the extended
  scalar sector of the N2HDM. Depending on the particles $\phi ^0
  _{i,j}$ chosen, some of the diagrams shown may not exist in the
  N2HDM.}
\label{fig:CPoddSelfEnergies}
\end{figure} 
In the neutral CP-odd sector the calculation of the
gauge dependences and of the pinch contributions is even more involved
than in the charged 
sector, since one has to take into account not only the mixing of the
$Z$ boson with $G^0$ and $A$, but additionally the mixing of the photon $\gamma$
with all other possible contributions. It is only the coherent sum of
all these mixing contributions which gives the correct gauge
dependences and pinch results. Due to these additional complications,
we restrict the presentation to the self-energies of two $A$ and the
mixing between $A$ and $G^0$. As in the charged sector, we focus on
the N2HDM Higgs contributions to the self-energies and pinch terms
only, since they form a gauge-independent subset on their own. The
relevant contributions are given by the Feynman diagrams in
Fig.~\ref{fig:CPoddSelfEnergies}. In total, the self-energies of this
subset read 
\begin{align}
i \Sigma ^\text{tad} _{AA} (p^2) &= \left. i \Sigma ^\text{tad} _{AA} (p^2) \right| _{\xi _V = 1} \\
&+ (1 - \xi _Z ) \frac{ig^2}{64 \pi ^2 \cos ^2 \Theta _W }  \left\{ - f_{AA} (p^2) \alpha _Z + \sum _{H_i} \mathcal{O} ^{(1)} _{H_i H_i} g_{AA} (p^2 , m_{H_i}^2 ) \beta _{Z H_i} (p^2) \right\}  \notag \\
&+ (1 - \xi _W ) \frac{ig^2}{32 \pi ^2 } \left\{ - f_{H^\pm H^\pm } (p^2) \alpha _W + g_{AA} (p^2 , m_{H^\pm }^2) \beta _{WH^\pm } (p^2) \right\}   \notag \\
i \Sigma ^\text{tad} _{AG^0} (p^2) &= \left. i \Sigma ^\text{tad} _{AG^0} (p^2) \right| _{\xi _V = 1} \\
&+ (1 - \xi _Z ) \frac{ig^2}{64 \pi ^2 \cos ^2 \Theta _W } \sum _{H_i} \mathcal{O} ^{(3)} _{H_i H_i} g_{AG^0 } (p^2 , m_{H_i}^2 ) \beta _{Z H_i} (p^2) ~. \notag 
\end{align}
As in the charged sector, these results, evaluated in the 2HDM limit,
reproduce the ones presented in Ref.~\cite{Kanemura:2017wtm} for the
2HDM. 

\subsection{Pinch contributions for the N2HDM \label{app:pinchself}}
The intricate gauge dependence of the scalar self-energies of the
N2HDM makes a gauge-independent definition of
the counterterms of the scalar mixing angles complicated. If one
considers instead an $S$-matrix element, \textit{e.g.}~a scattering
process of a pair of fermions, where these self-energies may appear as
intermediate states, the whole $S$-matrix element is gauge independent
by construction. Consequently, the gauge dependences cancel in an
intricate way between the self-energies and other contributions from
vertex and box corrections within the $S$-matrix element. \s

The main idea of the pinch technique
(\textit{cf.}~Refs.~\cite{Binosi:2004qe,Binosi:2009qm,Cornwall:1989gv,Papavassiliou:1989zd,Degrassi:1992ue,Papavassiliou:1994pr,Watson:1994tn,Papavassiliou:1995fq} for a detailed
exposition) is to isolate the gauge dependences of an arbitrary toy
scattering process, which features the to-be pinched self-energies in
a unique way. This is achieved by applying the elementary Ward
identities 
\begin{align}
	\slashed{k} \, P_{L/R} &= S^{-1} _1 (p+k) P_{L/R} - P_{R/L} S^{-1} _2 (p) + m_1 P_{L/R} - m_2 P_{R/L}  \notag \\
	P_{L/R} \, \slashed{k} &= P_{L/R} S^{-1} _1 (p+k) - S^{-1} _2 (p) P_{R/L} + m_1 P_{L/R} - m_2 P_{R/L} ~, \label{eq:pinchWardIdent}
\end{align} 
where $k$ denotes the loop momentum, $m_1$ and $m_2$ the masses
of the external fermions of the considered toy process and $S(p)$ the
fermion propagator  
\beq
	iS_k(p) = \frac{i(\slashed{p} + m_k)}{p^2 -
          m_k^2} = \frac{i}{\slashed{p} -
          m_k} ~. 
\eeq
It turns out that the gauge dependences are all similar in structure,
\textit{i.e.}~they are always self-energy-like, independently of their
origin within the scattering process. The isolation of all pinch
contributions from the toy scattering process then allows for a
manifestly gauge-independent definition of \textit{pinched
  self-energies}. Since these self-energies are considered to be
independent from the toy process chosen,
\textit{cf.}~\cite{Binosi:2009qm}, the pinched self-energies are
unique. 

\subsubsection{Pinch contributions for the CP-even sector}
\begin{figure}[b!]
\centering
\includegraphics[width=0.8\linewidth , trim = 0mm 0mm 0mm 6mm,
clip]{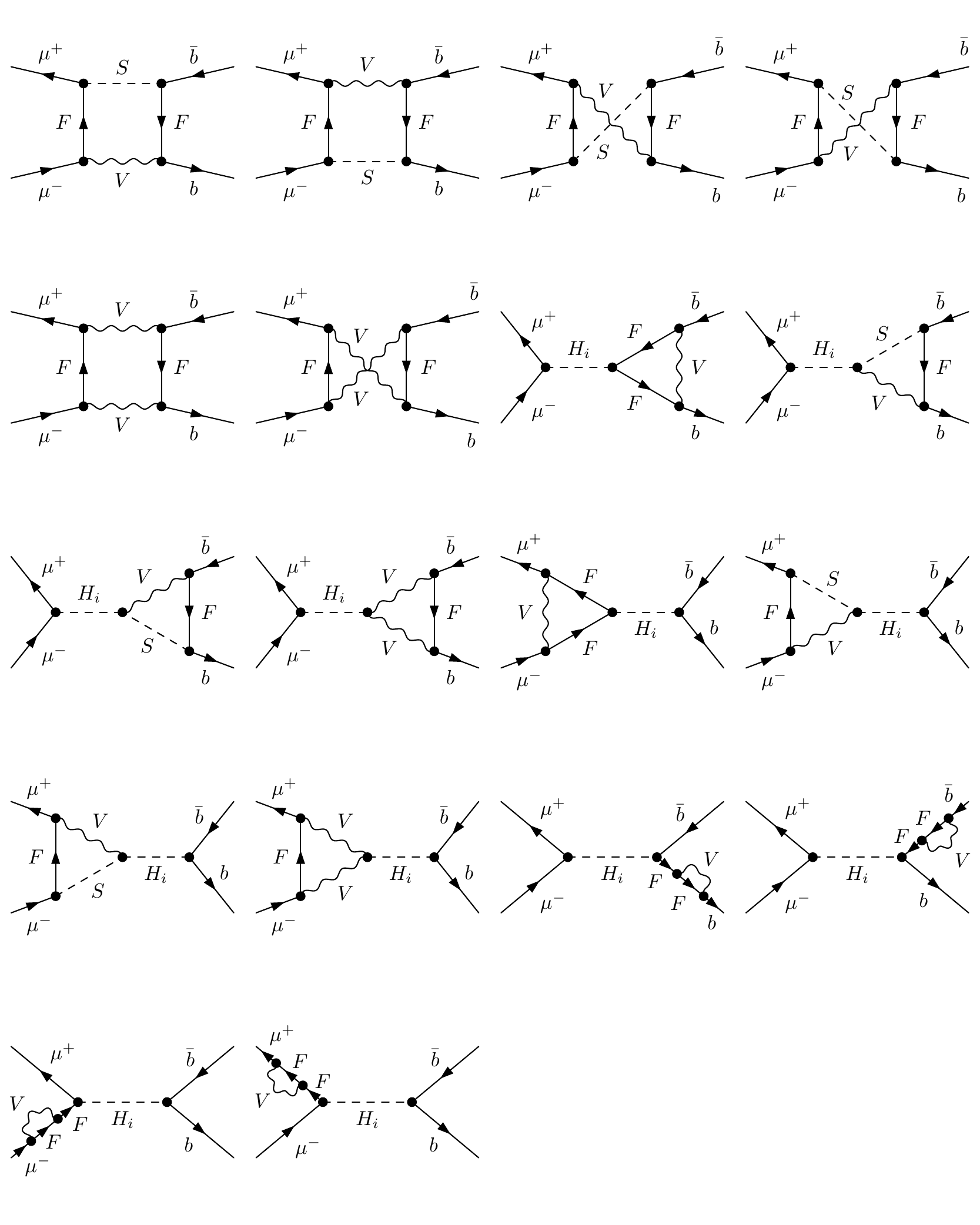} 
\caption{All generic Feynman diagrams contributing to the CP-even
  pinched self-energies.} 
\label{fig:CPevenpinched}
\end{figure} 
The full derivation of all pinch contributions for the N2HDM is beyond
the scope of this paper. We nevertheless present the derivation of the
pinch contributions for a few selected diagrams since we hope it is
instructive to the reader and since it demonstrates how the
pinch technique is applied. As the 
toy process for extracting the gauge dependences for the CP-even
sector we choose the process $\mu ^+ \mu ^- \rightarrow \bar{b}
b$. All Feynman diagrams yielding contributions for the CP-even
pinched self-energies are depicted in Fig.~\ref{fig:CPevenpinched}. It can be shown
that all pinch contributions stemming from these diagrams can be
brought into the form 
\begin{equation}
	\Gamma ^{H_i bb} \frac{i}{p^2 - m_{H_i}^2} i
        \Sigma^\text{PT} _{H_i H_j} (p^2) \frac{i}{p^2 -
          m_{H_j}^2} 
        \Gamma ^{H_j \mu \mu } \;,
\end{equation}
where $i \Sigma ^\text{PT} _{H_i H_j} (p^2)$ is a
relevant self-energy-like pinch contribution for the CP-even Higgs
bosons $H_i$ and $H_j$. Additionally, we define the contracted
vertices of a CP-even Higgs boson with a pair of external bottom
quarks or a pair of external muons as 
\begin{alignat}{5}
\Gamma ^{H_i bb} = \bar{u} (r_1) \frac{-igm_b \kappa _{H_i bb} }{2 m_W} v (r_2)
\qquad \mbox{and} \qquad
\Gamma ^{H_i \mu \mu } = \bar{v} (p_2) \frac{-igm_\mu 
\kappa _{H_i \mu\mu } }{2 m_W} u (p_1) \;,
\end{alignat}
where $u(p_1)$ and $v(r_2)$ and $\bar{u} (r_1)$ and $\bar{v} (p_2)$
are the (adjoint) spinors of the external on-shell fermions with their
respective momenta. \s

In order to derive the pinch contributions, we apply the elementary
Ward identities, \textit{cf.} Eq. (\ref{eq:pinchWardIdent}), and
insert additional CP-even Higgs boson propagators into the amplitude
via 
\beq
	1 = - \frac{i}{p^2 - m_{H_i}^2} i (p^2 - m_{H_i}^2) ~.
\label{eq:CPevenPropagators}
\eeq
Additionally, we make use of the sum rules of the N2HDM as given in
Eq.~(\ref{eq:twosumrules}) as well as of the coupling relation
\beq
	\kappa _{H_i ff} = \kappa _{H_i VV} - \tilde{\kappa} _{H_i VH} \kappa_{Aff} \;.
\eeq
The application of these formulae to
fermion-fermion-Higgs couplings enables the projection of the pinch
contributions onto the desired CP-even Higgs couplings to the
fermions, 
\beq
	\tilde{\kappa} \kappa _{All} =
        \sum_{H_j} \tilde{\kappa} \kappa _{H_j VV} \kappa _{All} 
        \kappa_{H_j ll} = - \sum _{H_j} \mathcal{O} ^{(1)} _{H_i H_j} 
        \kappa_{H_j ll} \hspace*{0.4cm} + \hspace*{0.4cm} \text{...} 
\label{eq:pinchSumRules}
\eeq
where ``$\text{...}$'' contains pinch contributions for other than the
CP-even Higgs self-energies. Consequently, we can neglect them for the
CP-even self-energies. \s

\begin{figure}[tb]
\centering
\includegraphics[width=0.7\linewidth , trim = 0mm 16mm 0mm 16mm, clip]{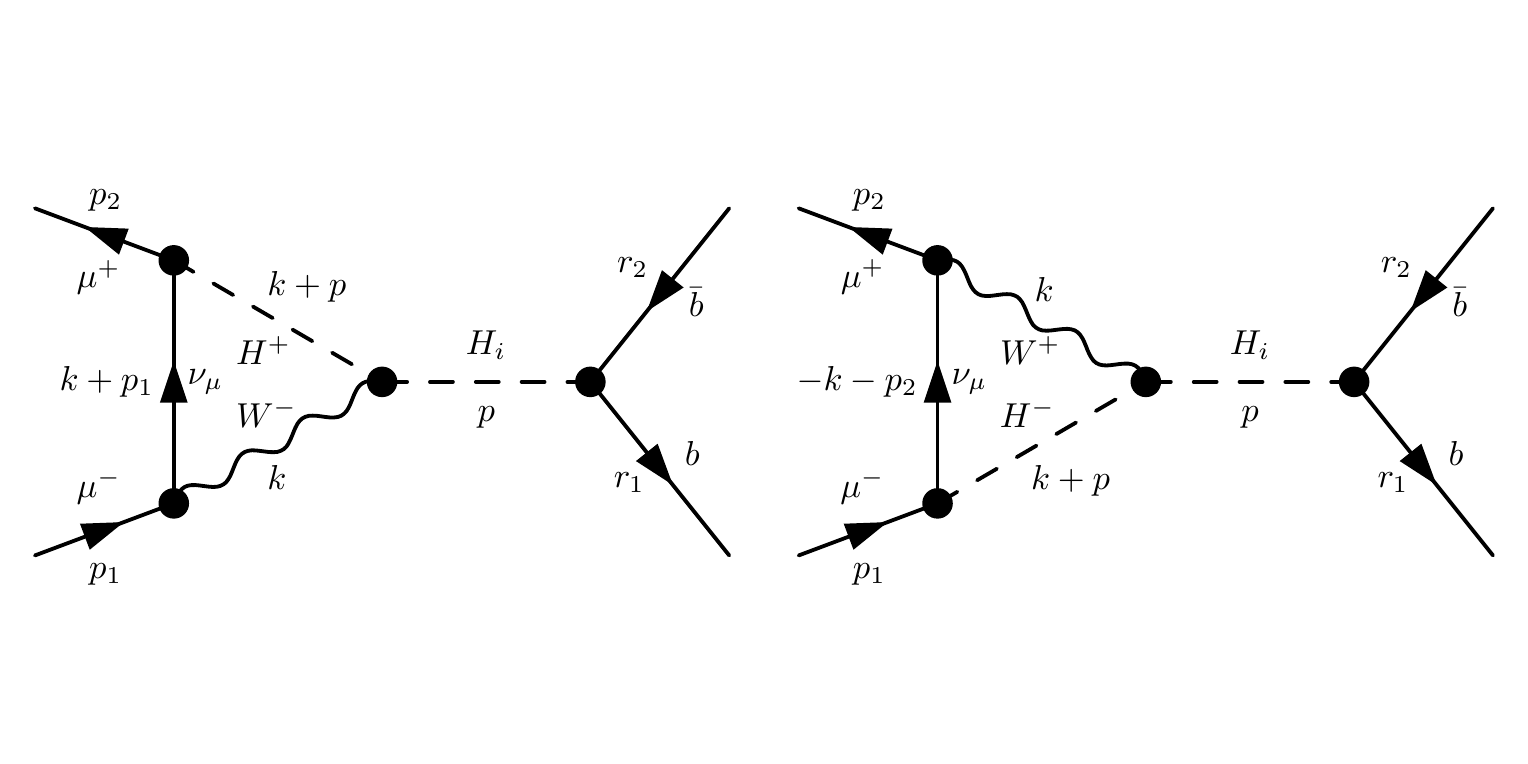}
\caption{Two Feynman diagrams for the toy process $\mu ^+ \mu ^-
  \rightarrow \bar{b} b$ involving scalar-scalar-vector vertices which
  give rise to gauge-dependent as well as additional gauge-independent
  pinch contributions for the CP-even
  self-energies. The momenta $p_1$ and $p_2$ are taken
  as incoming and the momenta $r_1$ and $r_2$ as outgoing, and
  $p=p_1+p_2$.} 
\label{fig:CPevenTwoExamples}
\end{figure} 
We consider the two contributions depicted in the Feynman diagrams in
Fig.~\ref{fig:CPevenTwoExamples}. The momenta are as defined in the
diagrams. With the definitions given above, the sum of both
diagrams reads\footnote{Note that the shift from four to $D$
  dimensions as well as the $+i\epsilon$ terms in the propagators are
  not explicitly stated here, but implicitly assumed to be set.} 
\begin{equation}
\begin{split}
& \sum _{H_i} \Gamma ^{H_i bb} \frac{i}{p^2 - m_{H_i}^2} \frac{g^2}{32\pi ^2} \int \frac{\text{d}^4 k}{i \pi ^2} \frac{1}{\left[ k^2 - m_W^2 \right] \left[ (k+p)^2 - m_{H^\pm }^2 \right] } \\
& \hspace*{1cm} \cdot \bigg\{ \bar{v} (p_2) \left[ P_L 
S_{\mu_\nu}(k+p_1) (\slashed{k} + 2\slashed{p} ) P_L
+ P_R (-\slashed{k} - 2 \slashed{p} )
S_{\mu_\nu}(-k-p_2) P_R \right] \frac{igm_\mu }{2m_W}
\kappa _{All} \tilde{\kappa} _{H_i VH} u (p_1) \\ 
& \hspace*{1cm} - (1-\xi _W) \bar{v} (p_2) \left[ P_L S_{\mu_\nu}(k+p_2) \slashed{k} P_L + P_R (-\slashed{k} ) S_{\mu_\nu}(-k-p_2) P_R \right] \frac{igm_\mu }{2m_W} \kappa _{All} \tilde{\kappa} _{H_i VH} u(p_1) \bigg\} \\
&\stackrel{(\ref{eq:pinchWardIdent})}{=} \sum _{H_i} \Gamma ^{H_i bb} \frac{i}{p^2 - m_{H_i}^2} \frac{g^2}{32\pi ^2} \bar{v} (p_2) \frac{igm_\mu }{2m_W} \kappa _{All} \tilde{\kappa} _{H_i VH} u(p_1) \Big\{ B_0 (p^2; m_W^2 , m_{H^\pm }^2 ) \\
	&\hspace*{1cm} - (1-\xi _W) \left[ \alpha _W - f_{H^\pm H^\pm } (p^2) \beta _{WH^\pm } \right] \Big\} \hspace*{0.4cm} + \hspace*{0.4cm} \text{...} \\
	&\stackrel{(\ref{eq:pinchSumRules})}{=} \sum _{H_i , H_j} \Gamma ^{H_i bb} \frac{i}{p^2 - m_{H_i}^2} \frac{-g^2}{32\pi ^2} \mathcal{O} ^{(1)} _{H_i H_j}  \Big\{ B_0 (p^2; m_W^2 , m_{H^\pm }^2 )  \\
	&\hspace*{1cm} - (1-\xi _W) \left[ \alpha _W - f_{H^\pm H^\pm } (p^2) \beta _{WH^\pm } \right] \Big\} \Gamma ^{H_j \mu \mu } \hspace*{0.4cm} + \hspace*{0.4cm} \text{...} \\
&\stackrel{(\ref{eq:CPevenPropagators})}{=} \sum _{H_i , H_j} \Gamma ^{H_i bb} \frac{i}{p^2 - m_{H_i}^2} \frac{-ig^2}{16\pi ^2} \left( \frac{p^2}{2} - \frac{m_{H_j}^2}{2} \right) \mathcal{O} ^{(1)} _{H_i H_j} \Big\{ B_0 (p^2; m_W^2 , m_{H^\pm }^2 ) \\
	&\hspace*{1cm}  - (1-\xi _W) \left[ \alpha _W - f_{H^\pm H^\pm } (p^2) \beta _{WH^\pm } \right] \Big\} \frac{i}{p^2 - m_{H_j}^2} \Gamma ^{H_j \mu \mu } \hspace*{0.4cm} + \hspace*{0.4cm} \text{...} \\
	&= \sum _{H_i , H_j} \Gamma ^{H_i bb} \frac{i}{p^2 - m_{H_i}^2} i \Sigma ^\text{PT} _{H_i H_j} (p^2) \frac{i}{p^2 - m_{H_j}^2} \Gamma ^{H_j \mu \mu } \hspace*{0.4cm} + \hspace*{0.4cm} \text{...}
\end{split}
\end{equation}
The first term of the right-hand side of the Ward identities in
Eq.~(\ref{eq:pinchWardIdent}) removes the internal fermion propagators
from the loops, \textit{i.e.}~the fermions are \textit{pinched out},
while the second term of the Ward identities vanishes due to the Dirac
equation. The third and fourth terms produce pinch contributions to
pinched vertices, but not to pinched self-energies. Consequently,
these terms are collected in ``$\text{...}$'', since they are of no
interest for the generation of a pinched
self-energy. The application of the sum rule in
Eq.~(\ref{eq:pinchSumRules}) produces additional pinch contributions
to other self-energies than the CP-even ones due to different
fermion-Higgs couplings. Consequently, these other terms are collected
in ``$\text{...}$'' as well. \s

As mentioned before, the pinch contributions take the form of a
self-energy and here they explicitly read
\begin{equation}
\begin{split}
i \Sigma ^\text{PT} _{H_i H_j} (p^2) = \frac{-ig^2}{16\pi ^2} \left( \frac{p^2}{2} - \frac{m_{H_j}^2}{2} \right) \mathcal{O} ^{(1)} _{H_i H_j} \Big\{ & B_0 (p^2; m_W^2 , m_{H^\pm }^2 ) \\
& - (1-\xi _W) \left[ \alpha _W + (m_{H^\pm }^2 - p^2) \beta _{WH^\pm
  } \right] \Big\} \;.
\end{split}
\label{eq:pinchResult}
\end{equation}
The terms proportional to $(1-\xi _W)$ are gauge-dependent pinch
contributions which cancel against parts of the gauge dependence of
the CP-even self-energies. The other term which remains for $\xi _W =
1$ is an additional gauge-independent pinch contribution which is
specific to scalar-scalar-vector vertices in the vertex corrections
\cite{Binosi:2009qm, Espinosa:2002cd}. Repeating the calculation for
the vertex corrections of the bottom quarks containing $H^\pm $ and
$W^\pm $ bosons in the loop yields the same result as in
Eq.~(\ref{eq:pinchResult}), but with $m_{H_j}^2$ replaced by
$m_{H_i}^2$. The combination
of these results yields the first term in the second line of
Eq.~(\ref{eq:sigaddhh}). \s

All Feynman diagrams contributing to the pinch terms for the CP-even
sector are depicted in Fig.~\ref{fig:CPevenpinched}. Repeating the calculation as
demonstrated above and combining all results leads to the pinch
contributions to the CP-even sector, 
\begin{equation}
\begin{split}
	i\Sigma ^\text{PT} _{H_i H_j} (p^2) &= i\Sigma ^\text{add} _{H_i H_j} (p^2) \\
	&- \frac{ig^2 \left( 1 - \xi _Z \right) }{64 \pi ^2 \cos \Theta _W } \bigg[ g_{H_i H_j} (p^2, m_A^2 ) \mathcal{O}^{(1)}_{H_iH_j} \beta _{ZA} (p^2) - f_{H_i H_j} (p^2) \mathcal{O}^{(4)}_{H_iH_j} \alpha _Z \\
	&\hspace*{2.7cm} + \frac{1}{2} g_{H_i H_j} (p^2, 0) \mathcal{O}^{(2)}_{H_iH_j} \left( \beta _{ZZ} (p^2) + \beta _{Z\xi Z} (p^2) \right) \bigg] \\
	&- \frac{ig^2 \left( 1 - \xi _W \right) }{32 \pi ^2 } \bigg[ g_{H_i H_j} (p^2, m_{H^\pm}^2 ) \mathcal{O}^{(1)}_{H_iH_j} \beta _{ZH^\pm} (p^2) - f_{H_i H_j} (p^2) \mathcal{O}^{(4)}_{H_iH_j} \alpha _W \\
	&\hspace*{2.7cm} + \frac{1}{2} g_{H_i H_j} (p^2, 0) \mathcal{O}^{(2)}_{H_iH_j} \left( \beta _{WW} (p^2) + \beta _{W\xi W} (p^2) \right) \bigg] ~.
\end{split}
\label{eq:pcofhihj}
\end{equation}
By comparing this result with Eq.~(\ref{eq:gdofhihj}), we realize that
in the sum of the pinch contributions with the CP-even self-energies
all gauge-dependent terms proportional to $(1-\xi _W)$
and $(1-\xi_Z)$ precisely cancel, leading to
\beq
	\overline{\Sigma} _{H_i H_j} (p^2) = \Sigma ^\text{tad} _{H_i H_j} (p^2) + \Sigma ^\text{PT} _{H_i H_j} (p^2) = \left. \Sigma ^\text{tad} _{H_i H_j} (p^2) \right| _{\xi_V = 1} + \Sigma ^\text{add} _{H_i H_j} (p^2) ~.
\eeq
Due to the cancellation of all gauge-dependent terms, the pinched
self-energy $\overline{\Sigma } _{H_i H_j} (p^2)$ is gauge independent
by construction and equivalent to the self-energy evaluated in the 
Feynman gauge, together with the sum of all additional terms stemming
from diagrams with internal scalar-scalar-vector vertices, as given in
Eq.~(\ref{eq:sigaddhh}). 

\subsubsection{Pinch contributions for the charged sector}
For the derivation of the pinch contributions of the charged sector we
use the toy process $\nu _e e^+ \rightarrow \nu _e e^+ $. The
calculation is analogous to the CP-even sector, \textit{i.e.}~we apply
the elementary Ward identities from Eq.~(\ref{eq:pinchWardIdent}) and
use the N2HDM sum rules to identify the correct couplings between the
external fermions and the scalar or vector particles of interest. In
the case of the self-energies involving the $H^\pm$ particles, we
again insert the corresponding propagator by 
\beq
	1 = - \frac{i}{p^2 - m_{H^\pm }^2} i (p^2 - m_{H^\pm }^2) ~.
	\label{eq:PTPropagatorInsert}
\eeq
For the self-energies involving $G^\pm$ or $W^\pm$, the corresponding propagators
\begin{alignat}{5}
\Delta _{\mu \nu } ( p ) \equiv \frac{-i}{p^2 - m_W^2} \left[ g_{\mu \nu} - (1- \xi _W ) \frac{p_\mu p_\nu }{p^2 - \xi _W m_W^2 } \right]
\qquad \mbox{and} \qquad
D (p) \equiv \frac{i}{p^2 - \xi _W m_W^2 } \, 
\end{alignat}
are included into the pinch contributions by
applying the identities \cite{Papavassiliou:1994pr}
\begin{align}
	g_{\nu \alpha } &= i \left\{ \Delta _{\nu \mu} (p) \left[ (p^2 - m_W^2) g^\mu _\alpha - p^\mu p_\alpha \right] - p_\nu p_\alpha D(p) \right\} \\
	ip_\mu &= p^2 D(p^2) p_\mu + m_W^2 p^\nu \Delta _{\nu \mu} ~. \label{eq:PTIdentitiyTrigger}
\end{align}
Due to these identities, the pinch contributions of the charged sector
have to be correctly assigned to all possible self-energy
combinations of $H^\pm$, $G^\pm$ and $W^\pm$. Consequently, the
analysis of the charged sector is significantly more involved than the
one of the CP-even sector. Taking into account all Feynman diagrams
contributing to the pinched self-energies of the charged
sector\footnote{These diagrams are obtained 
  analogously to the CP-even case. Since they are numerous, we show 
  exemplary only those for the CP-even sector.}, the collocation of
all pinch contributions for the 
various combinations of $W^\pm$ and $G^\pm$ yields 
\begin{align}
	i\Sigma ^\text{PT} _{WW, \mu \nu} (p^2) &= (1-\xi _W ) \frac{ig^2 m_W^2  }{64\pi ^2} p_\mu p_\nu  \sum _{H_i} \mathcal{O}^{(2)}_{H_iH_i} \left\{ \beta _{WH_i} (p^2) + 4 C_2 ^{WH_i} (p^2) \right\}  \\
	i\Sigma ^\text{PT} _{WG^\pm , \mu } (p^2) &= - (1-\xi _W ) \frac{ig^2 m_W  }{64\pi ^2} p_\mu \left\{ \alpha _W + \sum _{H_i} \mathcal{O}^{(2)}_{H_iH_i} \left[ m_{H_i}^2 \beta _{WH_i} (p^2) + 2 p^2 C_2 ^{WH_i} (p^2) \right]  \right\}  \\
	i\Sigma ^\text{PT} _{G^\pm G^\pm } (p^2) &= i\Sigma ^\text{add} _{G^\pm G^\pm } (p^2) \\
	&- (1-\xi _W ) \frac{ig^2 }{64\pi ^2} \left\{ -2 f_{G^\pm G^\pm } (p^2) \alpha _W + \sum _{H_i} \mathcal{O}^{(2)}_{H_iH_i} g_{G^\pm G^\pm} (p^2, m_{H_i}^2 ) \beta _{WH_i} (p^2) \right\}  \notag 
\end{align}
and for the combinations of $H^\pm$ and $G^\pm$ or $W^\pm$ results in
\begin{align}
	i\Sigma ^\text{PT} _{H^\pm H^\pm } (p^2) &= i\Sigma ^\text{add} _{H^\pm H^\pm } (p^2) \\
	& - (1-\xi _W ) \frac{ig^2}{64\pi ^2} \left\{ -2 f_{H^\pm
          H^\pm } (p^2) \alpha _W + g_{H^\pm H^\pm } (p^2, m_A^2 )
          \beta _{WA} \right. \notag \\
& \left. + \sum _{H_i} \mathcal{O}^{(2)}_{H_iH_i} g_{H^\pm H^\pm } (p^2, m_{H_i}^2 ) \beta _{WH_i} \right\}   \notag \\
	&- (1-\xi _Z ) \frac{ig^2 (\cos ^2 \Theta _W - \sin ^2 \Theta _W ) ^2 }{64\pi ^2 \cos ^2 \Theta _W } \left\{ - f_{H^\pm H^\pm } (p^2) \alpha _Z + g_{H^\pm H^\pm } (p^2, m_{H^+}^2 ) \beta _{ZH^\pm } (p^2) \right\} \notag \\
	&- (1 - \xi _\gamma ) \frac{ie^2}{16\pi ^2} \left\{ - f_{H^\pm
          H^\pm } (p^2) \alpha _\gamma + g_{H^\pm H^\pm } (p^2 ,
          m_{H^+}^2 ) \beta _{\gamma H^\pm } (p^2) \right\}  \notag
\end{align}
\begin{align}
	i\Sigma ^\text{PT} _{H^\pm G^\pm } (p^2) &= i\Sigma ^\text{add} _{H^\pm G^\pm } (p^2) \\
	&- (1-\xi _W ) \frac{ig^2}{64\pi ^2} \sum _{H_i} \mathcal{O}^{(3)}_{H_iH_i} g_{H^\pm  G^\pm } (p^2, m_{H_i}^2 ) \beta _{W H_i} (p^2)  \notag \\
	i\Sigma ^\text{PT} _{H^\pm W^\pm , \mu } (p^2) &= (1- \xi _W ) \frac{ig^2 m_W}{64\pi ^2} p_\mu \sum _{H_i} \mathcal{O}^{(3)}_{H_iH_i} \left\{ f_{H^\pm H^\pm } (m_{H_i}^2) \beta _{WH_i} (p^2) \right.\notag\\
& \left.+ 2f_{H^\pm H^\pm } (p^2) C_2 ^{WH_i} (p^2) \right\} ~.
\end{align}
By adding the pinch contributions to the gauge-dependent charged self-energies, the pinched self-energies of the charged sector read
\begin{align}
	\overline{\Sigma} _{WW, \mu \nu} (p^2) &= \left. \Sigma ^\text{tad} _{WW, \mu \nu} (p^2) \right| _{\xi_V = 1}  \\
	\overline{\Sigma} _{WG^\pm , \mu} (p^2) &= \left. \Sigma ^\text{tad} _{WG^\pm , \mu} (p^2) \right| _{\xi_V = 1}  \\
	\overline{\Sigma} _{G^\pm G^\pm } (p^2) &= \left. \Sigma ^\text{tad} _{G^\pm G^\pm } (p^2) \right| _{\xi_V = 1} + \Sigma ^\text{add} _{G^\pm G^\pm } (p^2)  \\
	\overline{\Sigma} _{H^\pm H^\pm } (p^2) &= \left. \Sigma ^\text{tad} _{H^\pm H^\pm } (p^2) \right| _{\xi_V = 1} + \Sigma ^\text{add} _{H^\pm H^\pm } (p^2)  \\
	\overline{\Sigma} _{H^\pm G^\pm } (p^2) &= \left. \Sigma ^\text{tad} _{H^\pm G^\pm } (p^2) \right| _{\xi_V = 1} + \Sigma ^\text{add} _{H^\pm G^\pm } (p^2)  \\
	\overline{\Sigma} _{WH^\pm , \mu } (p^2) &= \left. \Sigma ^\text{tad} _{WH^\pm , \mu } (p^2) \right| _{\xi_V = 1} ~.
\end{align}
The additional gauge-independent pinch contributions for $\Sigma
^\text{tad} _{H^\pm G^\pm } (p^2)$ are
stated\footnote{For the derivation of all additional pinch contributions we took into
  account all possible diagrams, not only the ones containing only the extended
  scalar sector of the N2HDM. This is consistent since the gauge
  dependence is cancelled already in the pinched
  self-energies.} in
Eq.~(\ref{eq:sigaddghpm}). The remaining additional contributions are
analogously derived from Feynman diagrams involving internal
scalar-scalar-vector vertices and explicitly read 
\begin{align}
	\Sigma ^\text{add} _{G^\pm G^\pm } (p^2) &= \frac{-g^2}{32\pi ^2} p^2 \bigg\{ B_0 (p^2; m_W^2 , m_W^2 ) + \sum _{H_i} \mathcal{O} ^{(2)} _{H_i H_i} B_0 (p^2; m_{H_i}^2 , m_W^2) \\
	&+ \frac{( \cos ^2 \Theta _W - \sin ^2 \Theta _W )^2 }{\cos ^2 \Theta _W } B_0 (p^2; m_W^2 ,m_Z^2) + 4 \sin ^2 \Theta _W B_0 (p^2; 0, m_{H^\pm }^2 ) \bigg\}  \\
	\Sigma ^\text{add} _{H^\pm H^\pm } (p^2) &= \frac{-g^2}{32\pi ^2} ( p^2 - m_{H^\pm }^2 ) \bigg\{ B_0 (p^2; m_A^2 , m_W^2 ) + \sum _{H_i} \mathcal{O} ^{(1)} _{H_i H_i} B_0 (p^2; m_{H_i}^2 , m_W^2) \\
	&+ \frac{( \cos ^2 \Theta _W - \sin ^2 \Theta _W )^2 }{\cos ^2 \Theta _W } B_0 (p^2; m_{H^\pm }^2 ,m_Z^2) + 4 \sin ^2 \Theta _W B_0 (p^2; 0, m_{H^\pm }^2 ) \bigg\} ~.
\end{align}
Note that self-energies involving the gauge boson $W^\pm$ as an
external particle do not receive additional gauge-independent pinch
contributions. 

\subsubsection{Pinch contributions for the CP-odd sector}
For pinching the CP-odd sector we choose the same process as for the
CP-even sector, \textit{i.e.}~the process $\mu ^+ \mu ^- \rightarrow
\bar{b} b$. The derivation of the pinch contributions is exactly
analogous to the CP-even neutral and to the charged
sector. By inserting the propagators and applying the elementary
identities Eq.~(\ref{eq:pinchWardIdent}) and
Eqs.~(\ref{eq:PTPropagatorInsert})-(\ref{eq:PTIdentitiyTrigger}), we
isolate all pinch contributions from the Feynman diagrams for the
corresponding CP-odd self-energies. In total, the 
pinch contributions read  
\begin{align}
i \Sigma ^\text{tad} _{AA} (p^2) &= \left. i \Sigma ^\text{add} _{AA} (p^2) \right| _{\xi_V = 1} \\
& - \frac{ig^2}{64 \pi ^2 \cos ^2 \Theta _W } (1 - \xi _Z ) \left\{ - f_{AA} (p^2) \alpha _Z + \sum _{H_i} \mathcal{O} ^{(1)} _{H_i H_i} g_{AA} (p^2 , m_{H_i}^2 ) \beta _{Z H_i} (p^2) \right\}  \notag \\
i \Sigma ^\text{tad} _{AG^0} (p^2) &= \left. i \Sigma ^\text{add} _{AG^0} (p^2) \right| _{\xi_V = 1} \\
& - \frac{ig^2}{64 \pi ^2 \cos ^2 \Theta _W } (1 - \xi _Z ) \sum _{H_i} \mathcal{O} ^{(3)} _{H_i H_i} g_{AG^0 } (p^2 , m_{H_i}^2 ) \beta _{Z H_i} (p^2) ~.  \notag
\end{align}
Adding the pinch contributions to the gauge-dependent self-energies
allows for the generation of the
pinched self-energies of the CP-odd sector: 
\begin{align}
	\overline{\Sigma} _{AA } (p^2) &= \left. \Sigma ^\text{tad} _{AA } (p^2) \right| _{\xi_V = 1} + \Sigma ^\text{add} _{AA } (p^2)  \\
	\overline{\Sigma} _{AG^0 } (p^2) &= \left. \Sigma ^\text{tad} _{AG^0 } (p^2) \right| _{\xi_V = 1} + \Sigma ^\text{add} _{AG^0 } (p^2) ~.
\end{align}
The additional pinch contribution for the self-energy $\Sigma
^\text{tad} _{AG^0 } (p^2)$ is given in Eq.~(\ref{eq:sigaddga}), and
the remaining additional gauge-independent pinch contribution explicitly
reads  
\begin{align}
	\Sigma ^\text{add} _{AA } (p^2) &= \frac{-g^2}{32\pi ^2 \cos ^2 \Theta _W } (p^2 - m_A^2 ) \bigg\{ 2 \cos ^2 \Theta _W B_0 (p^2; m_W^2 , m_{H^\pm }^2 ) + \sum _{H_i} \mathcal{O} ^{(1)} _{H_i H_i} B_0 (p^2; m_{H_i}^2 , m_Z^2) \bigg\} 
\end{align}

\end{appendix}
\vspace*{1cm}
\bibliographystyle{h-physrev}

\end{document}